\documentclass[longauth]{aa}  

\newcommand{\revise}[1]{#1}
\newcommand{\revisem}[1]{#1}

\usepackage{graphicx}
\usepackage[varg]{txfonts}
\usepackage[T1]{fontenc}
\usepackage{aas_macros}
\usepackage{amsmath}
\usepackage{amssymb}
\usepackage{url}
\usepackage{natbib}
\usepackage{subeqnarray}
\usepackage{color}
\usepackage{courier}
\usepackage{widetext}
\definecolor{linkcolor}{rgb}{0.6,0,0}
\definecolor{citecolor}{rgb}{0,0,0.75}
\definecolor{urlcolor}{rgb}{0.12,0.46,0.7}

\def\ie{\emph{i.e.}}
\def\eg{\emph{e.g.}}

\def\T{\emph{(Top)}}
\def\C{\emph{(Center)}}
\def\B{\emph{(Bottom)}}
\def\L{\emph{(Left)}}
\def\M{\emph{(Middle)}}
\def\R{\emph{(Right)}}

\newlength{\thsize}
\newlength{\hhsize}
\newlength{\khsize}
\newlength{\qhsize}
\def\nHinit{n_{\rm H,0}}
\def\nH{n_{\rm H}}
\def\ni{n_{\rm i}}
\def\Gi{\Gamma_{\rm i}}

\def\mi{m_{\rm i}}
\def\ne{n_{\rm e}}
\def\me{m_{\rm e}}

\def\Ge{\Gamma_{\rm e}}

\def\tff{\tau_{\rm ff}}
\def\cs{c_{\rm s}}
\def\vth{v_{\rm th}}
\def\kB{k_{\rm B}}

\def\cmc{{\rm cm}^{-3}}
\def\s{\,s$^{-1}$}

\def\Vlim{\mathcal{V}}
\def\Vlimcoef{V_{0.1\,\mu{\rm m}}}

\def\Vsys{\mu}

\def\am{a_{-}}
\def\ap{a_{+}}

\def\mm{m_{-}}
\def\mp{m_{+}}
\def\mkm{{m_{k-}}}
\def\mkp{{m_{k+}}}

\def\sigmak{\revisem{\hat{\sigma}_k}}
\def\crosssection{\revisem{\hat{\sigma}}}
\def\mk{m_k}
\def\nk{n_k}
\def\mP{m_{\rm P}}
\def\mT{m_{\rm T}}

\def\nP{n_{\rm P}}
\def\nT{n_{\rm T}}

\def\aP{a_{\rm P}}
\def\aT{a_{\rm T}}

\def\mPm{m_{\rm P-}}
\def\mTm{m_{\rm T-}}

\def\mPp{m_{\rm P+}}
\def\mTp{m_{\rm T+}}
\def\mCp{m_{\rm C+}}

\def\DVTP{\Delta V_{\rm P,T}}
\def\coll{C}

\def\rhok{\rho_k}
\def\rhoP{\rho_{\rm P}}
\def\rhoT{\rho_{\rm T}}
\def\rhoC{\rho_{\rm C}}

\def\Nbins{N_{\rm bins}}
\def\Nspec{N_{\rm species}}

\def\betak{{\beta_k}}
\def\betaP{{\beta_{\rm P}}}
\def\betaT{{\beta_{\rm T}}}

\def\nsgasSECPHO{\langle n \hat{\sigma}\rangle_{\rm gas}^{\rm secpho}}
\def\nsdustSECPHO{\langle n \hat{\sigma}\rangle_{\rm dust}^{\rm secpho}}
\def\rgamma{{{\gamma_j}}}
\def\ralpha{{{\alpha_j}}}

\def\qabs{Q_{\rm abs}}

\def\EspZ{\langle Z \rangle}
\def\crate{\left(\frac{\d n}{\d t}\right)_{\rm P,T}}

\def\taudrag{\tau_{\rm drag}}
\def\taugyr{\tau_{\rm gyr}}
\def\taus{\tau_\eta}
\def\tauL{\tau_{\rm L}}
\def\tauT{\tau_{\rm T}}
\def\tauP{\tau_{\rm P}}
\def\Reynolds{{\rm Re}}

\newcommand{\integer}{\mathrm{int}\,}
\newcommand{\erf}{\mathrm{erf}\,}

\def\DUSTDAP{\texttt{DUSTDaP}}

\def\d{{\rm d}}

\def\DV{\Delta V}
\def\exp{{\rm e}}

\def\Bvec{\mathbf{B}}
\def\RC{R_{\rm C}}
\def\xaxis{\hat{x}}
\def\yaxis{\hat{y}}
\def\zaxis{\hat{z}}
\def\ey{\mathbf{e_\phi}}
\def\ez{\mathbf{e_r}}
\def\ex{\mathbf{b}}

\def\alphan{\alpha_{\rm n}}
\def\Vn{\mathbf{V}_{\rm n}}
\def\Vi{\mathbf{V}_{\rm i}}
\def\muN{\mu_{\rm n}}
\def\muI{\mu_{\rm i}}
\def\mud{\mu_{\rm dust}}
\def\rhoN{\rho_{\rm n}}

\def\vAD{V_{\rm AD}}
\def\VAD{\mathbf{V}_{\rm AD}}
\def\vADref{V_{\rm AD}^{\rm ref}}
\def\sigI{\hat{\sigma}_{\rm in}}

\def\Vdrift{\mathbf{v}}
\def\BrotB{\Bvec\times\left(\mathbf{\nabla}\times\Bvec\right)}

\def\Ggrav{\mathcal{G}}

\def\rO{\eta_\Omega}
\def\rH{\eta_{\rm H}}
\def\rAD{\eta_{\rm AD}}

\def\modelTurbOnly{G1_coMRN33_b0p3_1ff_CAMN_Turbx1}
\def\modelADOnly{G1_coMRN33_b0p3_1ff_FCAMN_noTurb_VAD2e3dyn}
\def\modelTurbAD{G1_coMRN33_b0p3_1ff_FCAMN_Turbx1_VAD2e3dyn}
\def\nodust{G1_std_nodust}
\def\truncmodelnoevol{G1_cotruncMRN_b0p3_1ff_noevol}
\def\stdmodelnoevol{G1_coMRN33_b0p3_1ff_noevol}

\begin{document}

\title{Dust coagulation feedback on  magnetohydrodynamic resistivities in protostellar collapse
}


   \author{V. Guillet\inst{1,2},
	P. Hennebelle\inst{3,4},
	G. Pineau des For\^ets\inst{1,5},
	A. Marcowith\inst{2},
	B. Commer\c{c}on\inst{6}, 
	P. Marchand\inst{7}
          }


   \institute{
1. Universit\'{e} Paris-Saclay, CNRS,  Institut d'astrophysique spatiale, 91405, Orsay, France \\
2. Laboratoire Univers et Particules de Montpellier, Universit\'{e} de Montpellier, CNRS/IN2P3, CC 72, Place Eug\`{e}ne Bataillon, 34095 Montpellier Cedex 5, France   \\
3. Universit\'{e} Paris-Diderot, AIM, Sorbonne Paris Cit\'{e}, CEA, CNRS, F-91191, Gif-sur-Yvette, France \\
4. Laboratoire de Physique de l’Ecole normale sup\'{e}rieure, ENS, Universit\'{e} PSL, CNRS, Sorbonne Universit\'{e}, Universit\'{e} de Paris, F-75005 Paris, France \\
5. Observatoire de Paris, PSL University, Sorbonne Universit\'{e}, LERMA, 75014 Paris, France \\
6. Universit\'{e} Lyon I, 46 All\'{e}e d'Italie, Ecole Normale Sup\'{e}rieure de Lyon, Lyon, Cedex 07, F-69364 France \\
7. American Museum of Natural History, CPW at 79th, 10024 NY, United-States \\
   }

   \date{}

 
  \abstract
{The degree of coupling between the gas and the magnetic field during the collapse of a core and the subsequent formation of a disk depends on the assumed dust size distribution.}
{We study the impact of grain-grain coagulation on the evolution of magnetohydrodynamic (MHD) resistivities during the collapse of a prestellar core.}
{We use a 1-D model to follow the evolution of the dust size distribution, out-of-equilibrium ionization state and gas chemistry during the collapse of a prestellar core. To compute the grain-grain collisional rate, we consider models for both random and systematic, size-dependent, velocities. We include grain growth through grain-grain coagulation and ice accretion, but ignore grain fragmentation.}
{\revise{Starting with a MRN \citep{MRN77} size distribution, we find that coagulation in grain-grain collisions generated by hydrodynamical turbulence is not efficient at removing the smallest grains, and as a consequence does not affect much the evolution of the Hall and ambipolar diffusion MHD resistivities which still severly drop during the collapse like in models without coagulation. The inclusion of systematic velocities, possibly induced by the presence of ambipolar diffusion, increases the coagulation rate between small and large grains, removing small grains earlier in the collapse and therefore limiting the drop in the Hall and  ambipolar diffusion resistivities. At intermediate densities ($\nH \sim 10^8\,\cmc$), the Hall and ambipolar diffusion resistivities are found to be higher by 1 to 2 orders of magnitude in models with coagulation than in models where coagulation is ignored, and also higher than in a toy model without coagulation where all grains smaller than $0.1\,\mu$m would have been removed in the parent cloud before the collapse.}
}
{\revise{When grain drift velocities induced by ambipolar diffusion are included, dust coagulation happening during the collapse of a prestellar core starting from an initial MRN dust size distribution appears to be efficient enough to increase the MHD resistivities to the values necessary to 
strongly modify the magnetically-regulated formation of planet-forming disk. A consistent treatement of the competition between fragmentation and coagulation is however necessary before reaching firm conclusions.}
}

\authorrunning{V. Guillet et al.}

\titlerunning{}

 \maketitle

%



\section{Introduction}


While it is now well established that stars form through the collapse of prestellar cores, understanding the exact outcome of this process remains 
a challenge \citep{li2014}. In particular, many studies found that the properties of the centrifugally supported disks that form around the protostars sensitively depend on the intensity of the magnetic field and even possibly on its orientation \citep{allen2003,mellon2008,hf2008,joos2012,li2013,gray2018}. This is due to the magnetic braking that can efficiently transport angular momentum from the inner part of the collapsing cloud to the surrounding envelope. In the most extreme case, it has even been found that the formation of a centrifugally supported disk can be entirely suppressed, a process known as catastrophic magnetic braking \citep{allen2003,mellon2008,hf2008}. 

The magnetic field evolution is a direct consequence of its coupling with the gas. If this coupling is perfect (ideal Magnetohydrodynamics), 
then magnetic intensity is typically expected to be $\propto \rho^\kappa$ where 
$\rho$ is the gas density and $\kappa \simeq 1/2-2/3$. On the other hand, in the extreme case where the magnetic field would be completely decoupled from the gas, the magnetic intensity would stay constant and the magnetic field would
have no or a much more limited influence on the gas evolution. It is therefore fundamental to understand with enough accuracy how magnetic field and gas are coupled together. Since the gas within molecular clouds is weakly ionised, with a ionisation fraction in the order of, or even below $10^{-7}$ \citep[e.g.][]{shu1987}, the coupling between the gas, mainly the neutrals, and the magnetic field is imperfect. The neutrals can slip through the field lines which are attached to the ions, a process known as ambipolar diffusion. This latter process largely dominates over the other non-ideal Magnetohydrodynamic (MHD) processes in the ISM such as the Hall effect and the Ohmic resistivity, which are due to the imperfect coupling between the magnetic field and the ions and the electrons respectively.

In the context of dense prestellar cores, it is believed that dust grains are playing an important role. As the recombination rate of ions on the surface of grains scales with the density, the ionisation is several orders of magnitude lower in collapsing cores than in the rest of the ISM, and the impact of non-ideal MHD processes is more important.
Many detailed calculations of the ionisation inside dense cores have been performed \citep[e.g.][]{nakano2002,kunz2009,zhao2016,wurster2016,marchand2016,dzyurkevich2017}. The resulting resistivities depend on the exact assumptions regarding the ionisation rate, the chemistry network and the grain properties. In particular, it has been concluded that the abundance of small grains is particularly critical \citep{nishi91,zhao2016,dzyurkevich2017,zhao2018,grassi2019}. For example \citet{zhao2016} assuming a 
MRN size distribution \citep{MRN77} found that the ambipolar and Hall resistivities reach a maximum when only grains of size above $0.1\,\mu$m and $0.04\,\mu$m are considered, respectively. Moreover when varying the size of the smallest grains considered from 0.005 to $1\,\mu$m, \citet{zhao2016} found that the resistivities vary by one to two orders of magnitude (depending on the considered gas density). Therefore a sufficient knowledge of the grain distribution appears to be a crucial issue. Indeed several studies have investigated the impact that the resistivities have on circumstellar disks confirming that disks form more easily when the resistivity is high which happens when small grains are removed \citep[e.g.][]{krasnopolsky2011,zhao2016}. \citet{hennebelle2016} proposed an analytical model in which the size of the disk at the early stage is predicted to be $\propto \eta_{AD}^{2/9}$, where $\eta_{AD}$ is the ambipolar resistivity. The influence of the Hall resistivity has also been investigated in a series of papers \citep{krasnopolsky2011,wurster2016,tsukamoto2015a,koga2019}. Because of the quadratic dependence on the 
magnetic field of the Hall term, changing the sign of $B$  does not lead to a
 physically identical situation and one must distinguish between the parallel and anti-parallel configurations depending on the respective orientation of $B$ and the angular momentum. In the anti-parallel configuration the Hall effect tends to produce bigger disks while in the parallel one, it tends to produce smaller ones. 

Knowing the grain distribution within dense cores therefore appears to be a major issue. As recalled above, many authors have assumed an MRN type distribution, with a maximal grain size $\sim 0.25\,\mu$m. However this particular size distribution is only valid for the diffuse phase of the ISM. In denser gas, the phenomenon of coreshine in the NIR \citep{Pagani2010} and the observed increase of FIR and submm emissivity \citep[\eg][]{Stepnik2003,DelBurgo2005,Ysard2013} all point toward the coagulation of grains in the envelop of molecular clouds, affecting the grain structure and composition \citep{Jones2013,Koehler2015} and increasing the grain size significantly \citep[$\sim 1 \,\mu$m,][]{Steinacker2015} or only modestly \citep[$<0.5\,\mu$m,][]{Ysard2016}, depending on the dust models used.
This indicates that dust coagulation must be an efficient process and that, given the grain size-dependence of the resistivities inferred in previous studies, it may play an major role in the coalescence of the cloud. 

In this article, we make use of the Paris-Durham shock code \citep{FPdF03} amended for the calculation of dust charging, dynamics and evolution \citep{Guillet2007,Guillet2011}, to study the coagulation of dust grains in a collapsing core, as well as its feedback on the evolution of the Ohmic, Hall and ambipolar diffusion MHD resistivities of the medium.

The article is structured as follows. In Sect. \ref{sec:DustEvolutionModel}, we present our model for the grain dynamics and coagulation in a protostellar collapse. Section \ref{sec:Charge_Ionization_Resist} deals with the calculation of grain charge, ionization equilibrium and MHD resistivities. In Sect. \ref{sec:Results}, we present our results for a collapse without dust, a collapse with dust but no coagulation and a collapse with dust and coagulation. We discuss in Sect. \ref{sec:Discussion} some limitations in our study and recall our main results in Sect. \ref{sec:summary}.


\section{The dust evolution model}\label{sec:DustEvolutionModel}



The Paris-Durham shock code is a fortran code initially designed to solve the structure of stationnary shocks and predict the intensity of their emission lines \citep{FPdF85}. From the beginning, it included more than 100 chemical species and a large network of chemical reactions (> 1000). The impact of dust grains on the chemical composition and dynamics of the shock was later included \citep{FPdF03}, but only through a single, effective, grain fluid. In \cite{Flower2005}, the code was adapted to study grain-grain coagulation and its feedback on the ionization and depletion during the gravitational collapse of an elementary cell composed of gas and dust.


The Dust Dynamics and Processing code (\DUSTDAP) was developped from the Paris-Durham shock code to study the  detailed charge and dynamics of a full size distribution of dust grains in shocks \citep{Guillet2007}. 
It includes the physics of grain charging as well as grain-grain destruction, namely the shattering and vaporisation of grain cores, encountered in shocks propagating through dense clouds \citep{Guillet2009, Guillet2011}. 

We detail below how this code was amended to follow the evolution of the dust size distribution by grain-grain coagulation during the isothermal collapse of $1\,\cmc$ of gas.

\subsection{Initial conditions in the parent cloud}\label{sec:init}

Our gas-phase chemistry derived from \cite{FPdF85} and \cite{LeGal2014} includes 134 gas-phase species and $\sim700$ chemical reactions, with ion-neutral, neutral-neutral, and dissociative recombination reactions involving species containing H, He, C, N, O and S. The initial distribution of the elemental abundances across the gas phase and the solid phase is given in Table 1 of Flower et al. (2005) with the exception of S and Fe, whose fractional elemental abundances were $1.47\times10^{-5}$  \citep[][submitted]{HB2020} and $1.50\times10^{-8}$, respectively.

In dense clouds, grains are covered by icy mantles \citep{Hagen1983}. The thickness of this mantle is independent of the grain size and can be computed knowing the size distribution of grain cores \citep[see Appendix B from][]{Guillet2007}. For an MRN size distribution ($\alpha=-3.5$) with core radii ranging over $[5:250]$ nm, the mantle is 8.8 nm thick using the Table 2 from \cite{FPdF03} for the abundances of chemical species in icy mantles, assuming a specific density of 1\,g.cm$^{-3}$ for ices and 3\,g.cm$^{-3}$ for cores. The initial dust-to-gas mass ratio is initially of 0.9\%, and will increase up to 1.2\% through accretion during the collapse.

The initial proton density is $\nHinit = 10^4\,\cmc$ and the cosmic-ray ionization rate $\zeta=5\times10^{-17}\,$s$^{—1}$.
Under these conditions, we determine the steady-state abundances of the chemical species in the gas phase (neglecting any further depletion) which are then used as initial conditions of the collapsing core.


\subsection{Recipe for core collapse}\label{sec:collapse}


The free-fall timescale of a spherical cloud of uniform mass density $\rho_0$ is :
\begin{equation}\label{Eq-tff}
\tff = \sqrt{\frac{3\pi}{32 \Ggrav \rho_0}}\,, 
\end{equation}
where $\Ggrav$ is the gravitational constant.
For our initial proton density ($\nHinit = 10^4\,\cmc$) $\tff\simeq 430$ kyr. 
We will consider the collapse of a spherical core in a free-fall time.

If we state $x=R(t)/\RC$ the ratio of the core radius at time $t$ to its initial radius ($0 \le x < 1$), the dynamics of the free-fall follows the equation \citep{Flower2005}:
\begin{equation}\label{Eq-collapse}
\frac{\d x}{\d t} = -\frac{\pi}{2\revise{\tff}}\sqrt{\frac{1}{x}-1}\,.
\end{equation}
We must now make an assumption on the dependence of the gas density with $x$. In Appendix \ref{A-Larson}, we demonstrate that the assumption of an 
\revise{uniform compression of all fluids as per \cite{Flower2005}, \revise{meaning that the mass density $\rho(t) = \rho_0\,\left(\RC/R(t)\right)^3$} is a function of time only}, is a good approximation of what happens in the Larson compression scenario \citep{Larson1969} \revise{regarding the level of coagulation achieved  up to a given local density, even if it is a bad description for the density profile of the collapsing core expected at a given time}. Therefore, we state
\revise{
\begin{equation}\label{Eq-cons}
\frac{1}{\nH}\frac{\d \nH}{\d t}  =  
\frac{1}{\rho}\frac{\d\rho}{\d t} 
= -\frac{3}{x}\frac{\d x}{\d t} \,.
\end{equation}
}

\begin{figure}
\includegraphics[angle=-90,width=\hhsize]{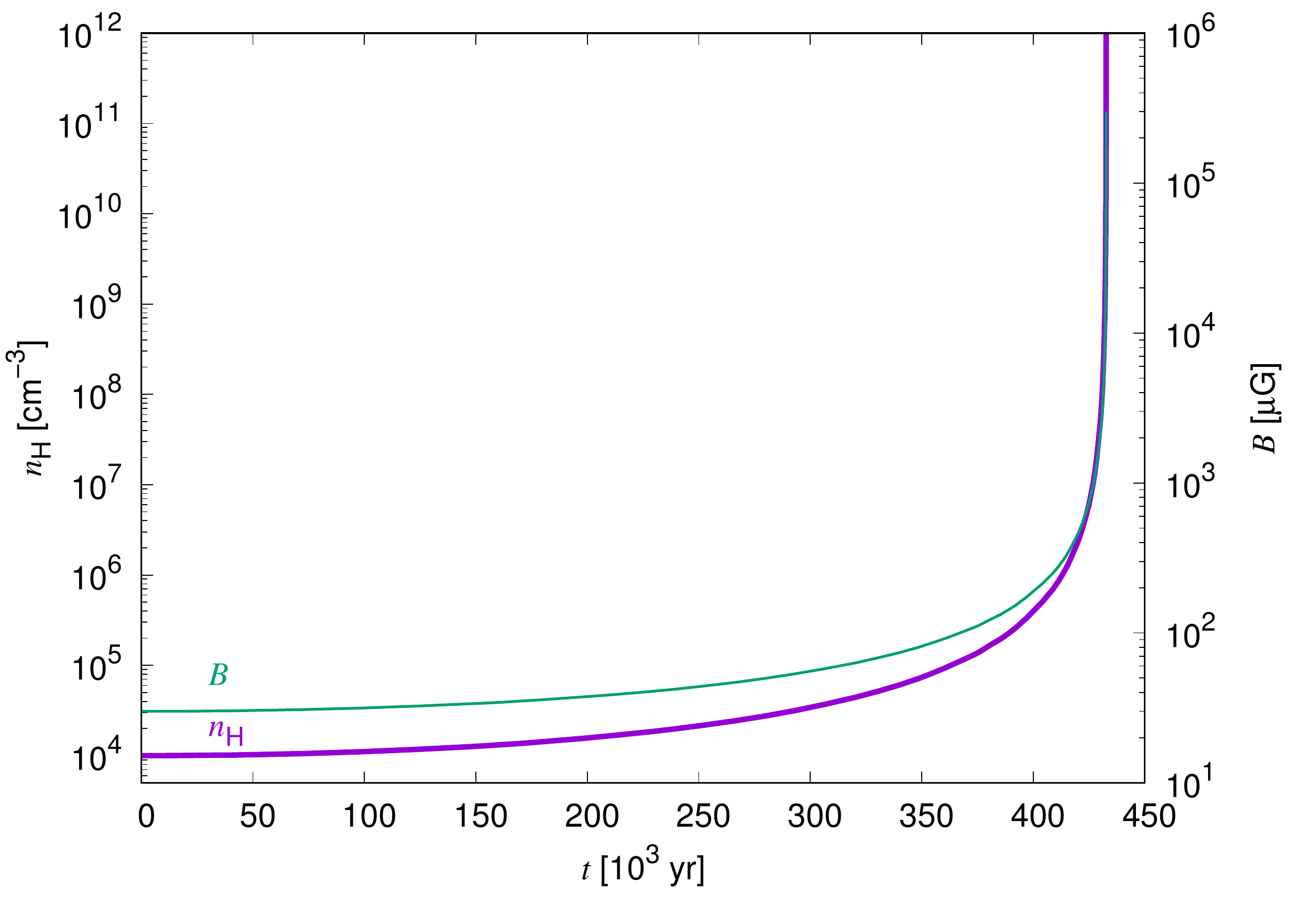}
\caption{\revise{Evolution of the cloud proton density $\nH$ and magnetic field intensity $B$ with time in our model. The collapse time is the free-fall time computed at the initial cloud density $\nHinit=10^4\,\cmc$: $\tff\simeq 430\times 10^3$ yr.}}
\label{fig:nH_t}
\end{figure}

Following \cite{marchand2016}, the magnetic field is assumed to be transverse to the direction of the collapse, and its intensity (in Gauss) is assumed to increase with the gas density during the collapse as follows 
\begin{equation}
B({\rm G}) = b \sqrt{\nH\,/\nHinit}\,,
\label{eq:Bscaling}
\end{equation}
with $b=30\,\mu$G taken as a reference value at the initial density $\nHinit$. This corresponds to a mass-to-flux ratio of 2.3 \citep{li2011}. \revise{For clarity, the evolution with time of the cloud density and magnetic field intensity are plotted in Fig.~\ref{fig:nH_t}.}


\subsection{Grain-grain coagulation}\label{sec:Smol}

In order to study the evolution of the grain size distribution during cloud core collapse, we have updated the \DUSTDAP\ code to include the coagulation of grains in low velocity grain-grain collisions. 

The evolution of the grain size distribution by grain-grain coagulation is controlled by the Smoluchowski equation \citep{Smol1916,Mizuno1988}
\begin{eqnarray}
\frac{\d\rho(m,t)}{\d t} &  =  & - \int_{0}^{\infty}\,m\,K(m,m')\,n(m,t)\,n(m',t) \,\d m' \nonumber \\ & + & \frac{1}{2}\int_0^m m\,K(m'-m,m')\,n(m'-m,t)\,n(m',t) \,\d m', \nonumber \\
\label{eq:coagkernel}
\end{eqnarray}
where $m$ is the grain mass, $n(m,t)$ (resp. $\rho(m,t)$) the density (resp. mass density) of grains of mass $m$ at time $t$, and $K$ ($\cmc$ s$^{-1}$) the collision kernel between grains of mass $m$ and $m'$.

Our numerical model for grain-grain coagulation is rudimentary: grains are assumed to be compact and spherical, and the product of grain-grain coagulation as well \citep{Hirashita2012}. 
The dust size distribution is modeled by $\Nbins$ logarithmic bins \citep{Guillet2007}.  For the code to run fast, the number of bins can not exceed a few tens because of the numerous integration variables associated to the grain charge distribution for each bin. In Appendix \ref{A-smol}, we detail the implementation and tests of our coagulation algorithm, and demonstrate that the convergence of the algorithm is obtained as soon as there is more than $\sim 7-8$ logarithmic bins per decade in grain radius, or equivalently 13 for the MRN 
size distribution.

\subsection{Handling of ice mantles covering grain cores}\label{sec:mantles}

The presence of icy mantles on the surface of grains is very important for grain-grain coagulation. Firstly, ice mantles greatly improve the sticking probability in grain-grain collisions \citep{C93}. Secondly, this mantle significantly enhances the radius of the smaller grains \citep{Guillet2007}, and therefore their coagulation rate.

Coagulation will modify the uniformity of mantle thickness by transferring the large volume of ices carried by small grains to the larger grains. The mass of icy mantles in each bin is therefore a variable that must be integrated.
The transfer of the mass of icy mantles through coagulation is handled together with the transfer of the mass of grain cores. At each step, knowing the mass of icy mantles and the properties of grain cores (average radius, cross-section and mass, number density) in each bin, we are able to compute the mantle thickness of grains using the equation detailed in Appendix B from \cite{Guillet2007}, and derive the effective radius, cross-section and mass of the core-mantle grains in each size bin.

Once the collapse has started, ices continue to accrete onto the surface of grains, increasing the mantle thickness uniformly. Note that mantle accretion can not  modify the number of grains.

\subsection{Velocity limit for grain-grain coagulation}\label{sec:Vlimit}

Laboratory experiments \citep{PB97} and theoretical studies \citep{C93} lead to a first conclusion that grain coagulate if their relative velocity does not exceed a velocity limit, and bounce on each other otherwise.
This velocity limit for coagulation is a function of the grain size \citep{C93}
\begin{equation}
\Vlim = \Vlimcoef \left(\frac{\hat{a}}{0.1\,\mu{\rm m}}\right)^{-5/6},
\label{eq:Vlim}
\end{equation}
where $\hat{a} = (a_1\times a_2)/(a_1+a_2)$ is the reduced grain radius of the two colliding grains, and $\Vlimcoef$ the velocity limit for two grains of 0.1\,$\mu$m, which lies between 0.1 and 0.4 km.s$^{-1}$ \citep{C93,PB97}. We will take the upper value $\Vlimcoef = 0.4$\,km.s$^{-1}$ as per \cite{PB97}, to take into account the increase of the velocity limit due to the presence of icy mantles on the surface of grain cores. 

\subsection{Grain dynamics}\label{sec:dynamics}

Grain dynamics is the fundamental ingredient of grain growth as it controls the collisional rates between grains \citep[\eg][]{Ossenkopf93}. In the context of core collapse, grain velocity dispersion can originate from the thermal agitation of the grain, from the acceleration of grains by HD or MHD turbulence, from the gravitational force, or from the electromagnetic force.

\subsubsection{Grain acceleration by Hydrodynamic turbulence}\label{sec:turbHD}


Since the pioneering work by \cite{Volk80}, the acceleration of dust grains by HD turbulence has been modeled by the stochastic acceleration of grains in a Kolmogorov cascade of turbulent eddies. In this framework, grains are accelerated by the eddies to which they can couple. 

We use the model of grain acceleration by HD turbulence, as revised by \cite{OC07}. The injection scale of the turbulence is assumed to be equal to the Jeans Length, and the injection velocity to the isothermal sound speed $\cs=\sqrt{\kB T/\mu}$, with $\mu$ the mean mass of gas particles. This defines an upper timescale for the turbulence cascade
\begin{equation}
\tauL=\frac{L_{\rm Jeans}}{\cs}=\frac{1}{2}\sqrt{\frac{\pi}{\Ggrav\rho}}.
\end{equation}
The dissipation scale of the turbulence is controlled by the Reynolds number \citep{Ormel2009}
\begin{equation}
\Reynolds = 6.2\times10^7 \sqrt{\frac{\rho/\mu}{10^5 \,\cmc}},
\end{equation}
corresponding to a timescale $\taus = \tauL / \sqrt{\Reynolds}$.

We chose not to use the model of grain acceleration by MHD  turbulence of \cite{YLD04} because the relations established in this article are mostly adapted to diffuse and moderately dense medium ($\nH \le 10^4\,\cmc$) and not the high densities encountered in our numerical collapsing core ($10^4 \le \nH \le 10^{12}\,\cmc$). In Sect. \ref{sec:Discussion}, we discuss the limitations of the HD models of grain dynamics and how the use of a model of grain acceleration by MHD turbulence would affect our conclusions.

In the theory of grain acceleration by hydrodynamical turbulence, the grain dynamic is controlled by the stopping time $\taudrag$ of the grain by collisions with gas particles. Large grains tend to couple to the large eddies, thereby acquiring strong kicks that accelerate them through a random walk in the velocity space. In the Epstein regime 
valid for sub-sonic grain velocities, the grain stopping time through collisions with the gas is 
\revise{\begin{equation}
\tau=\frac{3m}{4\,\rho\,\vth\,\crosssection}\propto a\,,
\label{Eq:taudrag}
\end{equation}}
where \revise{$a$, $\crosssection$ and $m$ are the grain radius, cross-section and mass}, respectively, $\rho$ is the mass density of the gas and $\vth=\sqrt{8\kB T/\pi\mu}$ the mean velocity of gas particles\footnote{Note the error in the formula for $\taudrag$ in \cite{Ormel2009} Eq. (A.1).}. 

Let us now consider in detail the expressions for the relative velocity between grains of different sizes. 
We call target (stopping time $\tauT$) the larger grain, and projectile (stopping time $\tauP$) the smaller grain of the two. 
Three different regimes can be distinguished for the rms relative velocity $\DVTP$ between the projectile and the target \citep{OC07}: 1) a regime of tight coupling where the stopping timescale of the target $\tauT$ is smaller than the dissipation timescale of the turbulence $\taus$, \ie\ where the target is so small that it can not escape even from the smallest eddies; 2) a regime for heavy particles where the target is so large that its stopping timescale $\tauT$ is much longer than injection timescale of the turbulence $\tauL$, thereby reducing the efficiency of the kicks by the largest eddies as the grain size increase; and 3) an intermediate regime where grains are optimally accelerated by an eddy of a particular size. Regarding the grains sizes involved in our study, small grains are in the tightly coupled particles regime, while large grains are in the intermediate regime. No grains from our size distributions fall into the heavy particles regime at any density as the tightly coupled and intermediate regimes becomes more and more important as the density increases. 
The level of turbulence, as well as the threshold between these two regimes, depends on the gas density through the stopping time (Eq.~\eqref{Eq:taudrag}). For clarity, we recall here the expressions for the mean quadratic relative velocity between the projectile and the target for the two first regimes \citep{OC07}:
\paragraph{Tightly coupled particles ($\tauT < \taus$)}
\begin{equation}
\DVTP^2 = \frac{3}{2}\cs^2 \sqrt{\Reynolds} \left(\frac{\tauT}{\tauL}-\frac{\tauP}{\tauL}\right)^2
\end{equation}
\paragraph{Intermediate regime ($\taus \le \tauT < \tauL$)}
\begin{eqnarray}
\DVTP^2 & = &\frac{3}{2}\cs^2 f\left(\frac{\tauP}{\tauL}\right)\frac{\tauT}{\tauL} \\
f(x)&  = & 3.2-(1+x)+\frac{2}{1+x}\left(\frac{1}{2.6}+\frac{x^3}{1.6+x}\right)
\end{eqnarray}
with $0 \le \tauP/\tauL \le 1$ and $1.97 \le f\le 2.97$ \citep{Ormel2009}.


\begin{figure}
\includegraphics[angle=-90,width=\hhsize]{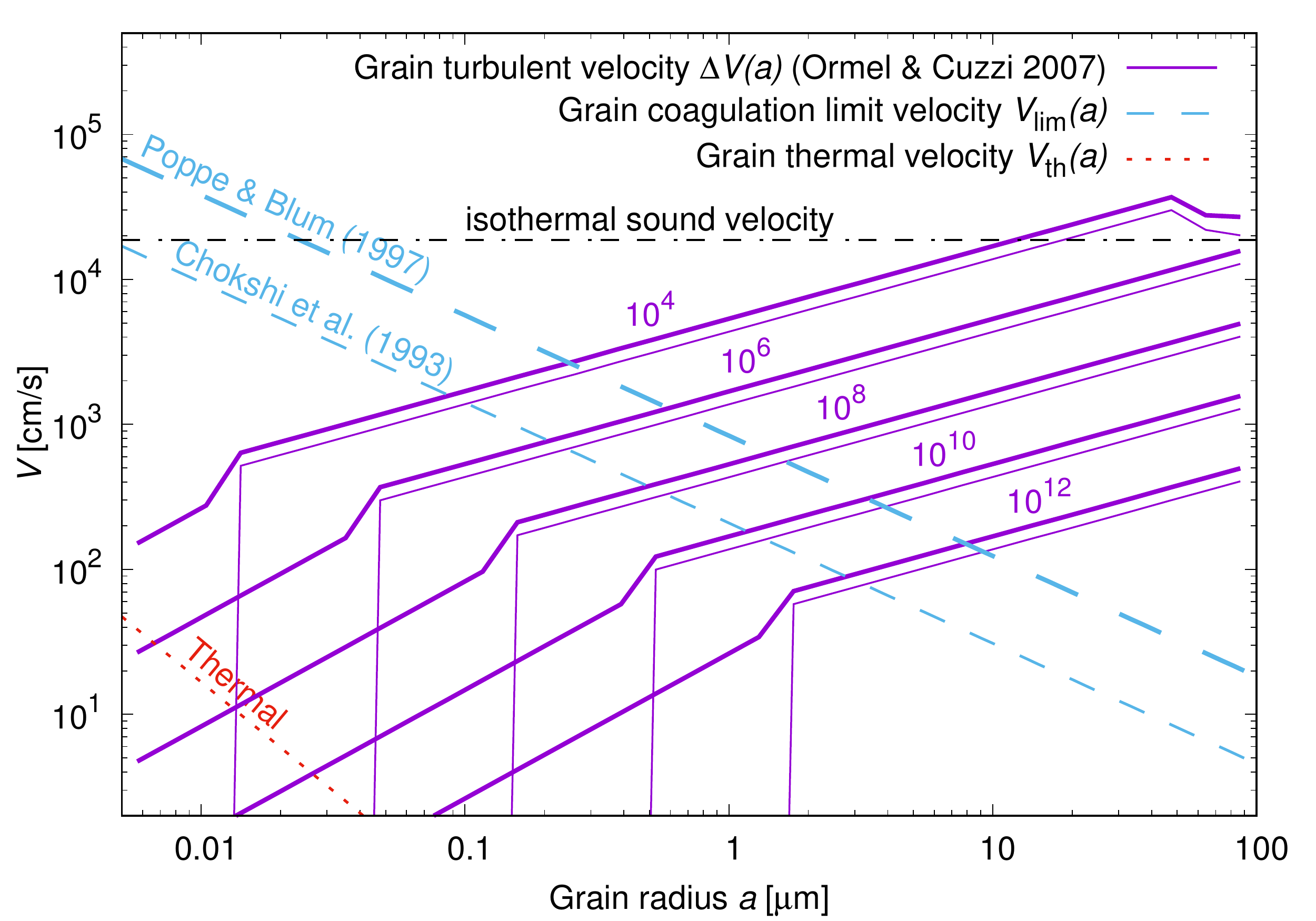}
\caption{Grain velocities as a function of the grain radius, for an homogeneous grain of specific density 3 g\,$\cmc$. The model of grain acceleration by turbulence \cite{OC07} is presented at different densities. Thick lines represent the relative rms velocity between the grain and the gas, while thin lines represent the relative rms velocity between grains of the same size. The contribution of Brownian motion to the velocity of grains is almost always negligible. The velocity limit for coagulation is presented in two versions : theory from \cite{C93} and experiments from \cite{PB97}. The sound velocity ($T = 10\,$K) in this isothermal phase is added for reference.}
\label{Vdist-Turb}
\end{figure}

Figure ~\ref{Vdist-Turb} shows how the rms relative velocity between grains depends on the grain radius, for increasing values of the gas proton density. The grain rms velocity increases with the grain radius and decreases with the gas density, and is in most cases strongly subsonic. If the projectile and target are of the same size, they can still present a relative velocity in the intermediate regime (thick lines), but not in the tightly coupled regime (thin lines).
A target and a projectile hitting each other with a relative velocity larger than velocity limit for coagulation (dashed blue lines) would not stick together but rather rebounce, the presence of ice coating  increasing this limit velocity (See Sect. \ref{sec:Vlimit}). The relative velocities generated by the thermal motion of grains (supposedly at the gas temperature $T=10\,$K) decrease strongly with the grain size: $v_{\rm thermal} \simeq 0.1\,(a/0.1\,\mu{\rm m})^{-3/2}$ cm.s$^{-1}$, for a grain specific density of 3\,g.cm$^{-3}$. They are small compared to the relative velocities generated by the gas turbulence, except at high densities. 

\subsubsection{Grain drift velocity through ambipolar diffusion}\label{sec:Vdrift}

Turbulence is not the only way to accelerate or decouple the different grain sizes from each other in dense clouds. Ambipolar diffusion, which appears as soon as the ionisation is too weak to couple the gas to the magnetic field, can also affect the grain dynamics, as it was demonstrated in the case of MHD C-shocks \citep{Guillet2007}. 

Grains are charged in the ISM, and as a consequence gyrate around magnetic fields \citep{Spitzer1941}. In dense clouds, their charge is usually negative and close to $-e$ \citep{FPdF03}, with $e$ the elementary electric charge. Grains are also coupled to the gas through the impact of gas particles. The Hall factor $\Gamma = \omega\,\tau$, defined as the product of the grain stopping time $\tau$ by the grain gyrofrequency $\omega$, characterizes the degree of coupling of the grain with the magnetic field and the gas. 
The Hall factor depends strongly on the grain size \citep[$\Gamma \propto a^{-2}$,][]{Guillet2007}. Small dust grains (those with $\Gamma \gg 1$) 
, are strongly coupled to the magnetic field and follow on average the dynamics of ions, spiraling around magnetic field lines and participating with ions to the coupling of the gas with the magnetic field through collisions with gas particles. Large grains (those with $\Gamma \ll 1$) 
follow the gas on average, being almost insensitive to the magnetic field despite their electric charge.

In a model where the magnetic field direction (along $\ex$) is transverse to the direction $\ez$ of the collapse, and ignoring the inertia of dust grains, we can express the drift velocity $\Vdrift_k$ of each grain size $k$ as a function of the ambipolar diffusion velocity \revise{$\VAD\equiv \Vn - \Vi$} \citep[Eq.~(21) from][]{Guillet2007} 
\begin{equation}
\Vdrift_k  = \Vn -\frac{\Gamma_k^2}{1+\Gamma_k^2}\,\VAD+\frac{\Gamma_k}{1+\Gamma_k^2}\,\VAD\times\ex \,,
\label{eq:Vdrift}
\end{equation}
where $\Vi$ is the velocity of ions and $\Vn$ that of neutrals along the direction $\ez$ of the collapse.
Note that the supplementary relative velocities generated by the turbulence between dust grains and gas particles were ignored in this derivation.



\begin{figure*}
\includegraphics[angle=-90,width=\thsize]{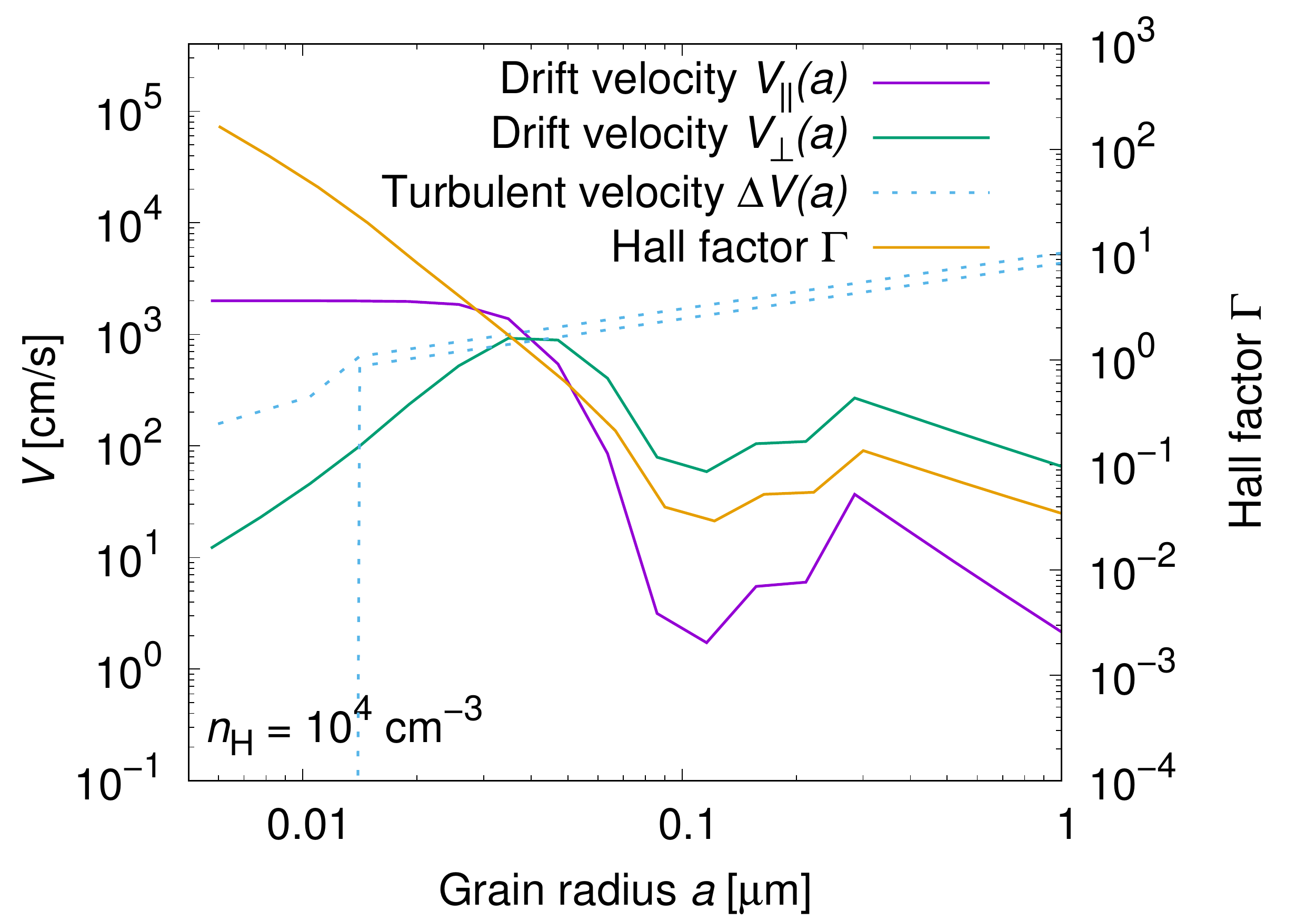}
\includegraphics[angle=-90,width=\thsize]{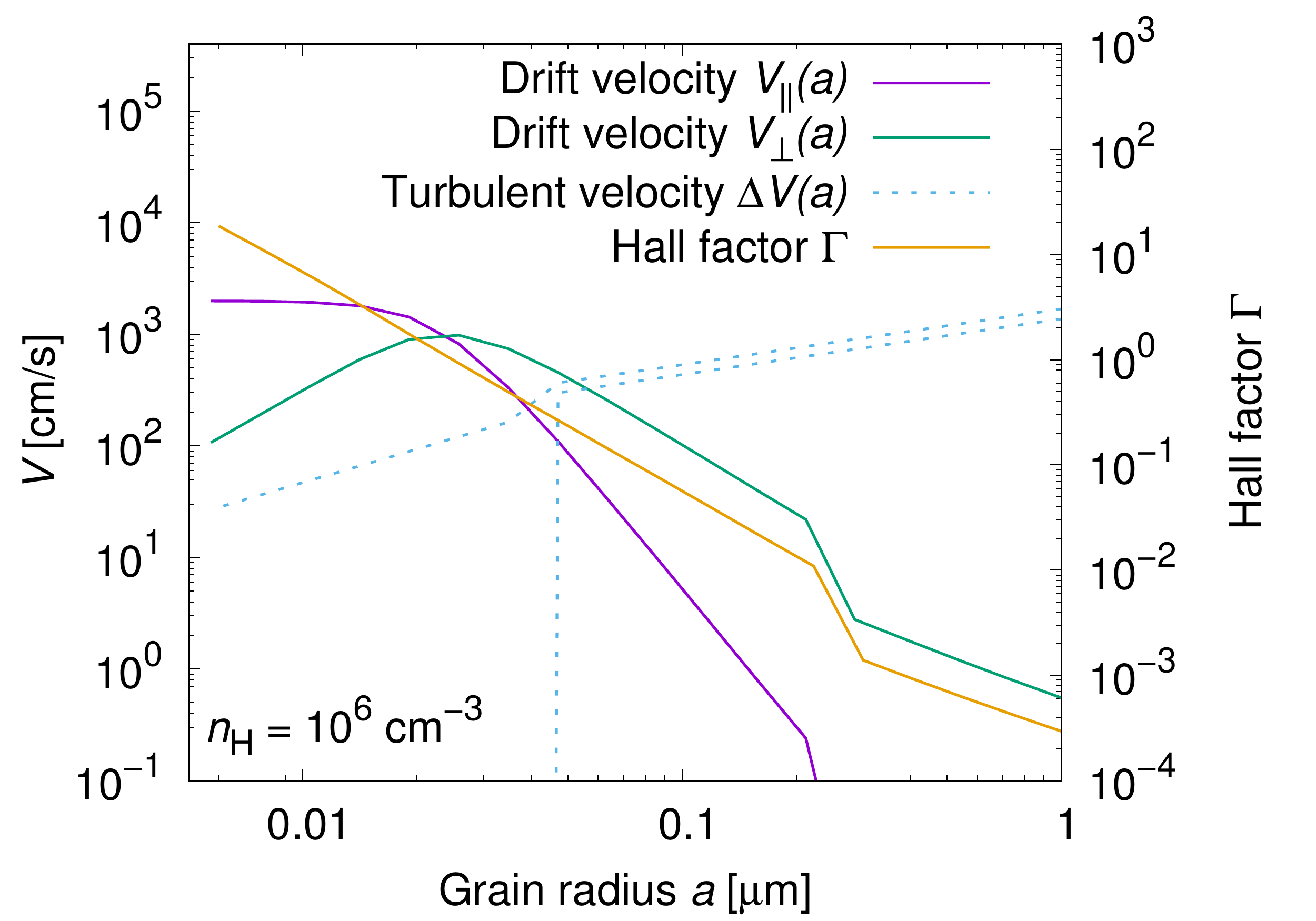}
\includegraphics[angle=-90,width=\thsize]{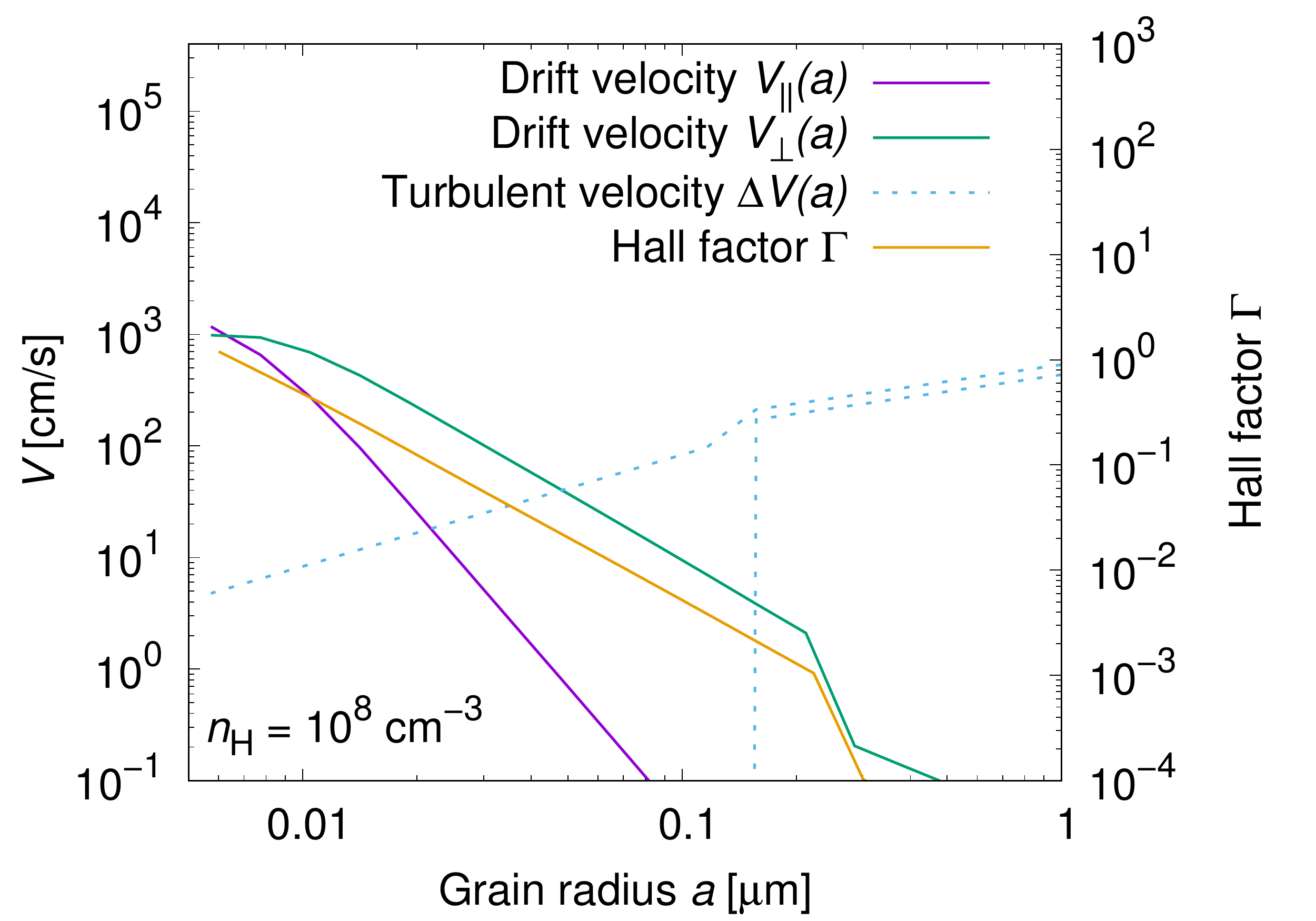}
\caption{\revise{Grain velocities \emph{(left axis)} and Hall factor \emph{(right axis)}, as a function of the grain radius for a gas proton density $\nH$ of $10^4\,\cmc$ \L, $10^6\,\cmc$ \M\ and $10^8\,\cmc$ \R. The ambipolar diffusion velocity $\vAD$ is fixed and equal to $\cs/10 = 20\,$m.s$^{-1}$. We present the components $V_\parallel$ parallel to the infall  (magenta) and $V_\perp$  perpendicular to the infall and to the magnetic field direction (green). For comparison, the turbulent relative velocities $\Delta V$ from Fig.~\ref{Vdist-Turb} are overplotted in dotted blue lines. The grain specific density is 3 g\,$\cmc$.}}
\label{fig:Vdist-AD}
\end{figure*}

The grain drift velocity induced by ambipolar diffusion is not a stochastic velocity, but a systematic velocity. It has two components : one, $V_\parallel$, along the collapse direction $\ez$, and the other, $V_\perp$, along $\ey$ which is perpendicular to the direction of the collapse and to the magnetic field direction.  Figure ~\ref{fig:Vdist-AD} presents how the component $V_\parallel$ and $V_\perp$ of the grain drift velocity depend on the grain radius, for increasing values of the proton density $\nH$, assuming a constant ambipolar diffusion velocity $\vAD=\cs/10=20\,$m.s$^{-1}$, \revise{as per \cite{BM1995}}. 
The parallel and perpendicular components of the relative velocity between the grain and the gas do not scale the same way with the Hall factor $\Gamma$ of the grain (see Eq.~\eqref{eq:Vdrift}): $V_\parallel-\Vn$ is maximal and equal in amplitude to $\vAD$ when $\Gamma \gg 1$, while $V_\perp-\Vn$ is maximal and equal in amplitude to $\vAD/2$ when $\Gamma = 1$. As $\Gamma$ decreases with increasing density, the systematic velocities generated by ambipolar diffusion tend to disappear with the collapse. 

\revise{
\subsection{Calculation of the ambipolar diffusion velocity $\vAD$}

The amplitude $\vAD$ of the ambipolar diffusion velocity depends on the coupling of charged particles with the gas, and therefore on the ionization degree of the gas and on the size distribution and charge of dust grains.

In high-density regions protected from UV radiations, the ionization can only result from the energetic electrons produced by the interaction of cosmic rays with the gas \revise{\citep{Pado2018}}. As the density increases, the ionization fraction $\ni/\nH$ decreases because the recombination rate of ions with electrons or charged grains scales as $\nH^2$, while the ionization rate scales with $\zeta\,\nH$, with $\zeta$ the cosmic ionization rate. The low ionization does not allow a strong coupling between the gas and the magnetic field, and a velocity difference appears at small scales between the ion and neutral fluids, a phenomenon called ambipolar diffusion. Studying the Class 0 protostar B335 with ALMA observations, \cite{Yen2018} inferred an upper limit of 0.3\,km.s$^{-1}$ for the ion-neutral drift in this particular object.  

The amplitude $\vAD$ of the ion-neutral drift velocity can be calculated by equalizing the Lorentz Force acting on charged particles (namely ions and dust grains) and the drag force exerted by the gas particles onto these charged particles. This yields:
\begin{equation}
\frac{|\BrotB|}{4\pi}  =  \rhoN\left(\vAD \,\ni \,\sigI+\sqrt{\frac{8\kB T}{\pi\muN}} \sum_k n_k \,\sigmak\, |\Vn-\Vdrift_k|\right) \,,
\end{equation}
where $\rho_n$ is the mass density of the gas, 
$T$ is the gas temperature, 
$n_k$ and  \revisem{$\sigmak$} are the number grain density and \revise{mean grain cross-section} in bin $k$, respectively,  
\begin{equation}
\sigI = 2.41\,\pi \,e \,\sqrt{\frac{\muI\,\muN}{\muI+\muN}\alphan}\,,
\end{equation}
is the ion-electron collisional cross-section, $e$ the electric charge of ions, $\mu_i$ and $\mu_n$ the mean mass of ions and neutral particles, respectively, and $\alpha_n$ the polarizability of neutrals \citep{FPdF03}, 

Based on Fig.~\ref{fig:Vdist-AD}, we neglect the second term in the RHS of Eq.~\eqref{eq:Vdrift} to remain linear\footnote{\revise{$|\Vn~-~\Vdrift_k|~\simeq~|\Gamma_k|/\sqrt{1+\Gamma_k^2}\,\vAD$ if we do not neglect this second term of the velocity, perpendicular to the direction of the collapse. Calculations show that our results are little affected by this approximation that decreases the value of $\vAD$ when $\Gamma  < 1$,  once small grains have already been removed by coagulation.}
}  in $\VAD$ and obtain
\begin{equation}
|\Vn - \Vdrift_k|   \simeq \frac{\Gamma_k^2}{1+\Gamma_k^2}\,\vAD\,.\label{eq:Vn-Vk}
\end{equation}

As a consequence, $\vAD$ can be approximated by :
\begin{equation}
\vAD  \simeq  \frac{|\BrotB|}{4\pi\,\rhoN\left(\ni\,\sigI+\sqrt{\frac{8\kB T}{\pi\muN}} \sum_k n_k\,\sigmak\,\frac{\Gamma_k^2}{1+\Gamma_k^2}\right)} \,.
\label{eq:VAD}
\end{equation}

The value of $\vAD$ scales as $b^2$, \ie\ with the square of the intensity of the magnetic field in the parent cloud (Eq.~\eqref{eq:Bscaling}). It also depends, through the gradient, on the geometry of the magnetic field, as well as on its evolution with the compression of the gas during the infall. 
For simplicity, we will start the collapse with $\vAD=\vADref=\cs/10 = 20\,$m.s$^{-1}$ at $\nH=\nHinit$ \citep{BM1995}, and assume $\BrotB \propto B^2/L_{\rm Jeans} \propto \nH^{3/2}$. This scaling results in two asymptotic regimes : if the ion term dominates over the dust term at the denominator of Eq.~\eqref{eq:VAD}, then $\ni\propto \sqrt{\nH}$ and $\vAD$ will be approximately constant; on the contrary, if the dust term dominates, then $\vAD \propto 1/\sqrt{\nH}$ as long as the dust size distribution does not change and remains strongly coupled to the magnetic field ($\Gamma_k \gg 1$)  .
}

\subsection{The coagulation kernel}\label{kernel}

It is the coagulation kernel $K$, \revise{and therefore in our case the mean relative velocity $\left\langle \Delta v\right\rangle$ between the projectile and the target grains,} that control the coagulation rate (see Sect.~\ref{sec:Smol} and Eq.~\eqref{eq:coagkernel}). 
\revise{Only those collisions at a relative velocity lower than the velocity limit $\Vlim$ will lead to coagulation, as grains are expected to bounce, and not to stick, at higher velocities (see Sect.~\ref{sec:Vlimit})}.

\subsubsection{Turbulence alone}

To express the coagulation kernel in the case where only turbulence accelerates dust grains, we follow the approach by \cite{Flower2005}. We assume that the grains relative velocities along $x$, and $y$ and $z$ are gaussian variables of 
\revise{variance $\DV_{\rm P,T}^2/3$ where $\DV_{\rm P,T}$ is the rms relative velocity between the target and the projectile (see Sect.~\ref{sec:turbHD}), and only account for collisions with a relative velocity smaller than the velocity limit $\Vlim$}.
In this frame, the demonstration by \cite{Flower2005} can be extended to the general case of grains of different sizes (and therefore velocities), with the same result:
\revise{
\begin{equation}
\frac{\langle \Delta v \rangle}{\DV_{\rm P,T}}  
= \frac{1}{\DV_{\rm P,T}} \int_0^{\Vlim} v\,f(v)\,\d v  = \sqrt{\frac{8}{3\pi}} \left(1-\left(1+\chi^2\right)\,\exp^{-\chi^2}\right) \,,
\label{Eq-Kturb}
\end{equation}
with $\chi \equiv \sqrt{3/2}\,\Vlim / \DV_{\rm P,T}$.}

\subsubsection{Turbulence and systematic velocity added in quadrature}

If we assume for simplicity that the presence of ambipolar diffusion does not change the nature (hydrodynamic) of the turbulent cascade, Equation~\eqref{Eq-Kturb} can be generalized to the case of turbulence velocities in the presence of a systematic differential velocity $\Vsys$ added in quadrature \revise{to the turbulent velocity of standard deviation $\DV_{\rm P,T}$} (see the demonstration in Appendix \ref{A-shifted})
\begin{eqnarray}
\frac{\langle \Delta v \rangle}{\DV_{\rm P,T}} &= &
\frac{1}{\sqrt{3\pi/2}}
\left(
\exp^{-\xi^2}
-\frac{
\exp^{-\left(\chi-\xi\right)^2}
\left[1+\frac{\chi}{\xi}\right]
+\exp^{-\left(\chi+\xi\right)^2}
\left[1-\frac{\chi}{\xi}\right]
}{2}
\right) \nonumber \\
& + &\frac{1}{\sqrt{3/2}} 
\left(\xi+\frac{1}{2\xi}\right)\,h(\chi,\xi)\,,
\end{eqnarray}
with \revise{$\xi \equiv \sqrt{3/2}\,\Vsys / \DV_{\rm P,T}$}, $h(\chi,\xi) \equiv \erf{\left(\xi\right)} - \frac{\erf{\left(\chi+\xi\right)} -\erf{\left(\chi-\xi\right)}}{2}$ and 
 $\erf$ the error function.



\section{Grain charge, ionization balance and plasma MHD resistivities}\label{sec:Charge_Ionization_Resist}

The charging of grains is, in dense clouds, a prerequisite to solve the ionization of the plasma.

\subsection{Charging processes}\label{sec:graincharge}

As per \cite{Guillet2007}, the grain charge is computed accounting for:
\begin{enumerate}
\item the collection of thermal electrons, with sticking coefficient $s_e = 0.5$ \citep[a value which is not taken to be 1 owing to the non-planar nature of dust grains, see][]{DraineSutin87};
\item the collection (and subsequent recombination) of thermal ions, with sticking coefficient $s_i = 1$;
\item the photoelectric detachment of electrons by Cosmic Rays-induced secondary photons (see Sect.~\ref{sec:updateCR}).
\end{enumerate}
\begin{figure}
\includegraphics[angle=-90,width=\hhsize]{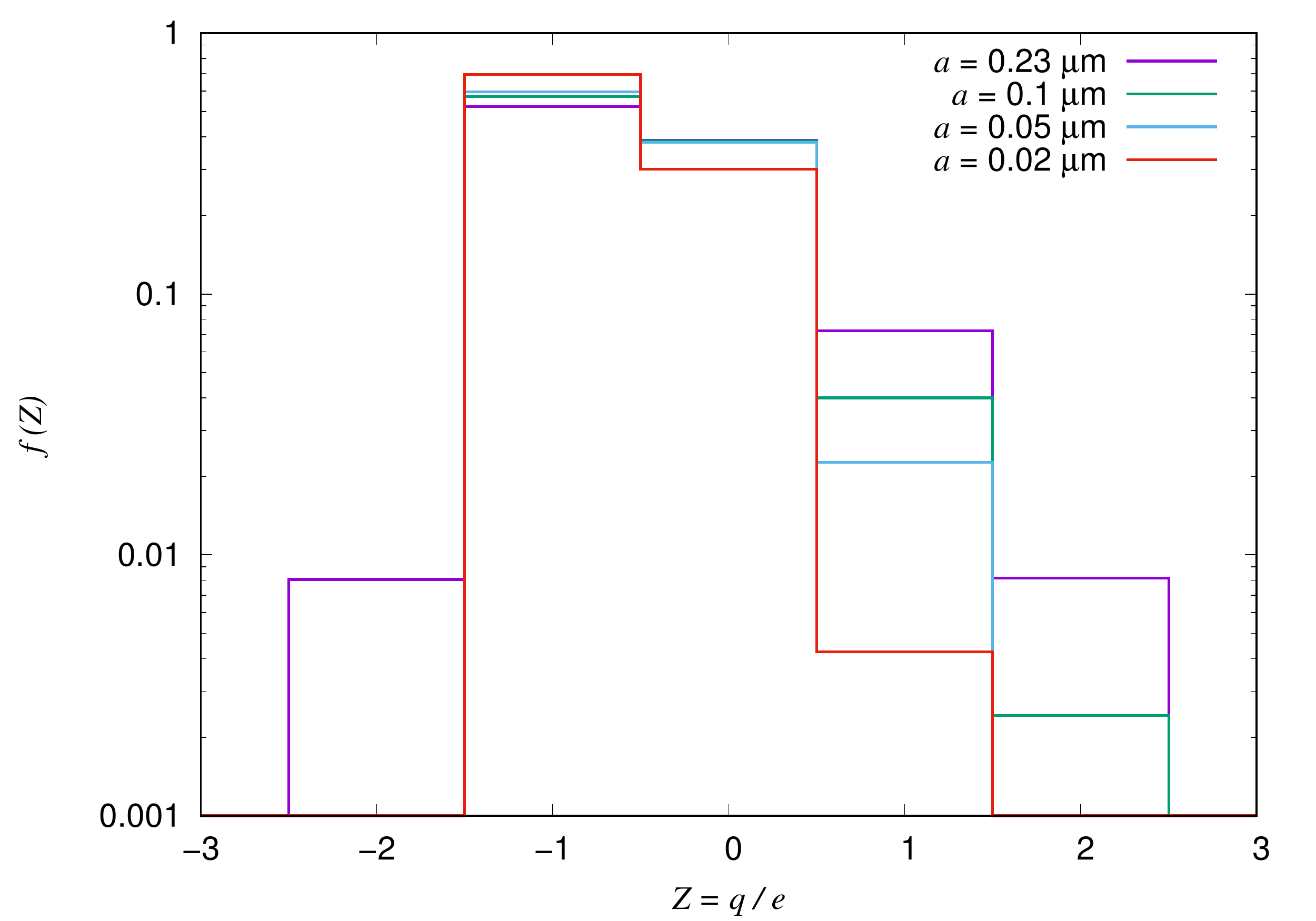}
\caption{Grain charge distribution $f(Z)$ for four representative grains sizes $a$ in the cloud envelop ($\nH = 10^4\,\cmc$, $T=10\,$K, $\zeta=5\,10^{-17}\,{\rm s}^{-1}$). The non-Gaussian tails at positive charges is the result of photoemission by CR-induced secondary UV photons (see Sect. \ref{sec:updateCR}).
}
\label{fig:fbyZ}
\end{figure}
If we momentarily ignore the photoelectric detachment of electrons, the mean grain charge is primarily controlled by the ratio of electrons and ions fluxes, by the ratio $\me\,\ne/(\mi\,\ni)$, where $\me$ is the electron mass and $\mi$ the mean ion mass. 

The \DUSTDAP\ code can integrate the charge distribution $f(Z)$ as well as the mean charge $\EspZ$ of a grain \citep{Guillet2007}. The charge distribution of large grains ($a> 0.25\,\mu$m) is almost gaussian and can therefore be approximated by its mean charge $\EspZ$. Small grains ($a < 0.25\,\mu$m), however, have a non-gaussian charge distribution, centered around $Z=-1$, and intrinsically discrete in nature. Being the dominant charge carriers among grains, their full charge distribution must be integrated \citep{Guillet2007}. 

In Fig.~\ref{fig:fbyZ}, we present the charge distribution of grains before the collapse starts ($\nH = 10^4\, \cmc$). Grains are on average negatively charged, with a mean charge number $\left\langle Z \right\rangle \simeq -1$. The charge distribution is not gaussian, being skewed toward positive values due to the photoemission of electrons by impinging secondary UV photons (see the next section). \revise{The larger the grain, the more positive its charge number $Z$. This is due to the fact that the coulombian enhancement of the grain cross-section $\tilde{\sigma} = 1 + Ze^2/ a \kB T$ \citep{Spitzer1941,DraineSutin87} for the capture of an electron at temperature $T$ by a grain of radius $a$ and charge $Z$ decreases with $a$ when the grain is positively charged, while the photoemission yield is almost independent of $a$ and $Z$ \citep{WD01}.}

The density of ions is calculated as the sum of the density of positively charged (molecular and atomic) gas species. Knowing the charge of each grain size, the density of free electrons can then be derived from the electroneutrality equation: $\ni - \ne +\sum_k \nk Z_k = 0$, where $n_k$ and $Z_k$ are the grain density and the mean grain charge in bin $k$, respectively.

\subsection{Update of the ionization rates by cosmic-rays}\label{sec:updateCR}

The suprathermal electrons generated by cosmic-rays excite H$_2$ molecules, which relaxe by producing UV photons, which are called secondary to differentiate them from primary starlight UV photons. Since \cite{FPdF03}, we have updated the ionization rates by cosmic-rays using the new tables provided by \cite{Heays2017}.
The cross-section per unit volume to secondary photons for gas and dust particles are:
\begin{eqnarray}
\nsgasSECPHO & = & \sum_{j=1}^{\Nspec} \frac{\rgamma}{0.15}\,\left(\frac{T}{300}\right)^\ralpha\times 2\times10^{-21}\,, \label{eq:nsgasSECPHO}\\
\nsdustSECPHO & = & \sum_{k=1}^{\Nbins} n_k\,\sigmak\,\qabs(a_k,{\rm 10\,eV})\,.
\end{eqnarray}

The total cross-section to secondary photons, $\nsgasSECPHO+\nsdustSECPHO$, depends on the size distribution of dust grains, as well as on the gas content.
It is however dominated by dust, as usually assumed \citep{Gredel1989}.
Since the ionization rates were provided by these authors under the assumption of an MRN size distribution, each ionization rate $\rgamma$ for the gas specie $j$ (Eq.~\eqref{eq:nsgasSECPHO}) must be multiplied by the correcting factor $2\times10^{-21}\,\nH/\left(\nsgasSECPHO+\nsdustSECPHO\right)$, to adapt to the evolving size distribution in the collapsing core.
Similarly, for the grain charging, Eq.~(4) of \cite{Guillet2007} must be adapted by replacing $\langle n\hat{\sigma}\qabs\rangle$ at the denominator by $\nsgasSECPHO+\nsdustSECPHO$.

\subsection{Charge fluctuations}

Grain charge fluctuates due to the stochastic nature of the grain charging processes. The timescale of charge fluctuations is larger for small grains than for large grains, owing to the reduced geometric cross-section of the latter. 
According to \cite{YLD04}, it is
 \begin{equation}
\tau_Z = \frac{\text{Var}(Z)}{\sum_Z f(Z)\times\left(J_e(Z)+J_i(Z)+J_{\rm pe}(Z)\right)}
 \end{equation}
 where $Z$ is the grain charge number, Var($Z$) is the variance of the charge distribution, and $J_e(Z)$, $J_i(Z)$, and $J_{\rm pe}(Z)$ are the rates for electron collection, ion recombination on the surface of grains, and photoelectric detachment by secondary photons, respectively (See Sect.~\ref{sec:graincharge}). 
 
\begin{figure}
\includegraphics[angle=-90,width=\hhsize]{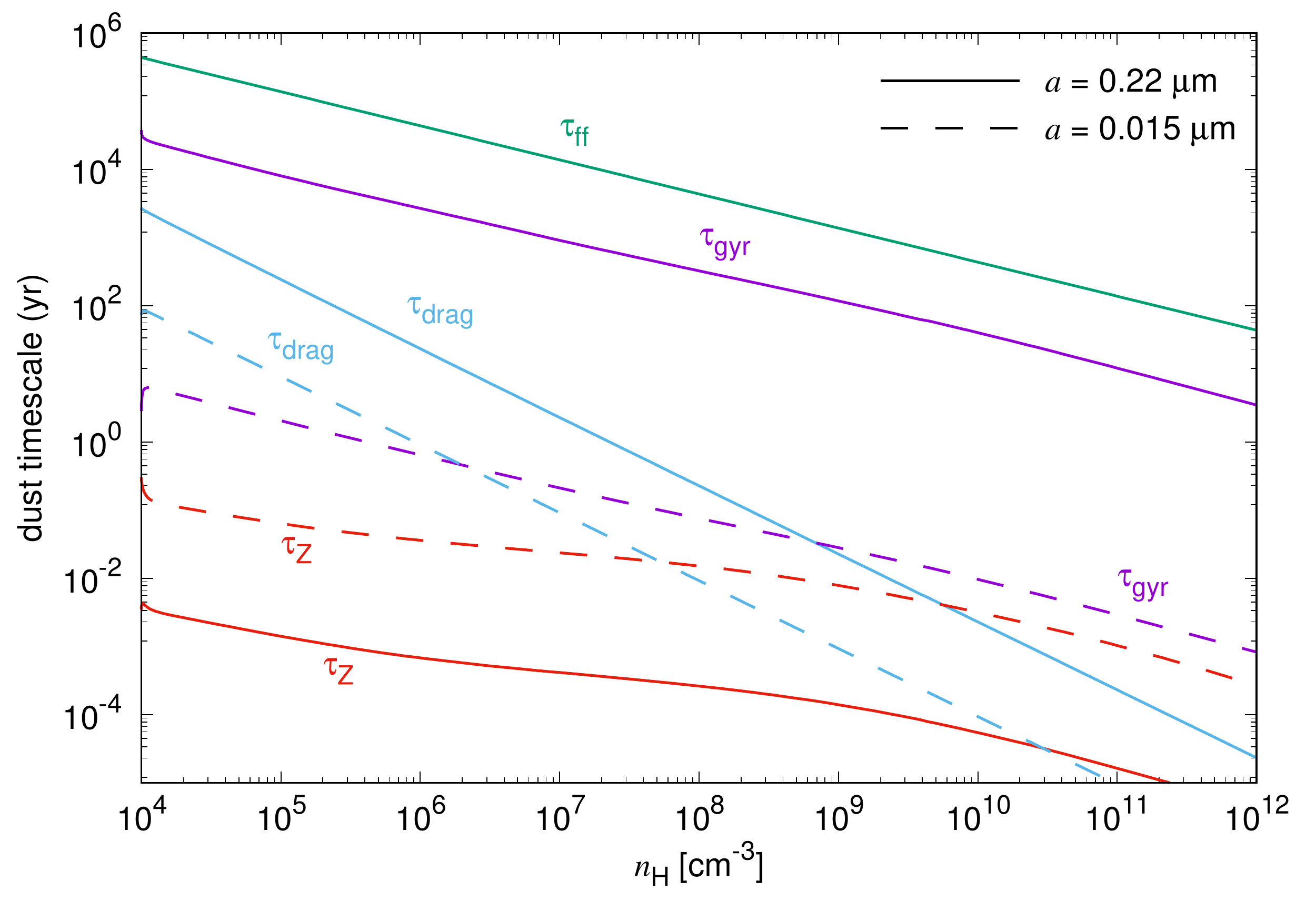}
\caption{Characteristical timescales as a function of the proton density, for the larger and smaller dust grain of the initial size distribution (of radius $0.22\,\mu$m and 0.015 $\,\mu$m respectively): gyration timescale ($\taugyr$), stopping timescale $\taudrag$, and charge fluctuation timescale ($\tau_Z$), compared to the free-fall timescale $\tff$. 
}
\label{taudust}
\end{figure}
 
In Fig.~\ref{taudust}, we show that the charge fluctuation timescale is smaller than the dynamical time of the collapse $\tff$, and also smaller than the grain Larmor timescale $\taugyr=1/\omega$ for all grain radius considered here and at all gas densities.
The charge fluctuation timescale is also smaller than the stopping timescale $\taudrag$ for large grains, but not for small grains ($a=10\,$nm)  at densities higher than $10^8\,\cmc$. As we shall see, those grains have already been removed by coagulation when this condition is not met anymore.
We will therefore ignore charge fluctuations in our study.

\subsection{Hall factor for electrons, ions and grains}

Let $j$ be a particular species (ion, electron, or grain), $\tau_j$ its stopping time, $\omega_j$ its Larmor pulsation, and $\Gamma_j = \omega_j\,\tau_j$ its Hall factor.
Following \cite{marchand2016}, we have:
\begin{eqnarray}
\tau_j & = & \frac{1}{a_{j\,{\rm He}}}\frac{m_j+m_{\rm H_2}}{m_{\rm H_2}}\frac{1}{\nH\,\langle\hat{\sigma} v\rangle_j} \,, \\
\omega_ j & = & \frac{q_j\,B/c}{m_j} \,,\\ 
\Gamma_j & = & \omega_j\,\tau_j \,.
\end{eqnarray}
The factor $a_{j,\rm He}$ accounts for collisions with helium atoms and is equal to 1.14 for ions, 1.16 for electrons and 1.28 for grains \citep{Desch2001}.

The Hall factor $\Gamma_j$ expresses the degree of coupling of the particle to the magnetic field. Its value is very high for electrons and ions ($\Ge \gg \Gi \gg 1$). 
For grains in our core collapse, $\Gamma_j \propto B/\left(a^2\,\nH\right)$ where $a$ is the grain size. Grains therefore tend to decouple from the magnetic field as $\nH$ increases during the collapse. Grain coagulation increases this trend by increasing the grain size $a$, and so does ice accretion by increasing $a$ and decreasing the grain specific density $\mud$.

\subsection{Plasma conductivities and resistivities}

We note\footnote{\revise{Not to be confused with the collisional cross-sections of gas and dust particles, which are denoted $\crosssection$, $\sigmak$ etc.}}  $\sigma$ the conductivity of a charged fluid.
The conductivity $\sigma_j$ of the species $j$ is by definition 
\begin{equation}
\sigma_j  =  n_j\frac{\,q_j^2}{m_j}\tau_j \,.
\end{equation}
We note the useful relation
\begin{equation}
\frac{\sigma_j}{\Gamma_j} = \frac{n_j\,q_j}{B/c}\,.
\end{equation}

The parallel, perpendicular and Hall conductivites of the plasma have the following expressions \citep[\eg][]{marchand2016}
\begin{eqnarray}
\sigma_\parallel & = & \sum_j \sigma_j \label{eq:sigpar}\\ 
\sigma_\perp & = & \sum_j \sigma_j\frac{1}{1+\Gamma_j^2} \\ 
\sigma_{\rm H} & = & -\sum_j \sigma_j\frac{\Gamma_j}{1+\Gamma_j^2} \,,
\label{eq:sigma_H}
\end{eqnarray}
where the sum is done over all charged species, namely electrons, ions and grains of all sizes.

The Ohmic (parallel) conductivity  is dominated by electrons, the most mobile particles. The perpendicular conductivity is dominated by the particles with low values of $\Gamma$, \ie\ by the particles the least coupled to the magnetic field, which are here dust grains.

We now derive approximate expressions for the Hall conductivity $\sigma_{\rm H}$ \revise{building on the fact that 
electrons and ions are strongly coupled to the magnetic field with $\Gamma_e\gg\Gamma_i\gg1$. In the absence of dust grains, using the electroneutrality equation Eq. ~\eqref{eq:sigma_H} becomes
\begin{equation}
\left(\sigma_{\rm H}\right)^{\rm no dust} \simeq  \frac{\ni\,e}{B/c}\,\frac{1}{\Gamma_i^2} \propto \ni\frac{\nH^2}{B^3}
\label{eq:sigmaH_nodust}
\end{equation}
} 
In the presence of charged grains, 
\begin{equation}
\sigma_{\rm H} 
 \simeq   
\frac{\sum_k n_k\,Z_k/\left(1+\Gamma_k^2\right)}{B/c}\,. \label{eq:sigmaHbis}
\end{equation}
The Hall conductivity is therefore controlled by the abundance and sizes of dust grains.


The Ohmic, Hall and ambipolar diffusion resistivities write \citep{marchand2016}
\begin{eqnarray}
\eta_\Omega & = & \frac{1}{\sigma_\parallel}\,,  \\
\eta_{\rm H} & = & \frac{\sigma_{\rm H}}{\sigma_{\rm H}^2+\sigma_{\rm \perp}^2}\,, \label{eq:rH} \\
\eta_{\rm AD} & = & \frac{\sigma_{\rm \perp}}{\sigma_{\rm H}^2+\sigma_{\rm \perp}^2} -  \frac{1}{\sigma_\parallel} \,.\label{eq:rAD}
\end{eqnarray}

\section{Results}\label{sec:Results}

We now present our results for the evolution of the grain size distribution, plasma ionization and MHD conductivities and resistivities during the collapse of a spherical core, for different scenarios of dust dynamics and evolution. 

In our standard model, the initial density is $\nHinit = 10^4\,\cmc$, the initial size distribution is the MRN covered by a 8.8 nm ice mantle, representing a total of 0.9\% of the gas mass (see Sect. \ref{sec:init}). The collapse timescale is the free-fall timescale $\tff$ computed at a density $\nHinit$, and the cosmic ray ionisation rate is $\zeta = 5\times10^{-17}$\s, taken to be constant through the collapse.
Before entering into the details of the feedback of dust coagulation on the MHD properties of the collapsing core, let us recall what would happen without dust.

\subsection{Collapse without dust}\label{sec:nodust}

\begin{figure*}
\includegraphics[angle=-90,width=\thsize]{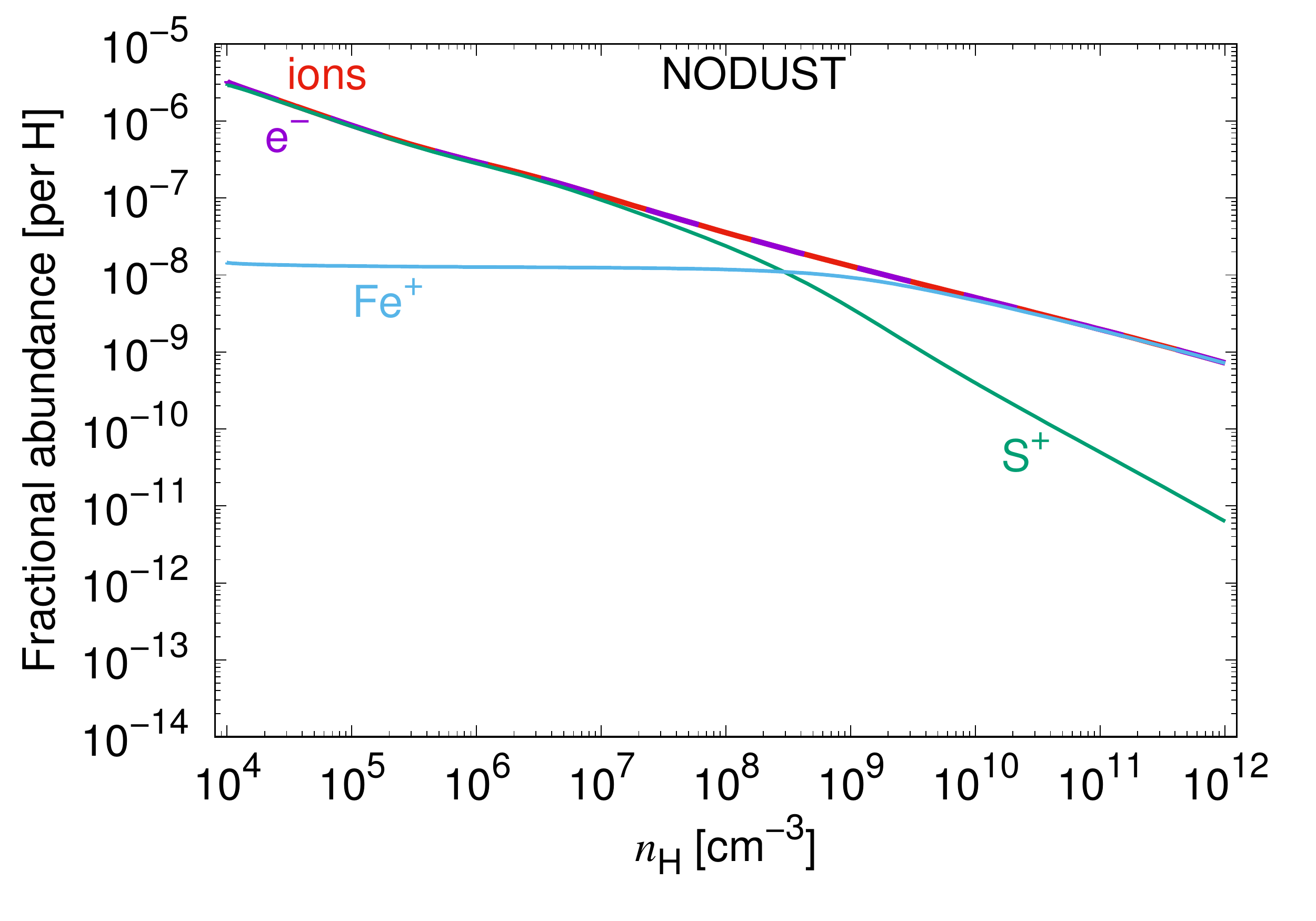}
\includegraphics[angle=-90,width=\thsize]{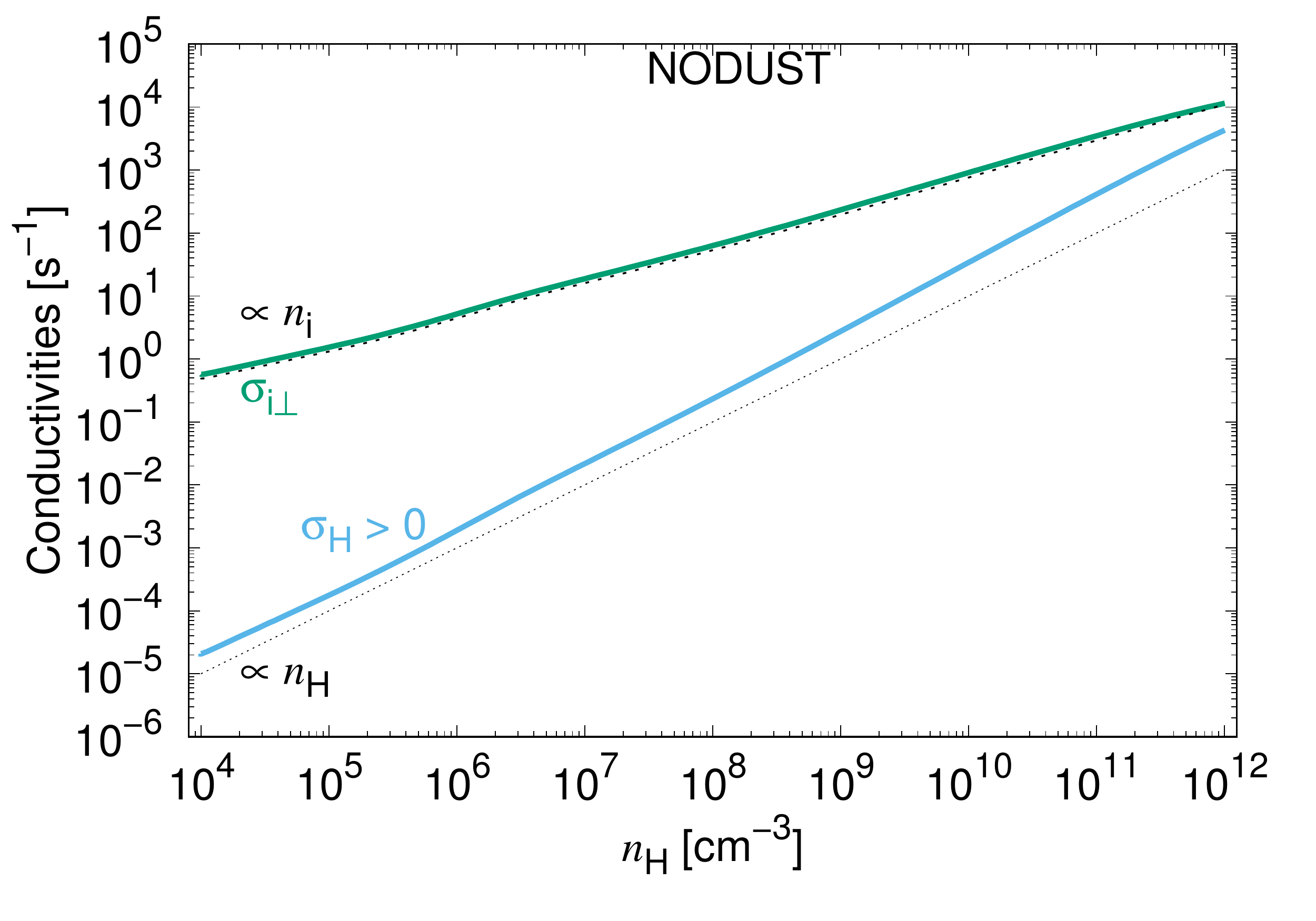}
\includegraphics[angle=-90,width=\thsize]{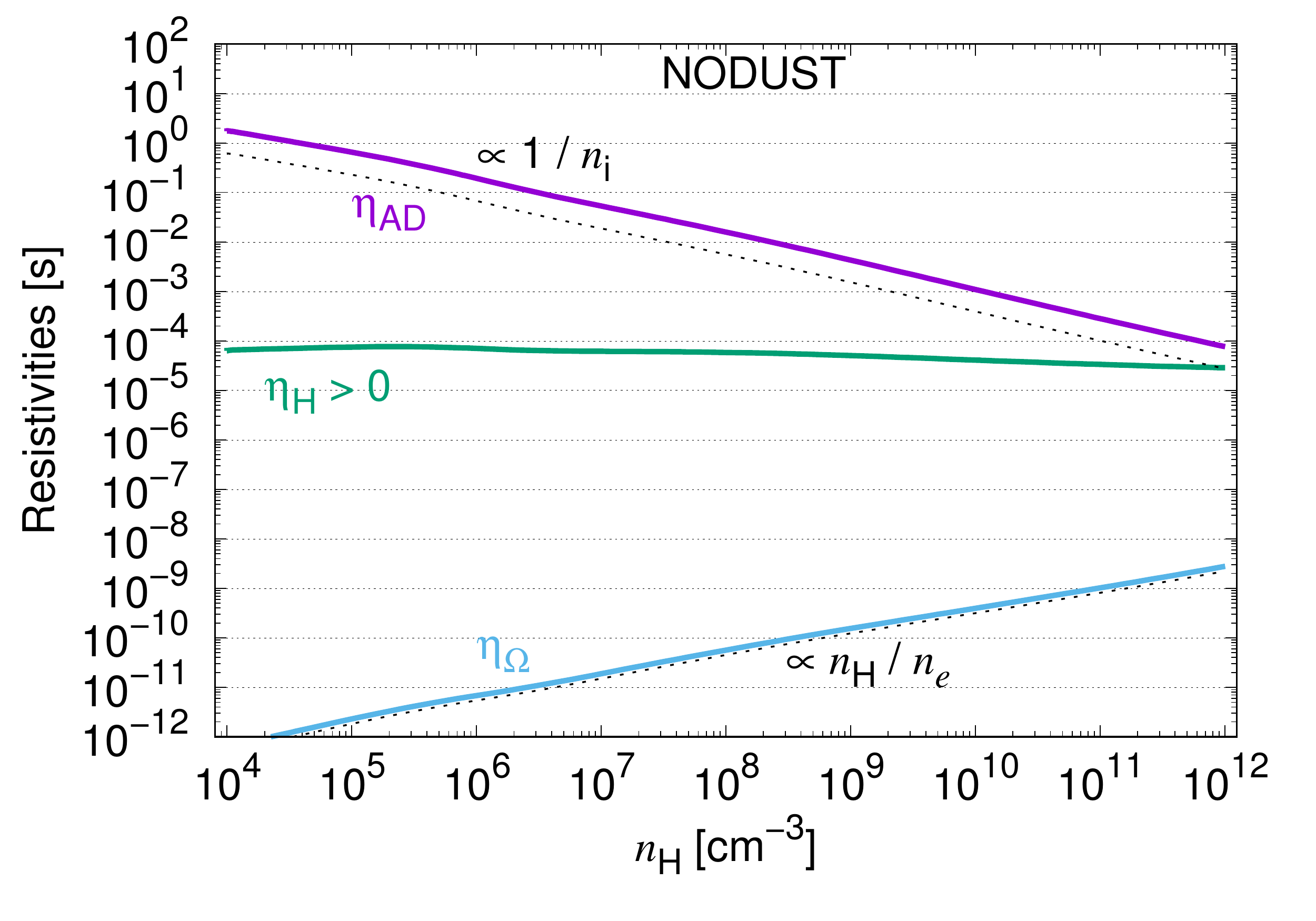}
\caption{Results for a model without dust. \L\ Fractional abundances of ions $\ni/\nH$ and electrons $\ne/\nH$ together with the fractional abundance of the dominant ions Fe$^+$ and S$^+$. \C\ Perpendicular ($\sigma_\perp$) and Hall ($\sigma_{\rm H}$) conductivities. \R\ Ohmic ($\eta_{\rm O}$), Hall ($\eta_{\rm H}$) and ambipolar diffusion ($\eta_{\rm AD}$) resitivities. Asymptotic trends are overplotted in dotted lines (see text for a derivation of those expressions). Horizontal lines are added to help the reader estimate the plasma resistivities for a given proton density.
}
\label{fig:nodust}
\end{figure*}

Figure~\ref{fig:nodust} presents the variation of the ionization fraction, MHD conductivities and resistivities with the gas density during the collapse, for a plasma without dust. The ionisation fraction (left panel) results from an equilibrium between the recombination of ions in the gas phase (at a rate $\propto \ni^2$) and the ionization by CRs (at a rate $\propto \zeta\,\nH$), leading to 
\revise{
\begin{equation}
\left(\frac{\ni}{\nH}\right)^{\rm no dust}  \propto  \sqrt{\frac{\zeta}{\nH}}\,.
\end{equation}
} 
Without grains, gas species can not deplete, and heavy ions like S$^+$ and Fe$^+$ remain the main ions through the collapse. 

The center panel of Fig.~\ref{fig:nodust} shows the monotonic evolution of the perpendicular and Hall plasma conductivities. As $\Gamma_{\rm e} \gg \Gamma_{\rm ions} \gg 1$, the plasma perpendicular conductivity is that of ions, and is therefore proportional to the ion density $\ni$. \revise{Combining equations \eqref{eq:Bscaling} and \eqref{eq:sigmaH_nodust}, the Hall conductivity $\sigma_{\rm H}$ is found to be proportional to the gas density $\nH$}. These scaling are well reproduced in the center panel of Fig.~\ref{fig:nodust}. The Hall conductivity is positive because the particles the less coupled to the magnetic field (here ions) are positively charged. We note that the perpendicular conductivity is higher than the Hall conductivity at all densities.

Regarding the plasma resistivities (Fig.~\ref{fig:nodust}, right panel), the ambipolar diffusion resistivity is dominating the Hall and Ohmic at all densities below $10^{10}\,\cmc$. Our model reproduces the expected relations inferred from Eqn. (\ref{eq:sigpar}) to (\ref{eq:rAD}) in the absence of dust grains : $\rO \propto \nH/\ne$, $\rH \propto \nH^2/(B^3\,\ni)$ (constant in our collapse model because $B \propto \sqrt{\nH}$ and $\ni\propto \sqrt{\nH}$), and $\rAD\propto 1/\ni$.

\subsection{Collapse without coagulation}\label{sec:noevol}

\begin{figure*}
\includegraphics[angle=-90,width=\thsize]{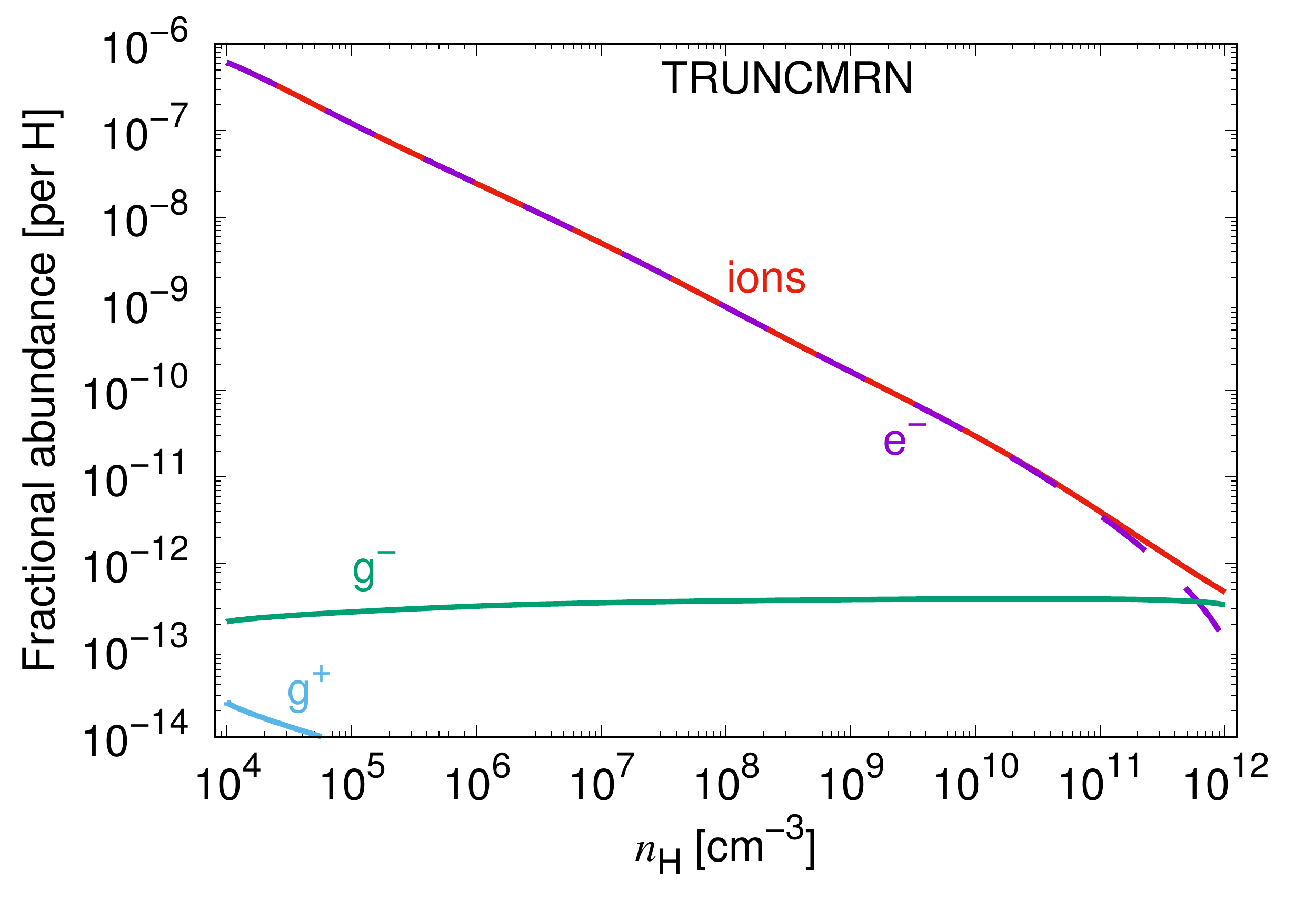}
\includegraphics[angle=-90,width=\thsize]{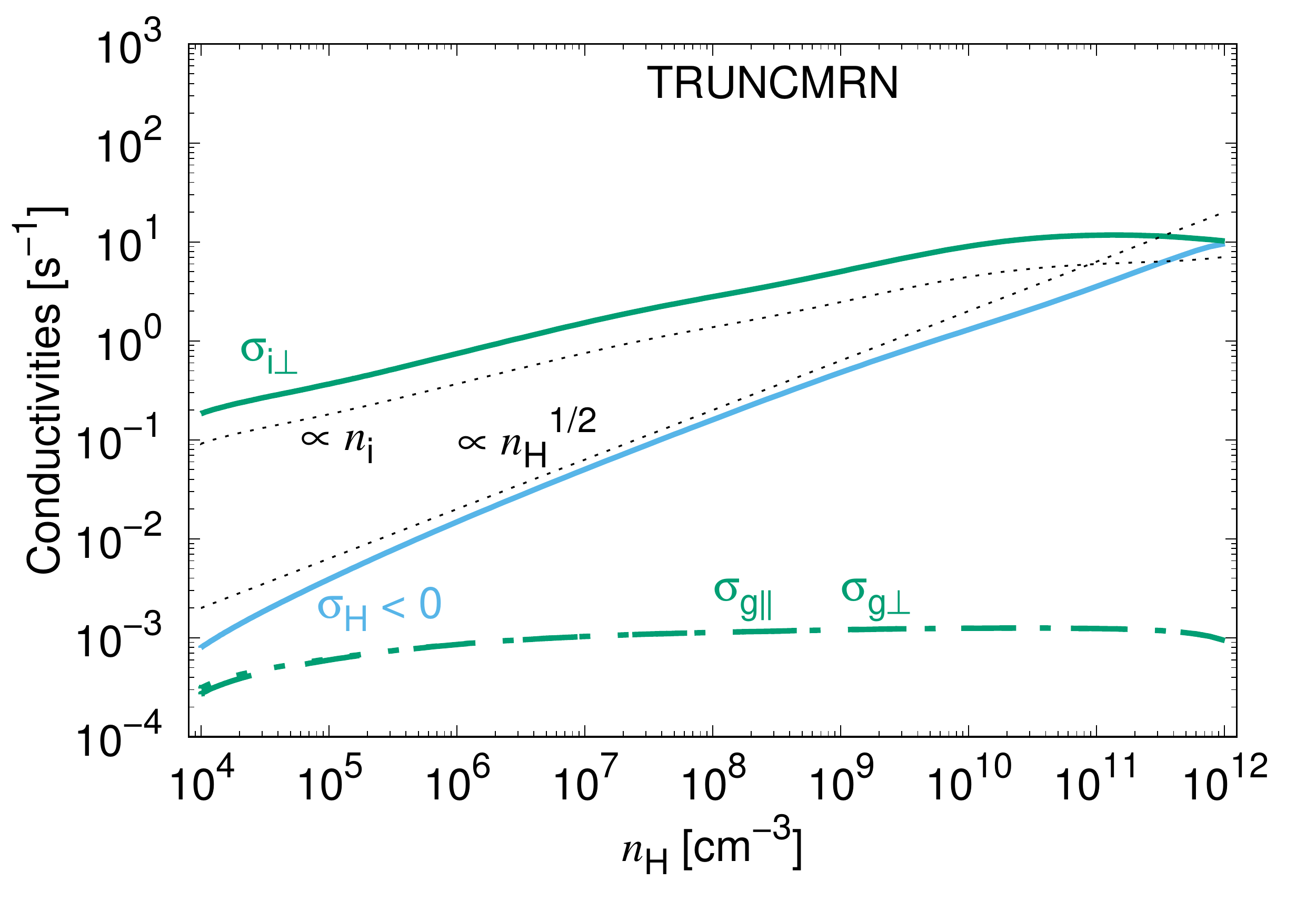}
\includegraphics[angle=-90,width=\thsize]{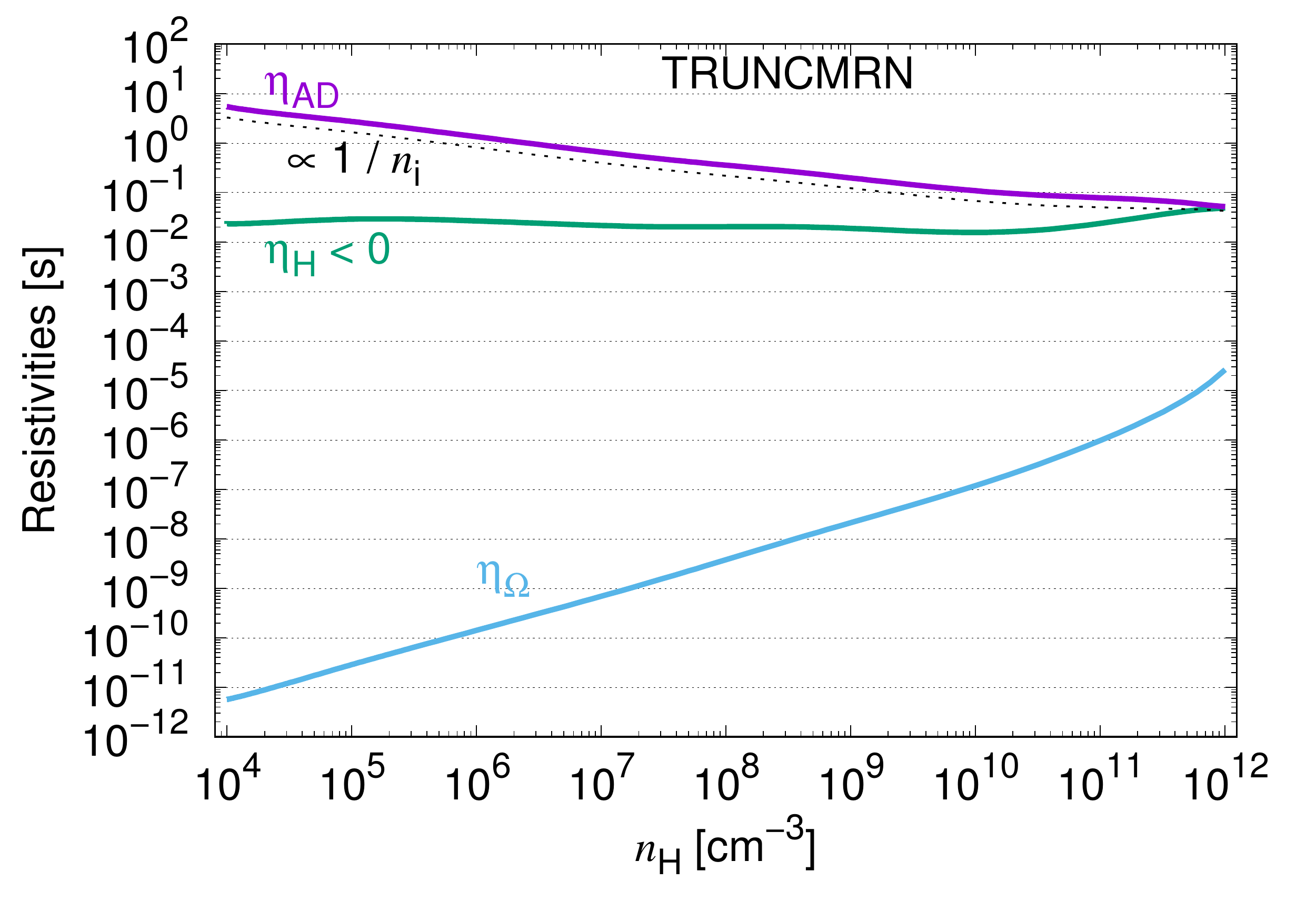}
\includegraphics[angle=-90,width=\thsize]{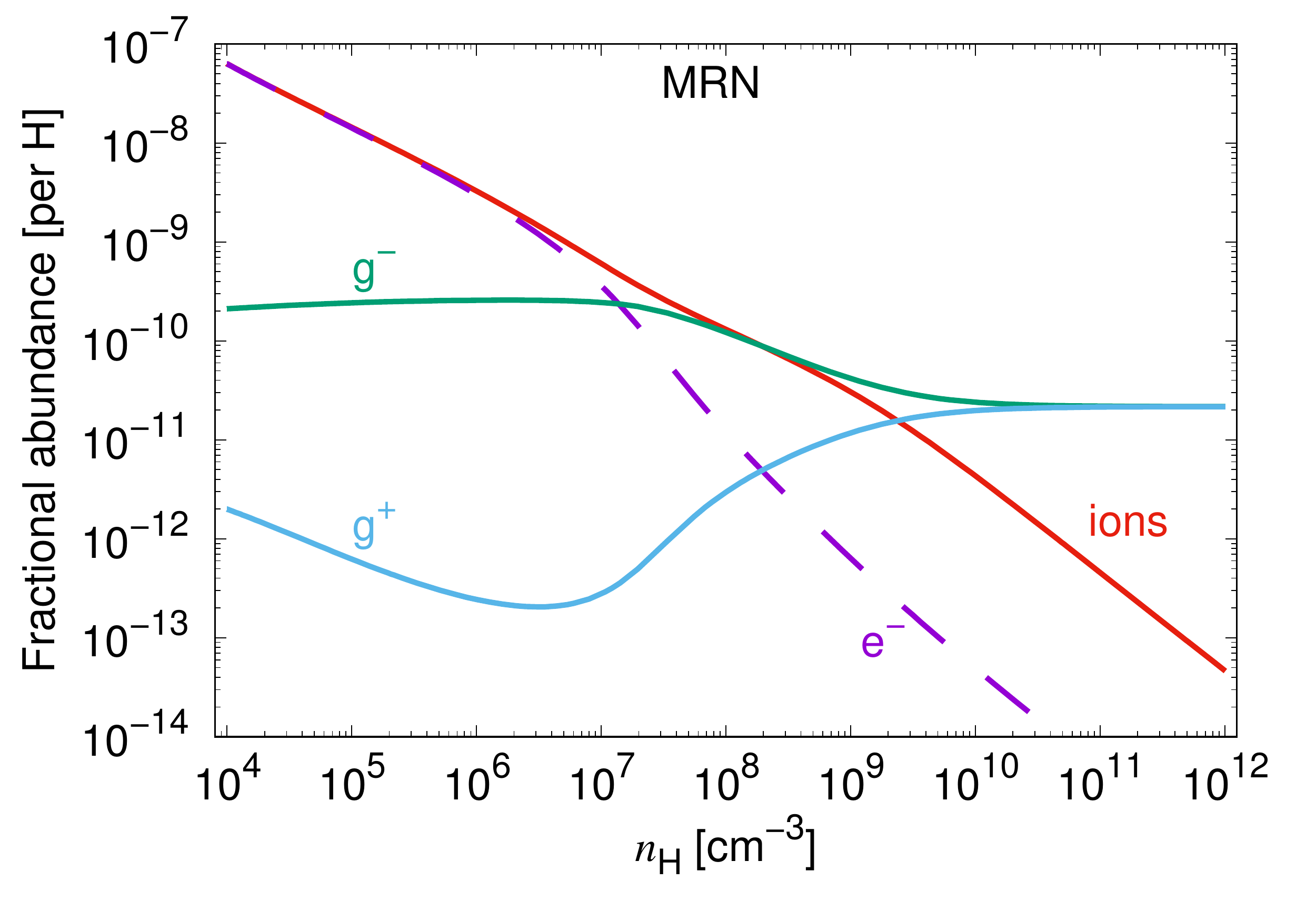}
\includegraphics[angle=-90,width=\thsize]{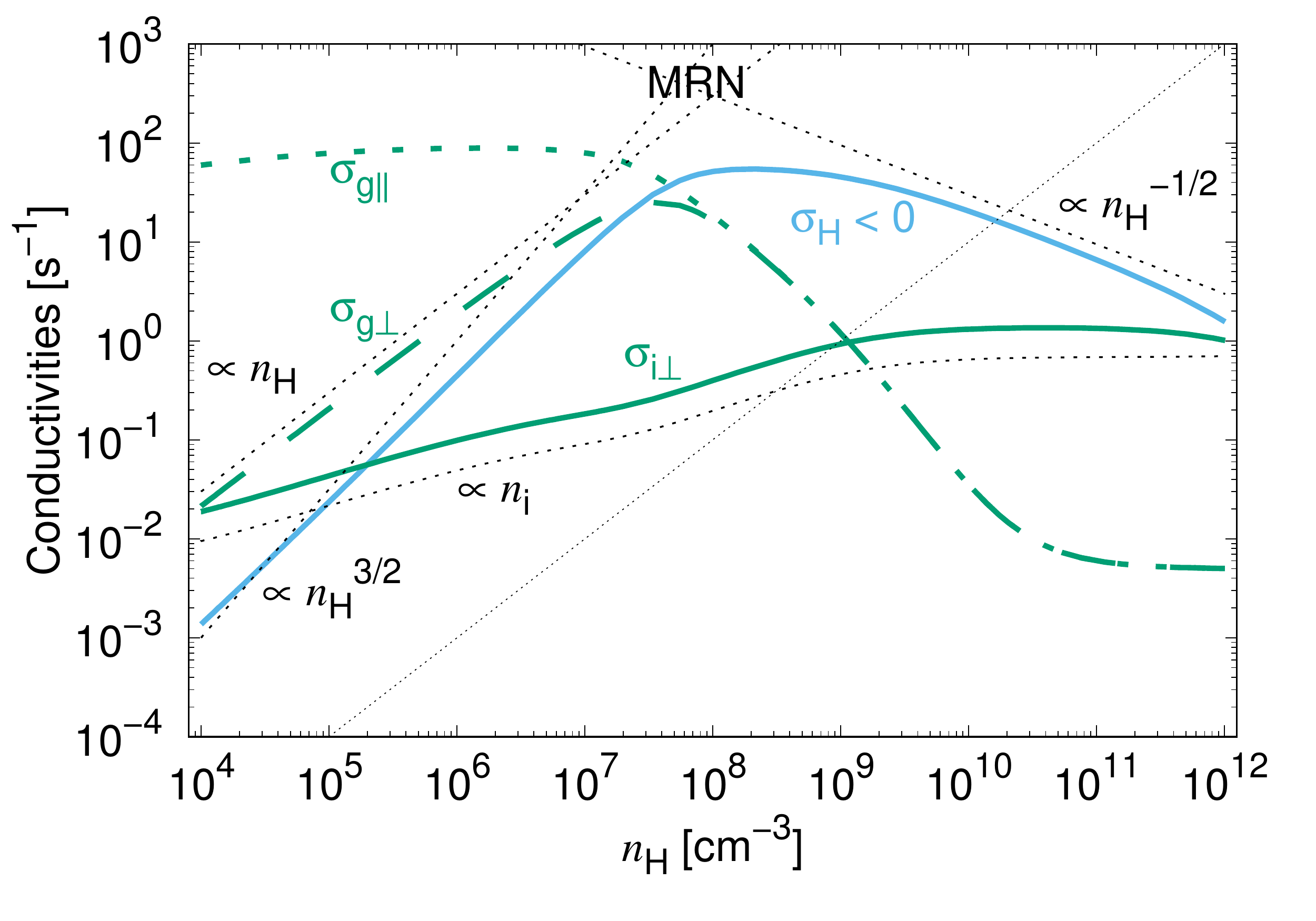}
\includegraphics[angle=-90,width=\thsize]{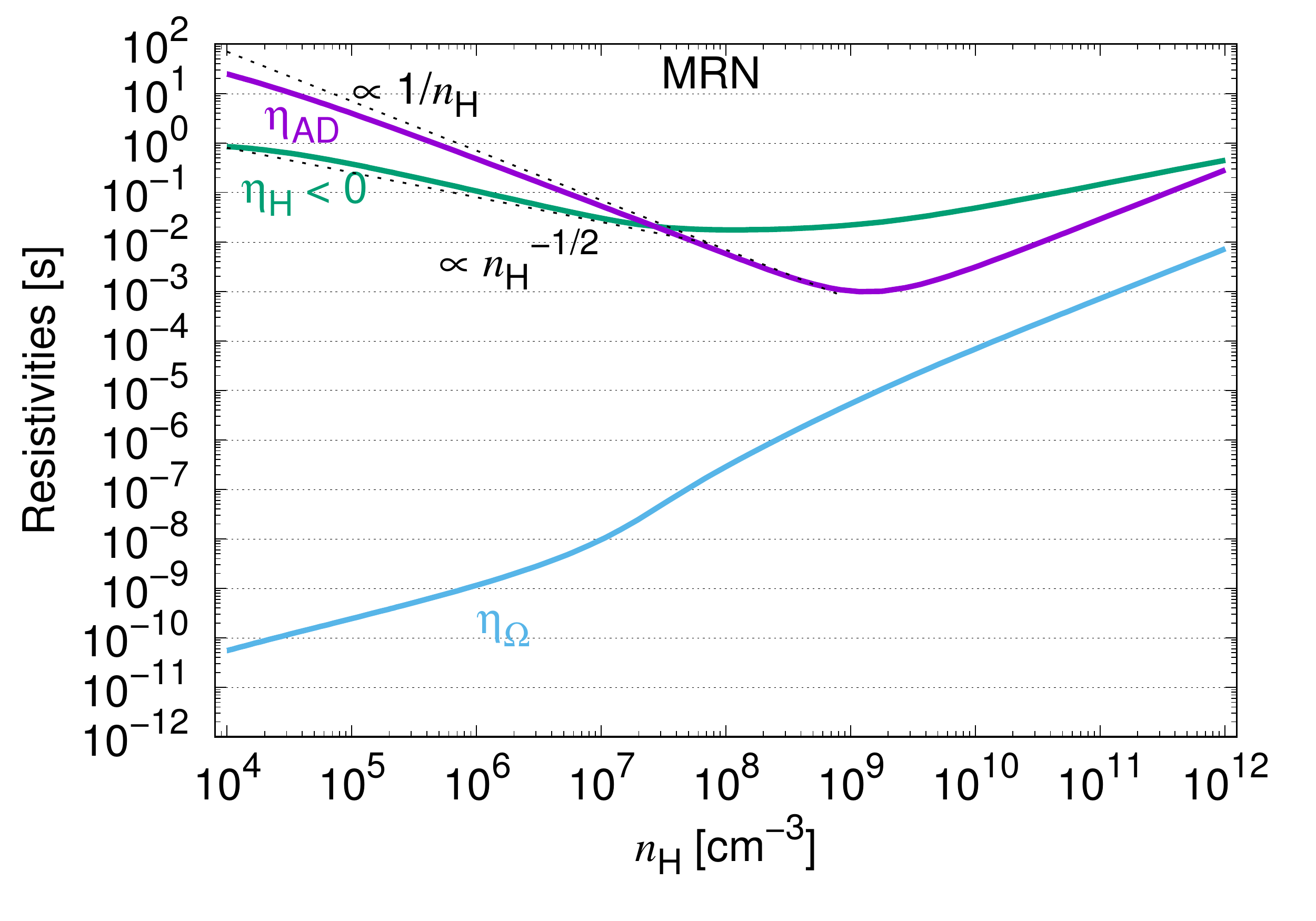}
\caption{Results for our models without dust evolution, with truncated-MRN ($100-250$ nm) \T\ and MRN ($5-250$ nm) \B\ size distributions, both covered by icy mantles. \L\ Evolution of the fractional abundances of charged species: ions, electrons, negative charges (\textit{g-}) and positive charges (\textit{g+}) on grains.
\C\ Evolution of the dust parallel conductivity ($\sigma_{g\parallel}$),  dust perpendicular conductivity ($\sigma_{g\perp}$), ion perpendicular conductivity ($\sigma_{i\perp}$) and Hall conductivity ($\sigma_{\rm H}$).
Scalings in powers of $\nH$ are overplotted in dotted lines for an easier comparison of our numerical results with the asymptotic trends described in Sect. \ref{sec:nodust}).
\R\ Evolution of the Ohmic ($\eta_{\Omega}$), Hall ($\eta_{\rm H}$), and ambipolar diffusion ($\eta_{\rm AD}$) resistivities.
}
\label{fig:stdnoevol}
\end{figure*}

We now compare the predictions for two kind of models \revise{with dust but} without coagulation: 1) an MRN size distribution ($5-250$ nm), and 2) a truncated-MRN ($100-250$ nm) which has $\sim 10^3$ times less grains than the MRN. \revise{All grains in the parent cloud are covered by icy mantles, 8.8 nm thick for the former, and 50 nm thick for the latter.} The truncated-MRN, which was proposed by \cite{zhao2016}, can be considered as a dust model where the removal of small grains would have already happened in the parent cloud before the collapse.

Figure ~\ref{fig:stdnoevol} presents our results for the MRN (\revise{bottom} row) and truncated-MRN (\revise{top} row) case. \revise{In the presence of dust grains, the ionization fraction (left column) is not only determined by the CR ionization rate and the density, but also by the total surface carried by dust grains which serve as a catalyst for ion recombination. This explains why the ionisation fraction is one order of magnitude smaller in the MRN case than in the truncated-MRN case, which is itself much smaller than the no-dust case.} The fractional abundance of negative grains remains almost constant with the density, while the ionization fraction decreases with the density. As a consequence, there appears a shortage of free electrons leading to a situation where dust grains become the main charge carrier and dominate the ionisation, a situation known as a dust-dust plasma where \revise{$\ne = \ni / \sqrt{\mi/\me}$ remains constant} \citep{Ivlev16}. The higher the number of grains, the earlier this shortage appears: from $\nH=10^6\,\cmc$ for the MRN case and from $\nH=10^{11}\,\cmc$ for the truncated-MRN case. In such a situation, the mean grain charge is not constant anymore and progressively converges toward zero.

\revise{The total perpendicular conductivity of the plasma, sum of that of ions and dust grains (Fig.~\ref{fig:stdnoevol}, center column), is a key ingredient to understand the evolution of non-ideal MHD effects during the collapse (see Eqn.~\eqref{eq:rH} and \eqref{eq:rAD}). 
Before the collapse starts, the total perpendicular conductivity of the plasma is lower in the MRN case than in the truncated-MRN case, which is itself lower than in the no-dust case. This is consequence of the fact that the total conductivity is dominated by ions, the density of which is anticorrelated with the abundance of small grains. To analyze the evolution of the plasma conductivities as the density increases during the collapse,} we must distinguish two asymptotic regimes in the coupling of dust grains to the magnetic field determined by their Hall factor $\Gamma \propto B/(\nH\,a^2)$: 
\begin{itemize}
\item strong coupling ($\Gamma \gg 1$, small grains or low density) 
\begin{eqnarray}
\left(\sigma_{g\perp}\right)^{\Gamma \gg 1} & \propto & \frac{\nH}{B^2}\sum_k n_k\,\sigmak\,Z_k  
\label{eq:sigmaperpGammagg1}\\
\left(\sigma_{\rm H}\right)^{\Gamma \gg 1} & \propto &  \frac{1}{B}\sum_k \frac{ n_k\,Z_k}{\Gamma_k^2} \propto  \frac{\nH^2}{B^3}\sum_k \frac{ n_k\,\sigmak^2}{Z_k}
\label{eq:sigmaH_Gammagg1}
\end{eqnarray}
\item weak coupling ($\Gamma \ll 1$, large grains or high density)
\begin{eqnarray}
\left(\sigma_{g\perp}\right)^{\Gamma \ll 1}  & \simeq&  \sigma_{g\parallel} \propto \frac{1}{\nH}\sum_k \frac{n_k\,Z_k^2}{\sigmak} \\ 
\left(\sigma_{\rm H}\right)^{\Gamma \ll 1} & \propto &  \frac{1}{B}\sum_k n_k\,Z_k \propto \frac{\ne - \ni}{B}.\label{eq:sigmaH_Gammall1}
\end{eqnarray}
\end{itemize}

The transition from the strong to the weak coupling regime happens when the mean, density-weighted, grain Hall factor is close to 1.
When dust grains are strongly coupled to the magnetic field ($\Gamma_k \gg 1$), the total perpendicular conductivity of dust grains $\sigma_{g\perp}$ approximately scales as $\nH$ while the ion perpendicular still scales as $\ni$,  more slowly with the density. When dust grains are weakly coupled to the magnetic field ($\Gamma_k \ll 1$), the dust perpendicular conductivity $\sigma_{g\perp}$ is equal to the dust parallel conductivity $\sigma_{g\parallel}$, and therefore only depends on the grain \revise{mean} charge if the size distribution is kept fixed. 

These two different regimes can explain the differences \revise{in the evolution of the plasma conductivities} observed between the MRN \revise{(where initially $\Gamma \gg 1$)} and truncated-MRN {\revise{(where $\Gamma \ll 1$)} cases. 
Overall, the conductivities of the plasma increase faster with the density in the MRN case than in the truncated-MRN case, \revise{becoming higher for $\nH \ge 10^6\,\cmc$ despite a lower initial value in the parent cloud.}
The large ($a \ge 0.1\,\mu$m) grains of the truncated-MRN case are already weakly coupled to the magnetic field at the beginning of the collapse. Furthermore, their charge remains almost constant during the collapse because the ratio $\ne/\ni$ is itself constant (see left panel and Sect. \ref{sec:graincharge}). As a consequence, the dust perpendicular conductivity remains constant in the truncated-MRN case and is totally negligible at all densities compared to the ions perpendicular conductivity. \revise{The conductivities and resistivities are therefore very similar to the no-dust case regarding their evolution trends with the gas density, only differing in their amplitude which is function of the ionization degree of the plasma}. On the contrary in the MRN case that entails a high abundance of small, strongly coupled, grains, the perpendicular conductivity \revise{initially dominated by dust grains} first increases proportionally to the density. Then, as soon as the depletion of electrons appears, the mean grain charge converges toward zero which severely decreases the perpendicular conductivity of dust grains (Fig.~\ref{fig:stdnoevol} for densities between $10^8$ and $10^{10}\,\cmc$). \revise{The dust perpendicular conductivity eventually becomes lower than the ions perpendicular conductivity, reaching a plateau when the charge distribution is stabilized (Fig.~\ref{fig:stdnoevol}, left panel, $\nH \ge 10^{11}\,\cmc$ ).} 
 
The Hall conductivity, like the perpendicular conductivity, is affected by the degree of coupling of dust grains with the magnetic field. When dust grains are strongly coupled to the magnetic field, the Hall conductivity strongly increases with the density, scaling approximately as $\nH^{3/2}$ (Eq.~\eqref{eq:sigmaH_Gammagg1} \revise{ignoring variations in the grain charge $Z_k$}). This is what is observed at low density in the MRN case. When grains decouple from the magnetic field at higher density, the Hall conductivity does not increase so fast, or even decrease. In the truncated-MRN case we get $\sigma_{\rm H} \propto \nH^{1/2}$ (Eq.~\eqref{eq:sigmaH_Gammall1} with $n_k\,Z_k \propto \nH$), while in the MRN case the dusty plasma regime yields $\sigma_{\rm H} \propto \nH^{-1/2}$ (Eq.~\eqref{eq:sigmaH_Gammall1} with $\ne - \ni$ \revise{remaining constant}, see Fig.~\ref{fig:stdnoevol} center bottom panel). 

\revise{The right panels of Fig.~\ref{fig:nodust} shows the evolution of the plasma MHD resistivities for the MRN and truncated-MRN cases. Before the collapse starts, the Hall and ambipolar diffusion resistivities are higher in the MRN case than in the truncated-MRN case (and are higher in the truncated-MRN case than in the no-dust case) because, as explained above, the presence of small grains tends to decrease the ion density and therefore the ion conductivity. 
As the collapse proceeds, the situation however reverses. 
Regarding the scaling of the resistivities with the density, the truncated-MRN case is similar to the no-dust case (Fig.~\ref{fig:nodust} and Sect. \ref{sec:nodust}) because the perpendicular conductivity is dominated by ions all through the collapse ($\sigma_\perp \propto \ni$). 
In the MRN-case at densities lower than $10^9\,\cmc$ ($\sigma_\perp \propto \nH$), the perpendicular conductivity is dominated by dust grains. As a consequence, the Hall and ambipolar diffusion resistivities decrease more steeply with the density: $\eta_{\rm AD}\simeq 1/\sigma_\perp \propto 1/\nH$ and $\eta_{\rm H} \simeq \sigma_H/\sigma_\perp^2 \propto 1/\sqrt{\nH}$. These distinct scalings result in values of the MHD resistivities at intermediate densities ($10^8\,\cmc$) that can be a factor 10 to 100 lower in the MRN case than in the truncated-MRN case despite a higher value in the parent cloud, with the value for the Hall resistivity eventually exceeding that of ambipolar diffusion for densities higher than $\sim 10^8\,\cmc$ in the MRN case.

Overall, these results are in good agreement with those of \cite{marchand2016}, an other model based on an MRN size distribution without dust evolution (see their figures 3 and 5), and also those of \cite{zhao2016} who found a reduced magnetic braking for a truncated-MRN distribution compared to a MRN distribution.}

\subsection{Collapse with accretion and coagulation}

\begin{figure*}
\includegraphics[width=\thsize]{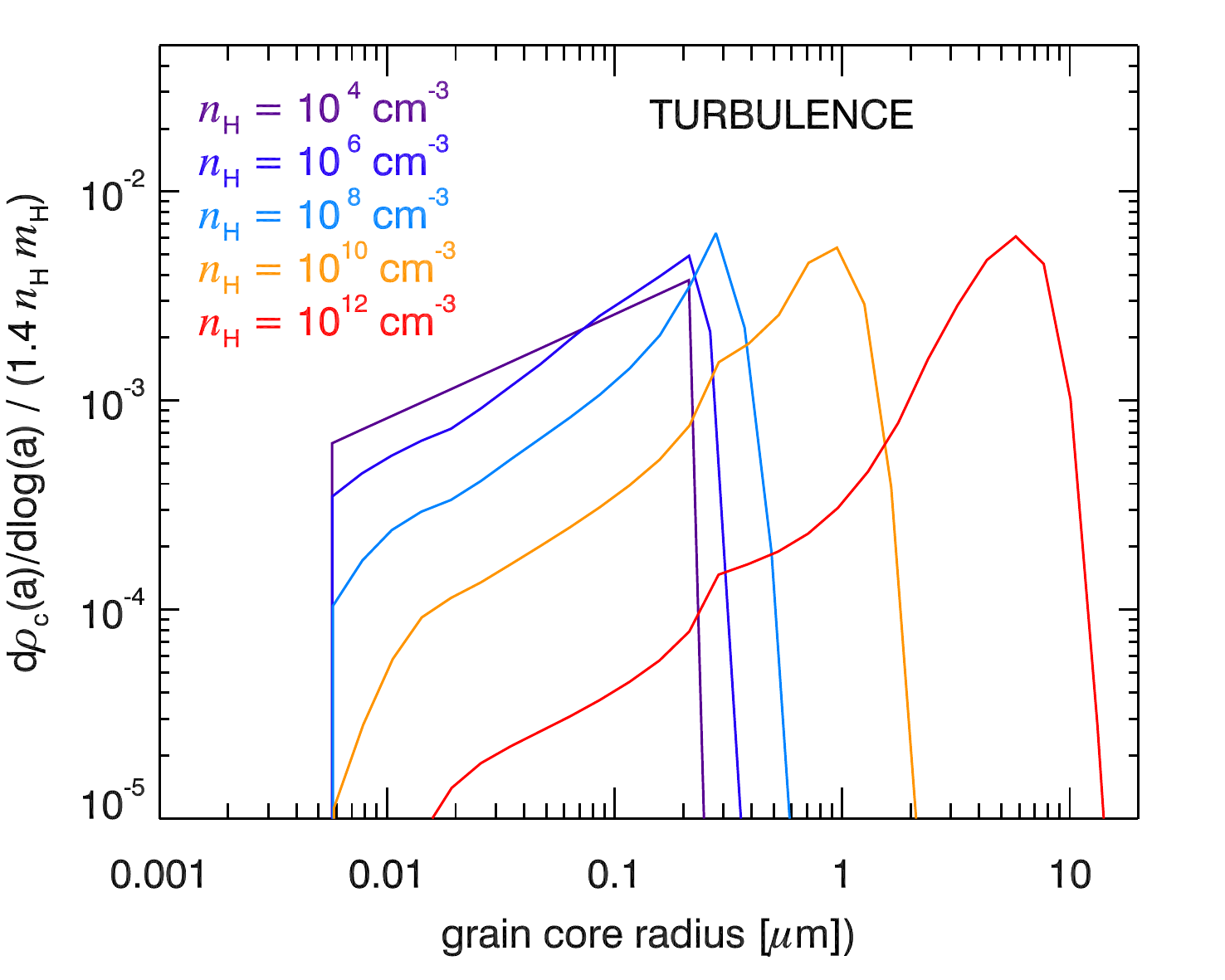}
\includegraphics[width=\thsize]{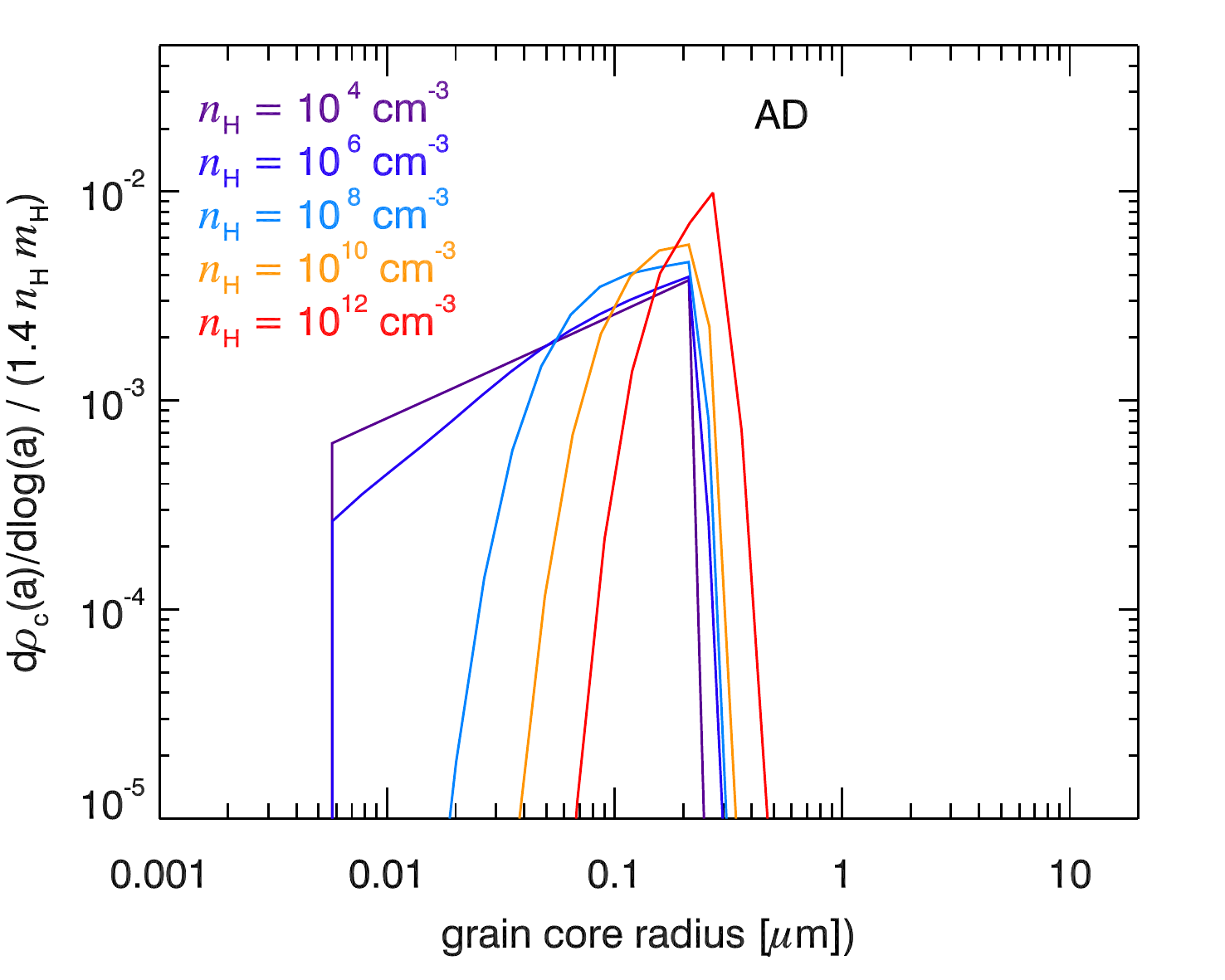}
\includegraphics[width=\thsize]{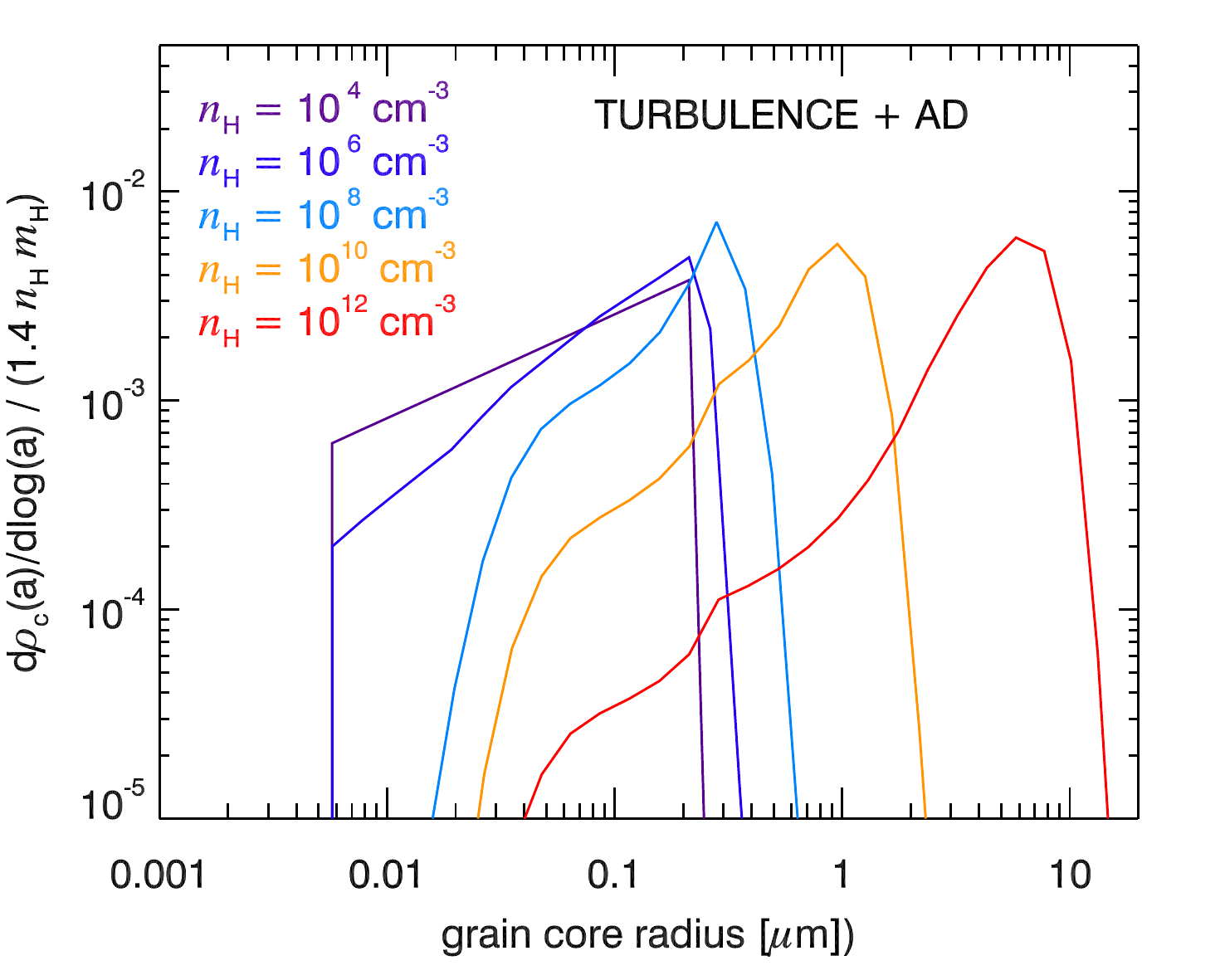}
\includegraphics[angle=-90,width=\thsize]{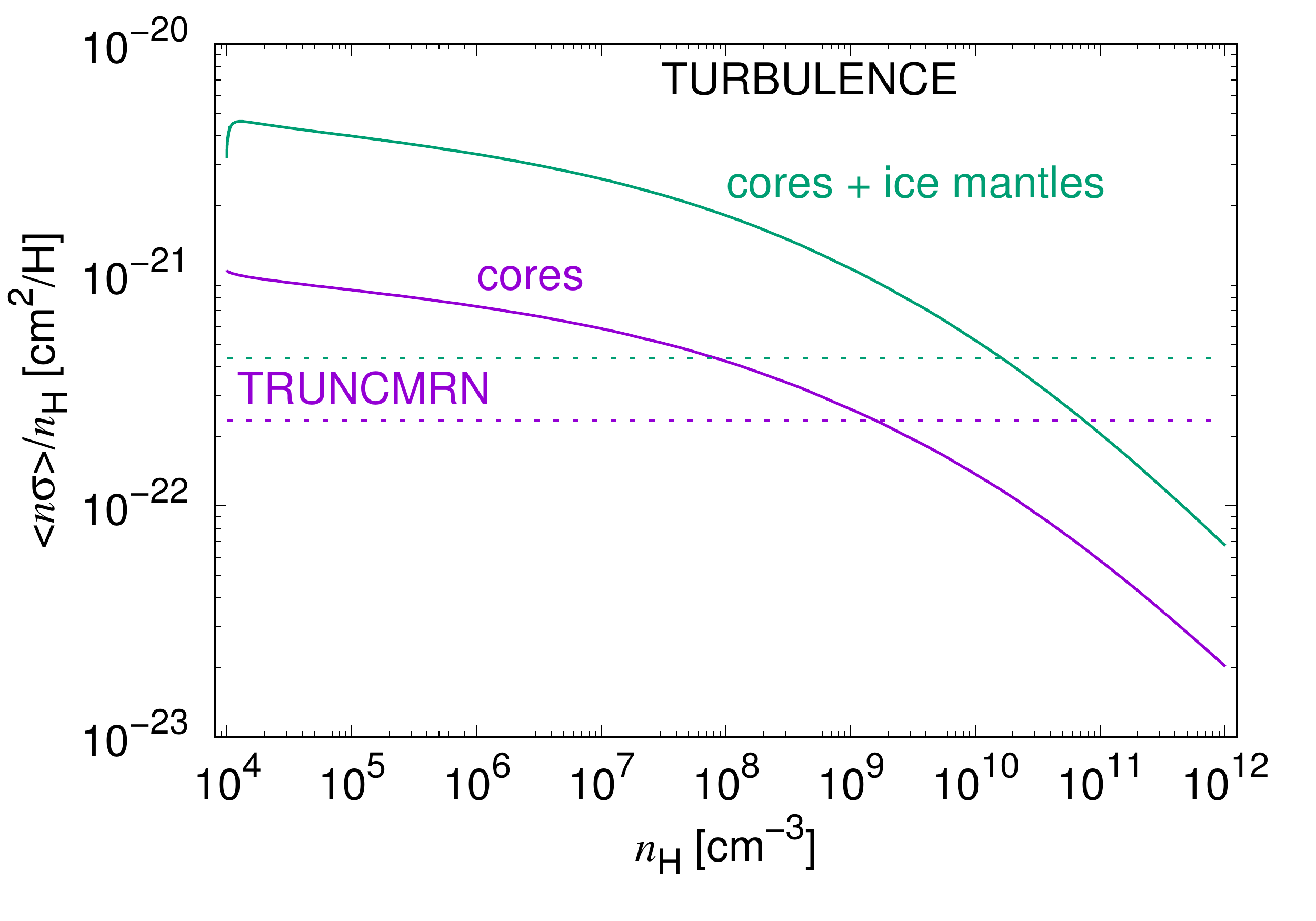}
\includegraphics[angle=-90,width=\thsize]{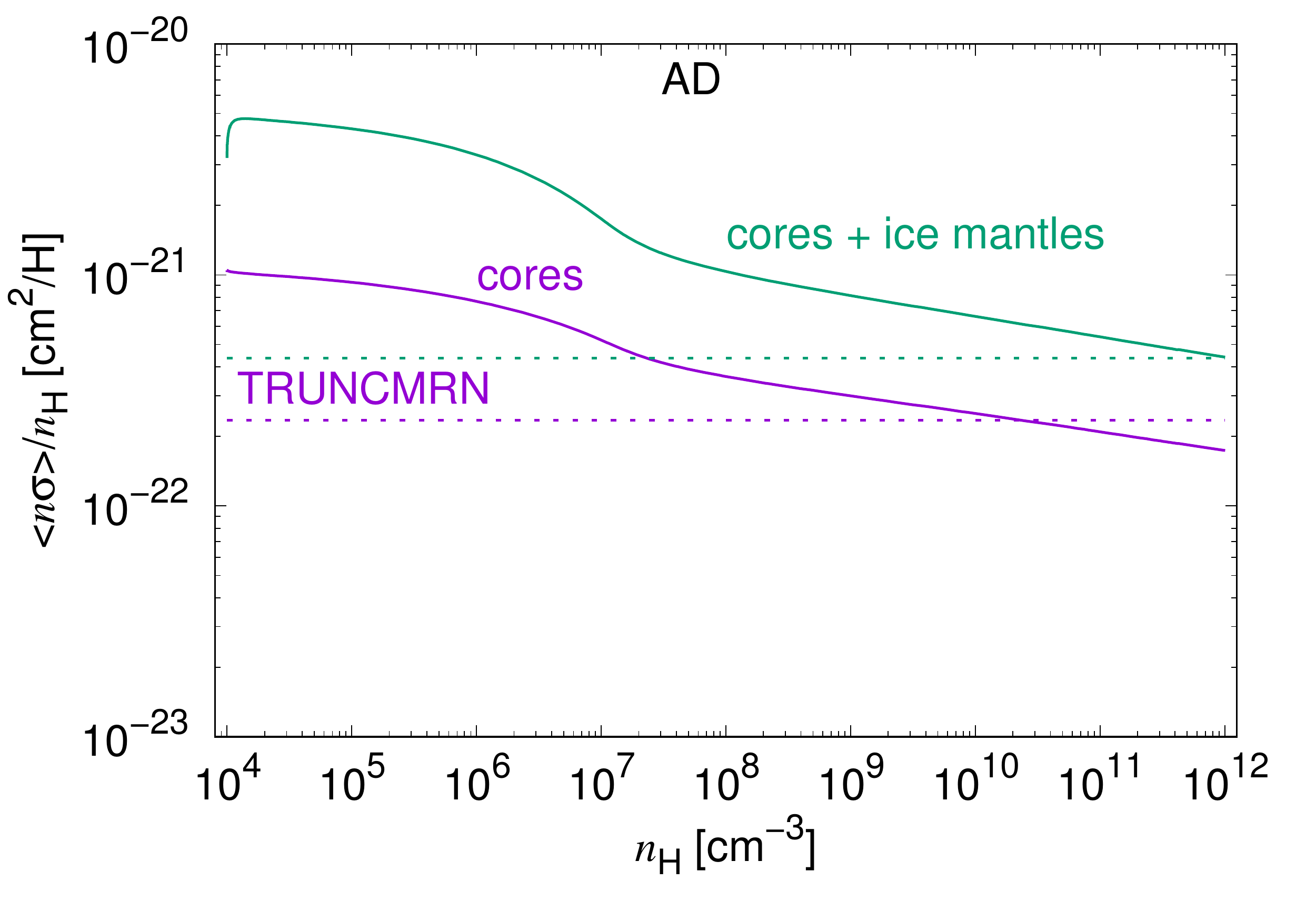}
\includegraphics[angle=-90,width=\thsize]{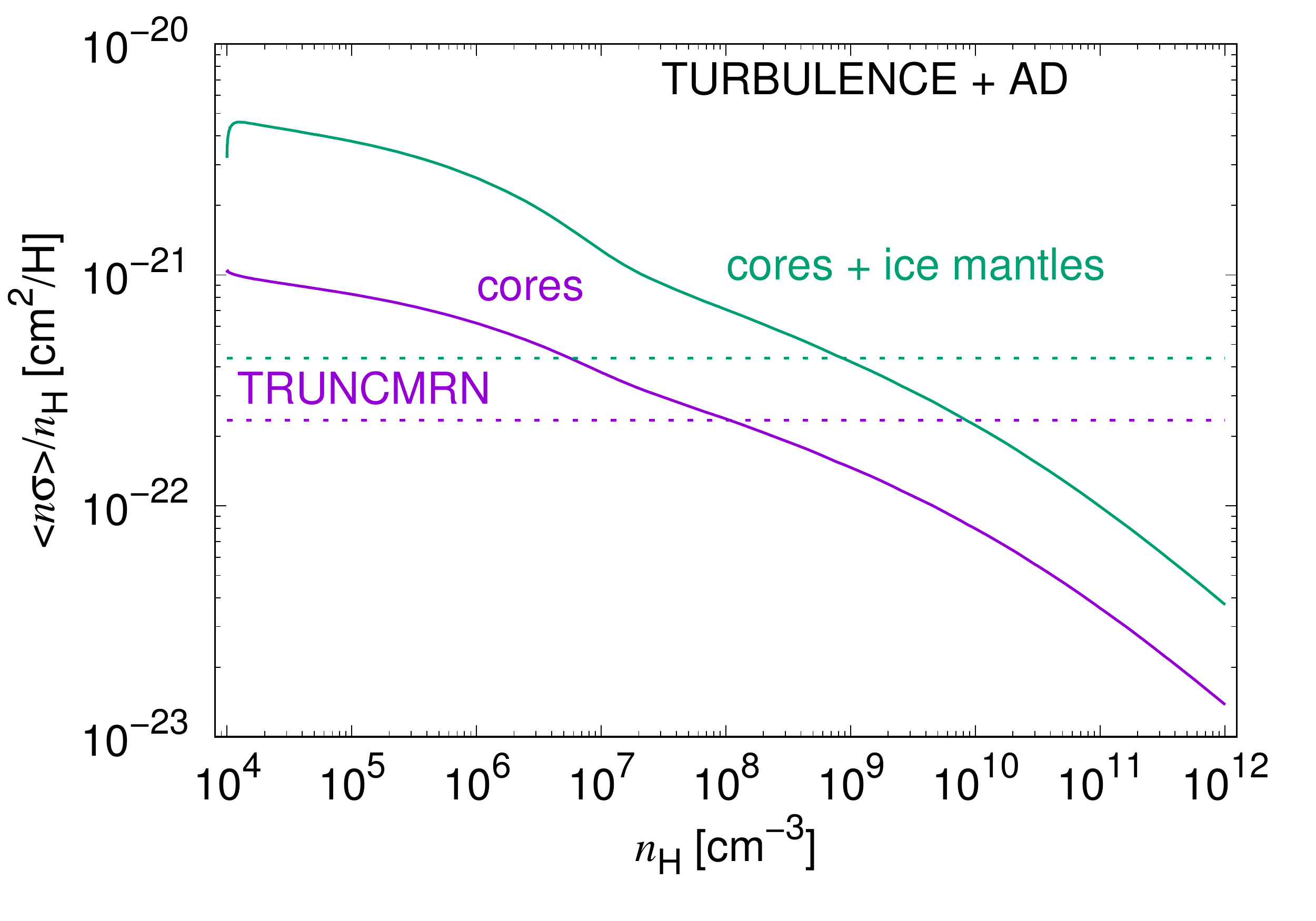}
\includegraphics[angle=-90,width=\thsize]{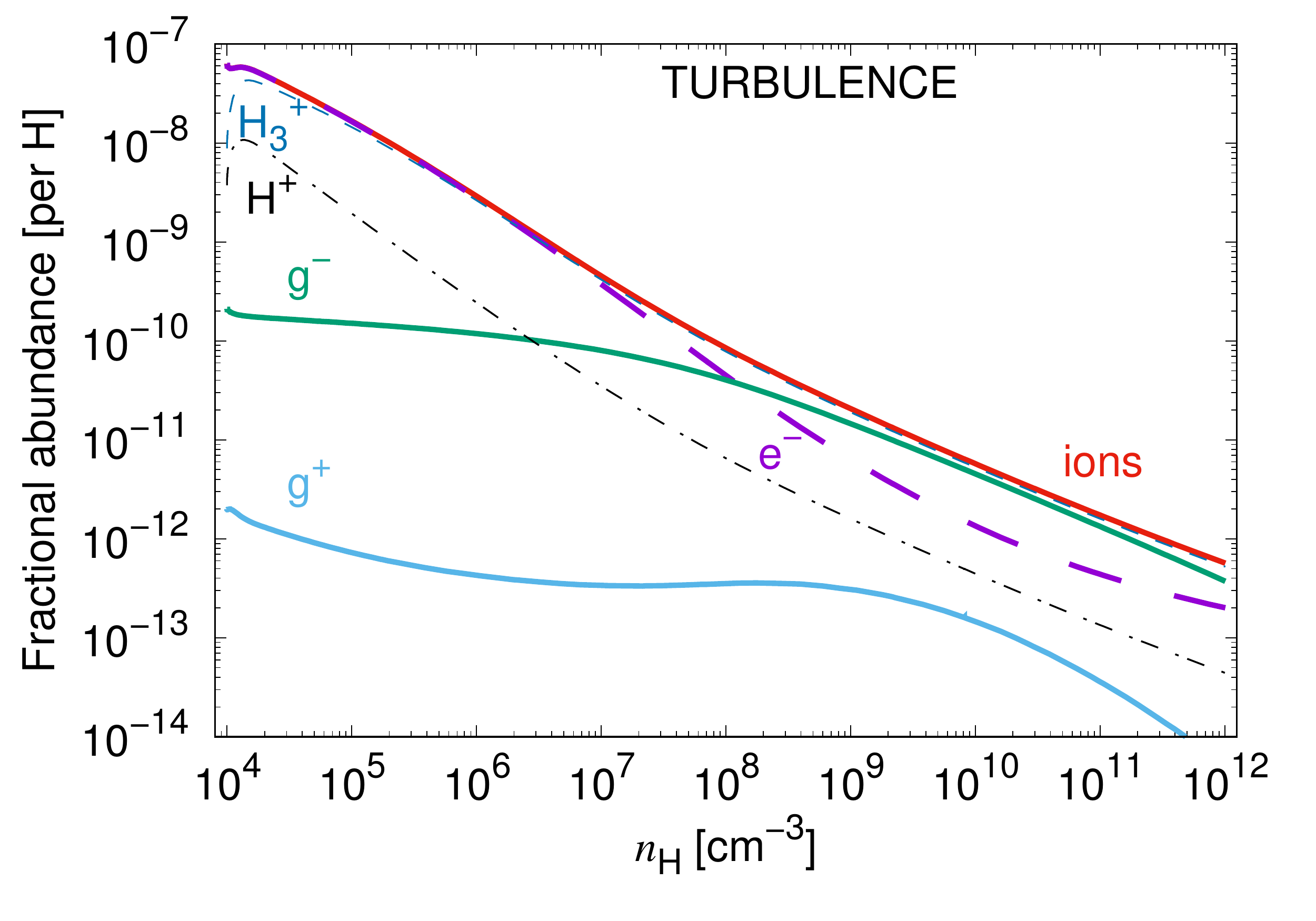}
\includegraphics[angle=-90,width=\thsize]{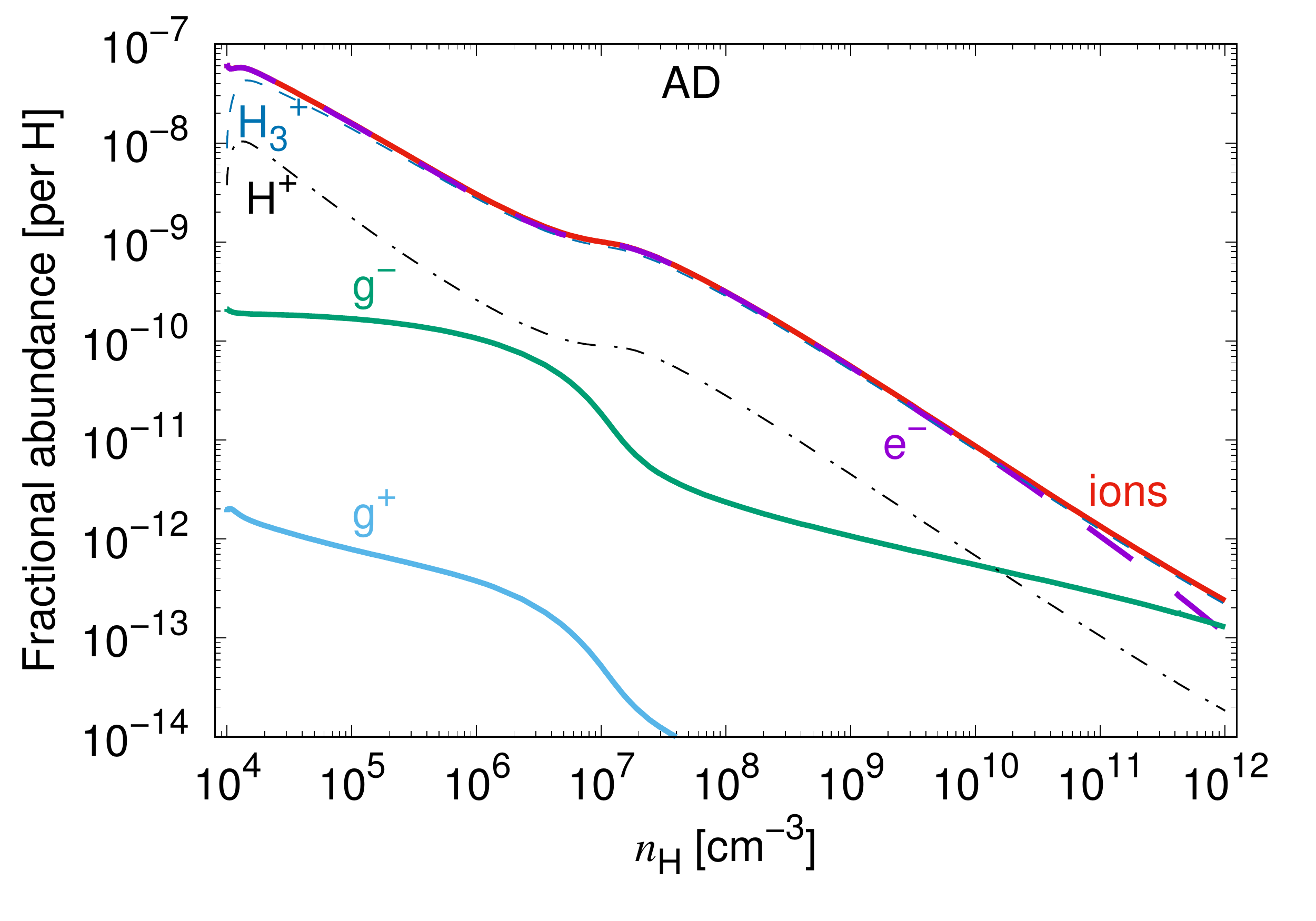}
\includegraphics[angle=-90,width=\thsize]{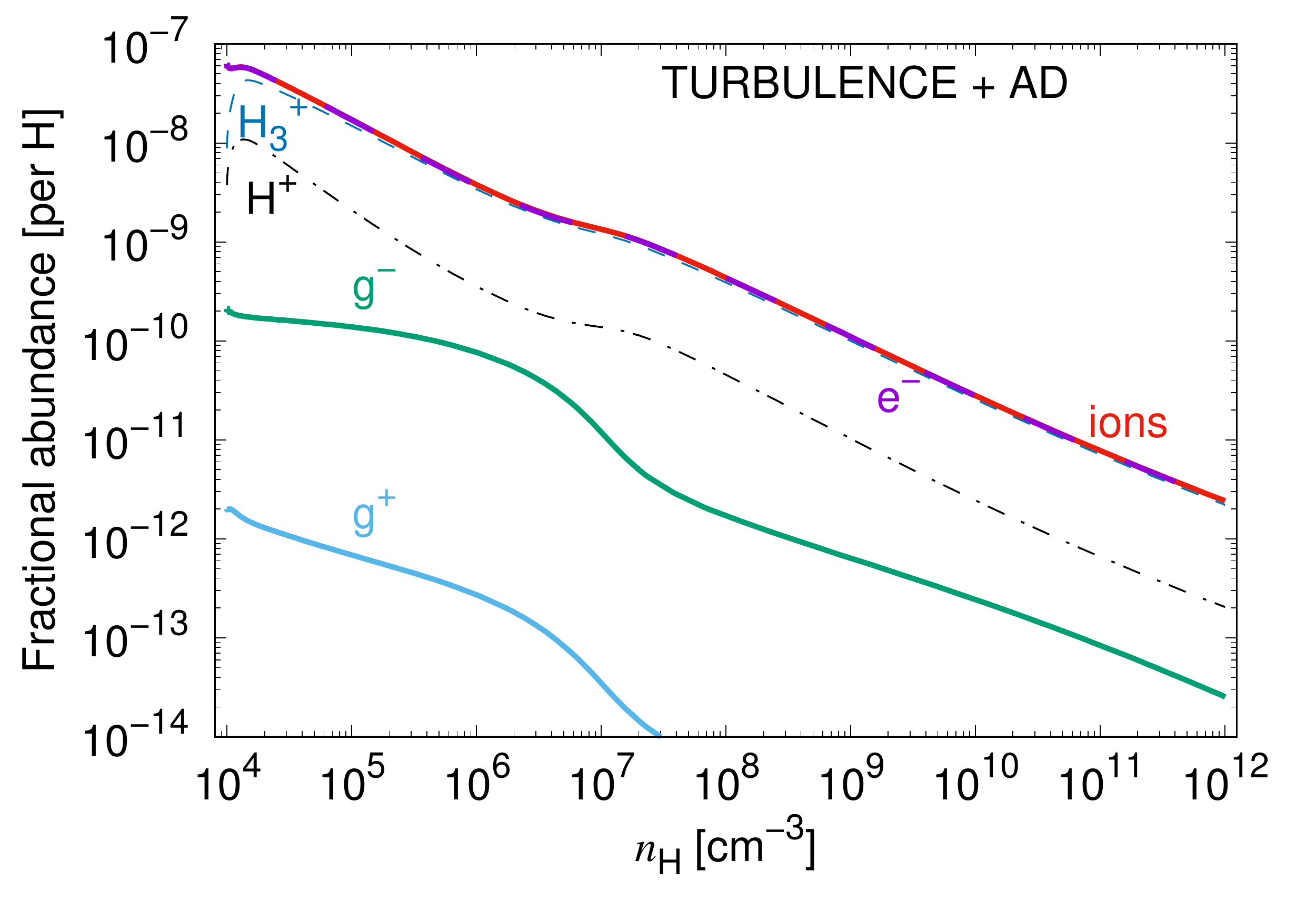}
\includegraphics[angle=-90,width=\thsize]{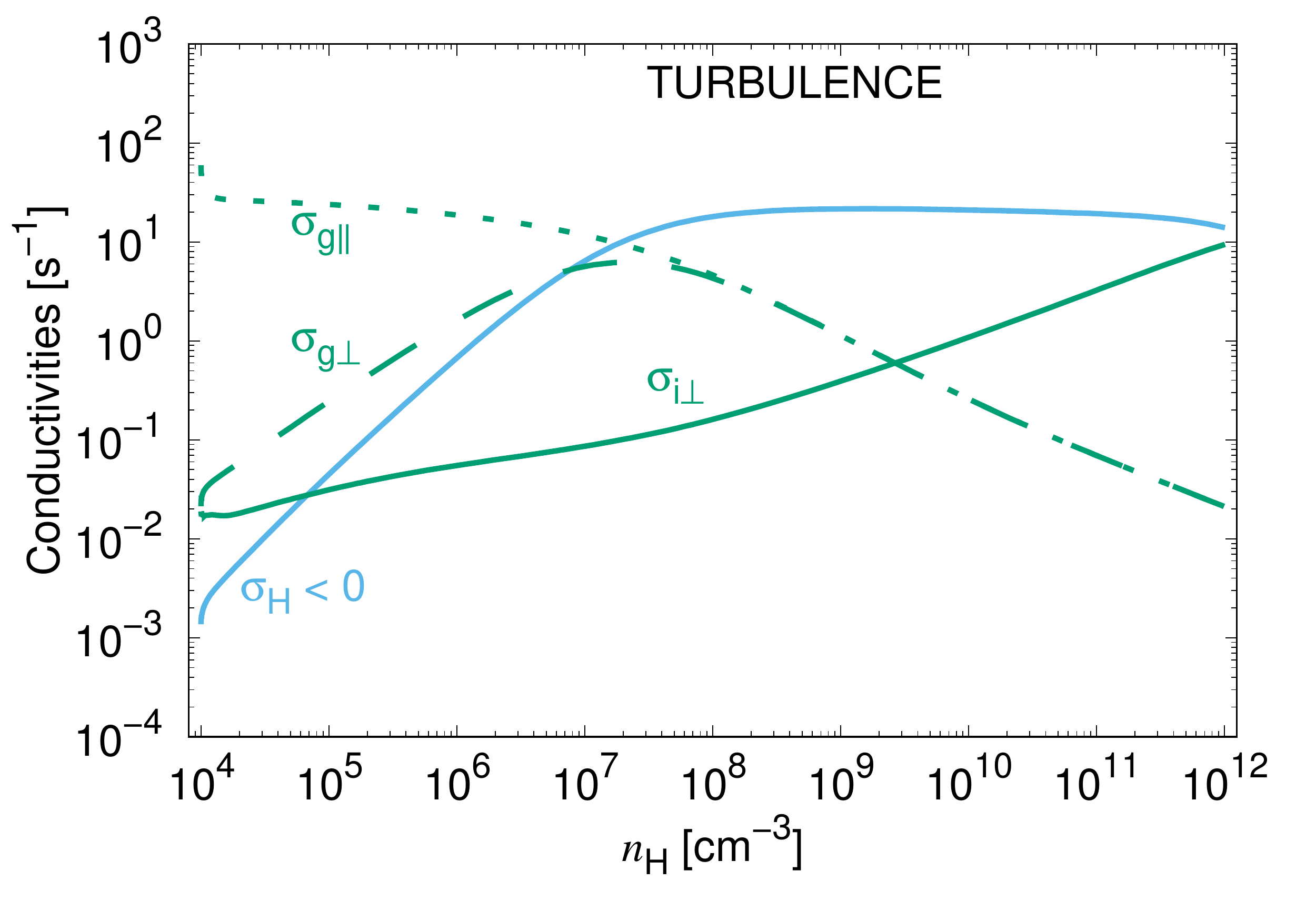}
\includegraphics[angle=-90,width=\thsize]{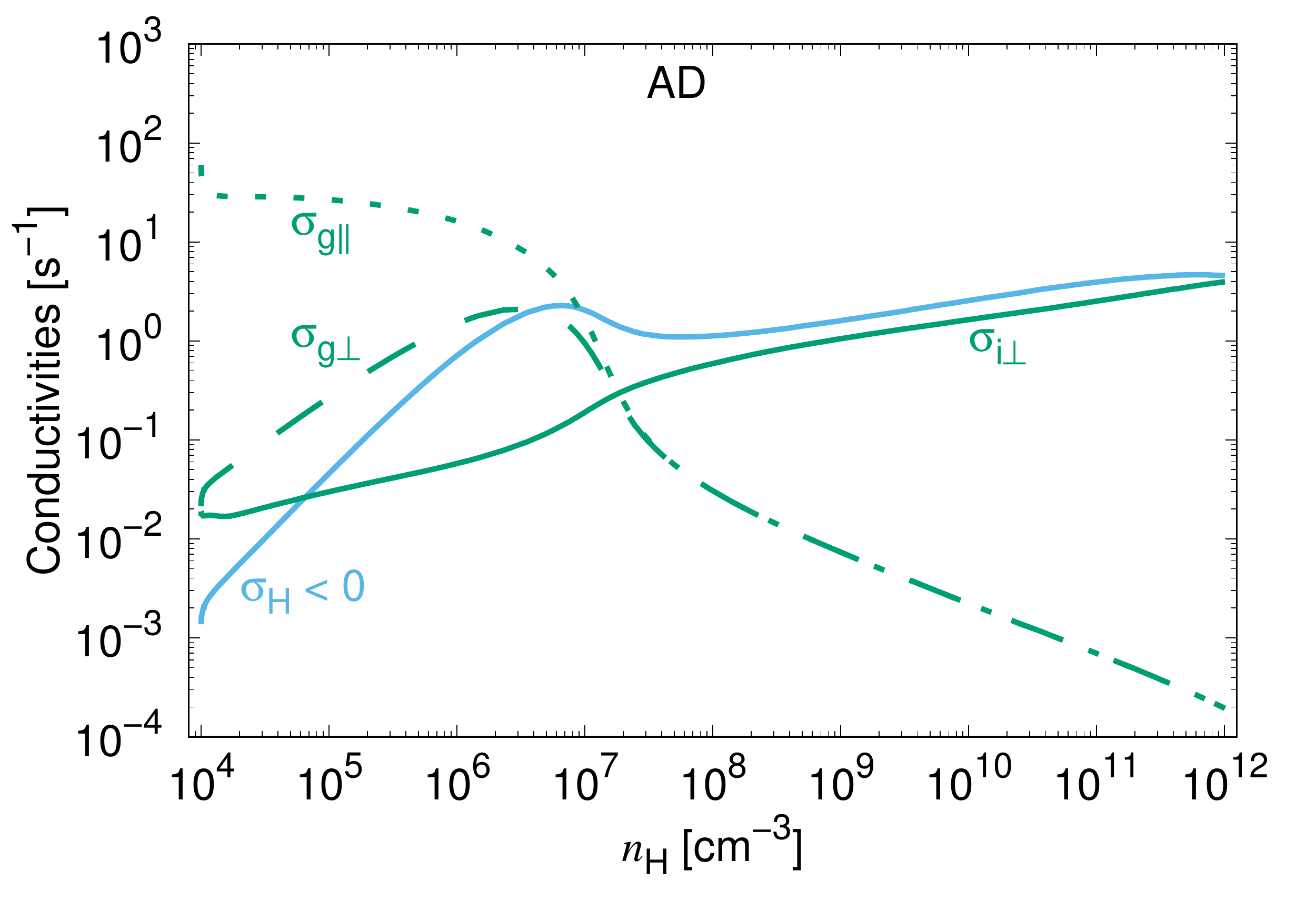}
\includegraphics[angle=-90,width=\thsize]{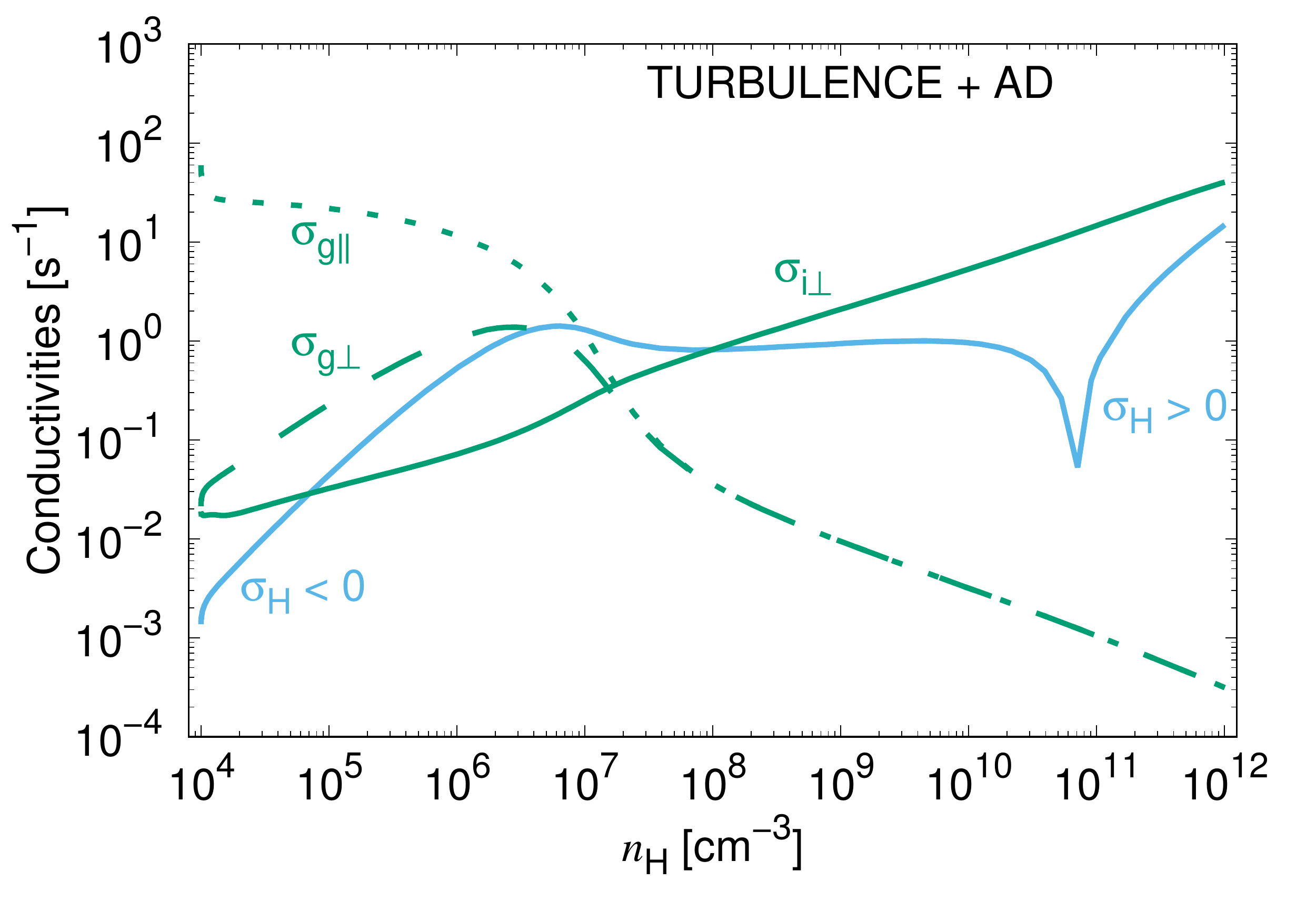}
\includegraphics[angle=-90,width=\thsize]{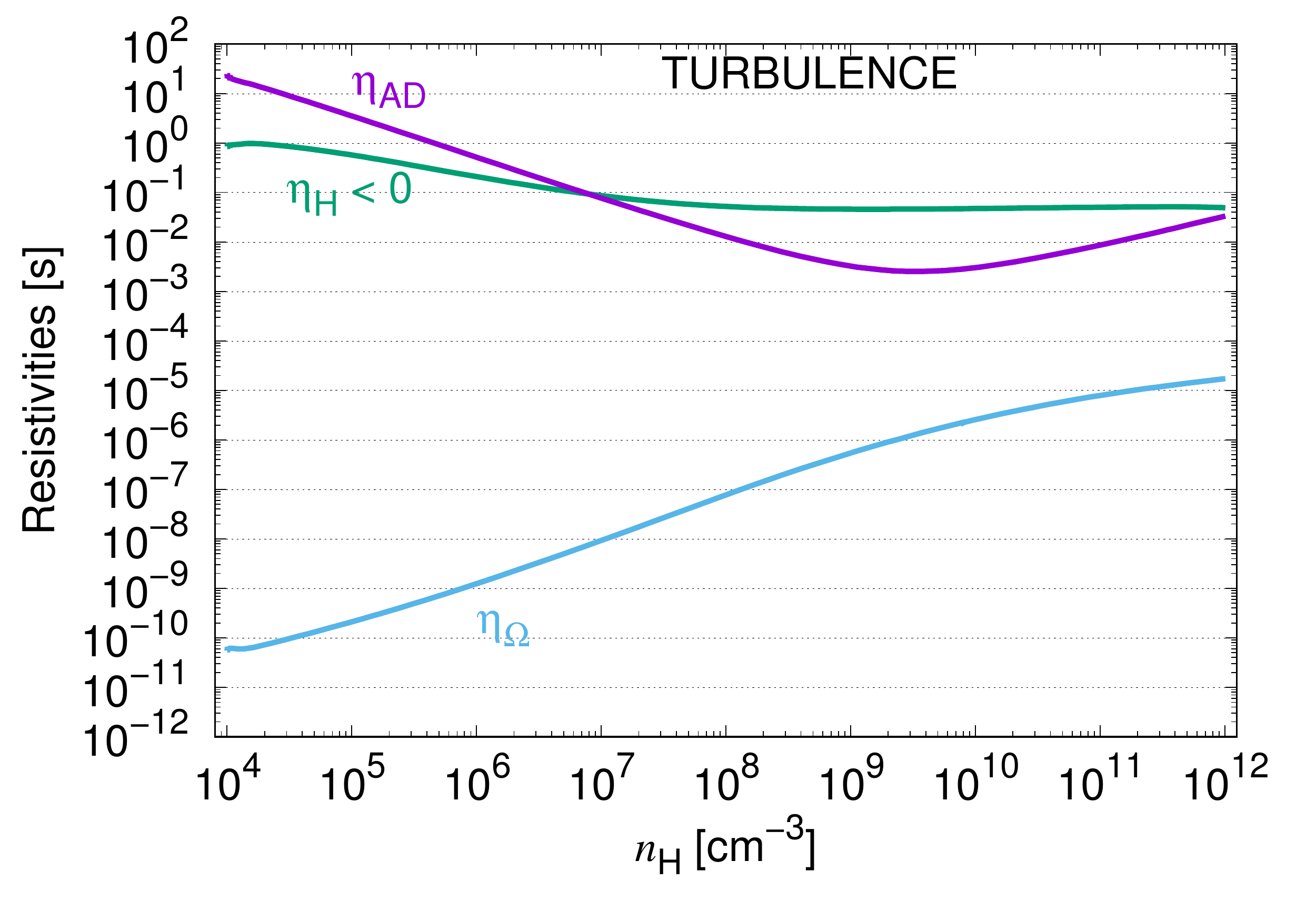}
\includegraphics[angle=-90,width=\thsize]{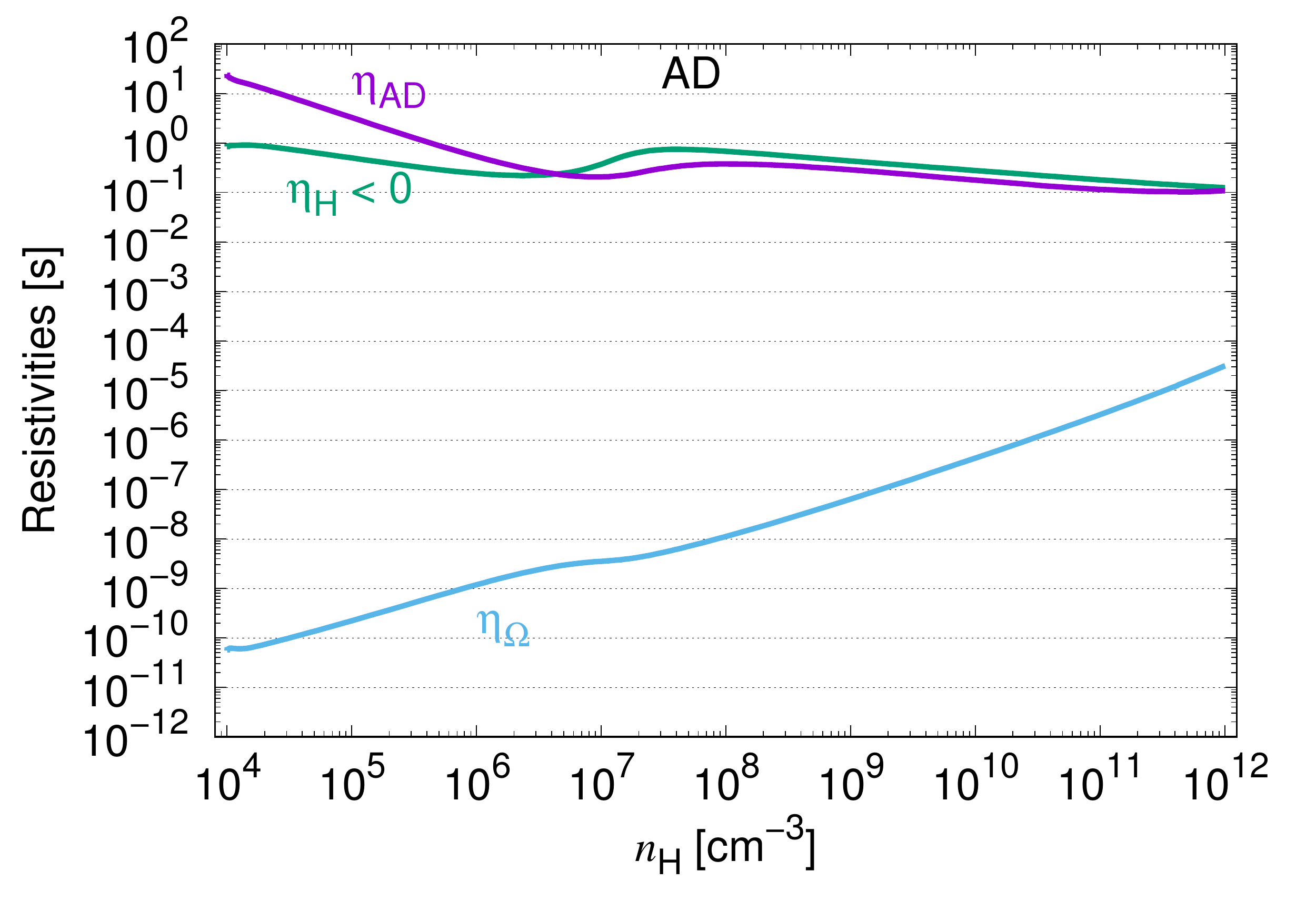}
\includegraphics[angle=-90,width=\thsize]{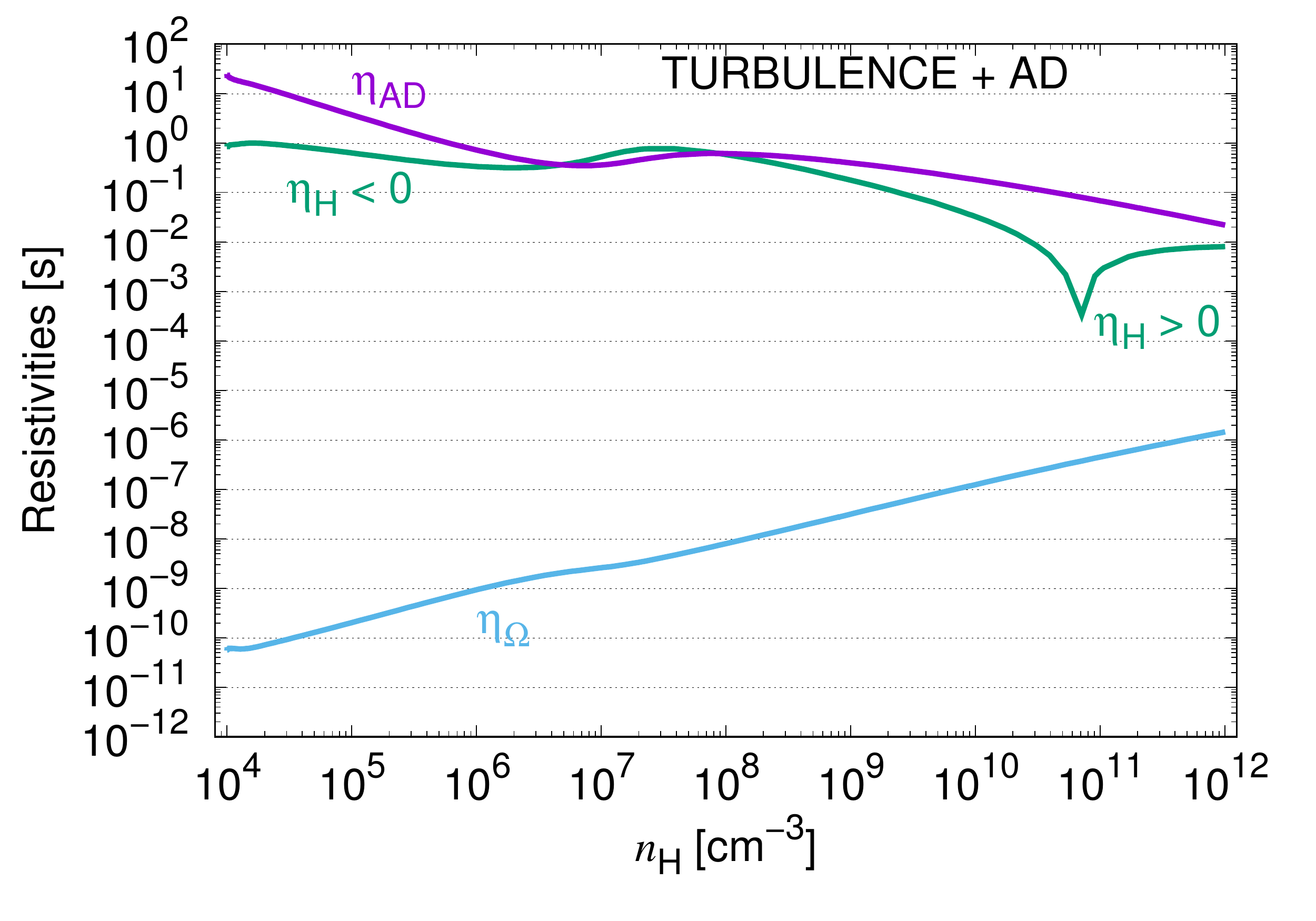}
\caption{Results for our standard model with grain accelerated by turbulence only \L\, ambipolar diffusion only \C, and both mechanisms \R. From top to bottom : dust size distribution at increasing densities, total dust cross-section per H \revise{(with our results for the truncated-MRN model overplotted)}, ionisation, conductivities and resistivities. See Fig.~\ref{fig:stdnoevol} for a description of the last three rows.}
\label{fig:models}
\end{figure*}

Let us now turn to the effect of grain evolution, namely ice accretion and grain-grain coagulation, onto the grain size distribution and onto the MHD properties of the plasma. Figure~\ref{fig:models} presents our results for three different scenarios of the dust velocity field : 1) turbulence only, 2) ambipolar diffusion only, and 3) turbulence and ambipolar diffusion (see Sect. \ref{sec:dynamics} for a description of grain dynamics by turbulence and ambipolar diffusion). 

The first row of Fig.~\ref{fig:models} presents the evolution of the size distribution of grain cores (\ie\ of the refractory part of the grain, covered by icy mantles) during core collapse when dust coagulation and ice accretion are activated.
In the early stage of the infall (Fig.~\ref{fig:models}, $\nH < 10^6\,\cmc$), coagulation removes small grains by sticking them onto large grains. This does not however significantly change the upper limit of the grain size distribution. 
The removal of small grains is faster when the effect of ambipolar diffusion on grain velocities is included (\textit{center} and \textit{right}) because in that case, unlike with turbulence, small grains decouple from the gas, and therefore also from the large grains which follow the gas. \revise{Depending on the density, this decoupling is more intense along B or perpendicular to B (see Fig.~\ref{fig:Vdist-AD}).
Note that, to compare our results with the truncated-MRN, the disappearing of grains smaller than $0.1\,\mu$m only happens late in the collapse ($\nH \ge 10^{10}\,\cmc$).}
When the acceleration of dust grains by turbulent eddies is included (\textit{left} and \textit{right}), grain growth proceeds rapidly, the grain size distribution becoming more and more dominated by a single size approaching $\sim 10\,\mu$m at $\nH=10^{12}\,\cmc$. 
To summarize, turbulence makes grain grow in size at later times but \revise{only slowly removes small grains, while ambipolar diffusion efficiently removes smaller grains at early times but does not allow to increase the grain size significantly}. In Sect.~\ref{sec:graingrowth}, we compare our results for the grain growth rate with the others studies.

The second row of Fig.~\ref{fig:models} presents how the total geometrical cross-section of grains evolves during the collapse. The total dust cross-section is a key ingredient of the plasma physics, because it controls the recombination rate of ions and the perpendicular conductivity of dust grains (see Eq.~\eqref{eq:sigmaperpGammagg1}), which usually dominates the total perpendicular conductivity. \revise{Starting with values characteristic for the MRN size distribution already covered by a 8.8 nm ice mantle (see Sect.~\ref{sec:init})}, the total dust cross-section first slightly increases through the growth of icy mantles on the surface of grains ($\nH \le 2.10^4\,\cmc$), a process that is fast enough to complete before any significant coagulation through grain-grain collisions has occurred. Once started, grain-grain coagulation systematically decreases the total dust cross-section. When drift velocities by ambipolar diffusion are included, the faster removal of small grains accelerates this process at low densities ($\nH \le 10^6\,\cmc$). Still, it is only when the acceleration of grains by turbulence is included that the total dust cross-section keeps on decreasing all through the collapse ($\nH \ge 10^6\,\cmc$). \revise{The dust size distribution reaches total cross-sections per H that are comparable to that for our truncated-MRN only late in the collapse ($\nH \sim 10^9 - 10^{10}\,\cmc$)}.

The third row of Fig.~\ref{fig:models} shows 
how the total number of negative charges carried by grains is affected by grain-grain coagulation. Unlike in the case without coagulation (Fig.~\ref{fig:stdnoevol}), grains never become the dominant charge carrier characteristic of a dusty plasma (see Sect. \ref{sec:noevol}). The ionization is always dominated by ions, though we can observe at high densities (from $\nH \ge 10^7\,\cmc$) a small depletion of free electrons, characteristic for a dust-ion plasma \citep{Ivlev16}, when ambipolar diffusion drift velocities are ignored.

The fourth row presents the evolution of the parallel, perpendicular and Hall conductivities of the plasma. The total parallel conductivity (not shown) 
is high and always controlled by the electron density. As explained in the previous section, the perpendicular conductivity is dominated by dust grains at low densities, and ions at high densities. The density threshold between these two regimes depends on the abundance of the smaller grains, which are the last grains to decouple from the magnetic field as the density increases. For an initial MRN distribution, this transition happens at low density when the removal of small grains is efficient, as it is the case when grain drift velocities are included ($\nH\simeq 10^6-10^7\,\cmc$), and at high densities ($\nH\simeq 10^9-10^{10}\,\cmc$) when only turbulence is considered or, similarly, when coagulation is ignored (Fig.~\ref{fig:stdnoevol}). From this analysis, we conclude that an efficient removal of small grains stops the increase of the Hall and ambipolar diffusion conductivities with the density, while the growth of large grains does not in itself affect these conductivities. When the density is high enough, or the mean grain size high enough, the Hall conductivity can change from being negative to being positive, implying that the plasma Hall conductivity is then dominated by ions.

The bottom row presents the evolution of the Ohmic, Hall and ambipolar diffusion resistivities with the density. The removal of small grains limits the drop of the Hall and ambipolar diffusion resistivities with the density, increasing their values by a factor 10-100 compared with models where the coagulation is only triggered by turbulent velocities (left panel), or with models with an MRN distribution and no coagulation at all (Fig.~\ref{fig:stdnoevol}, \revise{bottom} row). 
\revise{Comparing the amplitude of MHD resistivities in the MRN-case with coagulation and ambipolar diffusion drift velocities (Fig.~\ref{fig:models}) and in the truncated-MRN case without coagulation (Fig.~\ref{fig:stdnoevol}), the ambipolar diffusion resistivities have similar values at intermediate densities ($\nH \sim 10^8\,\cmc$) while the Hall resistivity remains at least one order of magnitude higher in the former than in the latter case at densities lower than $\sim 10^9\,\cmc$. At a density of $10^8\,\cmc$, the MHD properties of the model with coagulation and of the truncated-MRN model still differ significantly, being dominated by dust for the former, and by ions for the latter. To obtain high values for the Hall and ambipolar diffusion resistivities necessary to the formation of a larger disk, it may therefore not be necessary to assume, as per the truncated-MRN model, that all grains smalller than $0.1\,\mu$m have been removed by coagulation in the parent cloud before the collapse : removing only parts of those grains through grain-grain coagulation should also achieve it if this happens early in the collapse, as is the case when grain drift velocities generated by ambipolar diffusion are included. 
}

\begin{figure}
\includegraphics[angle=-90,width=\hhsize]{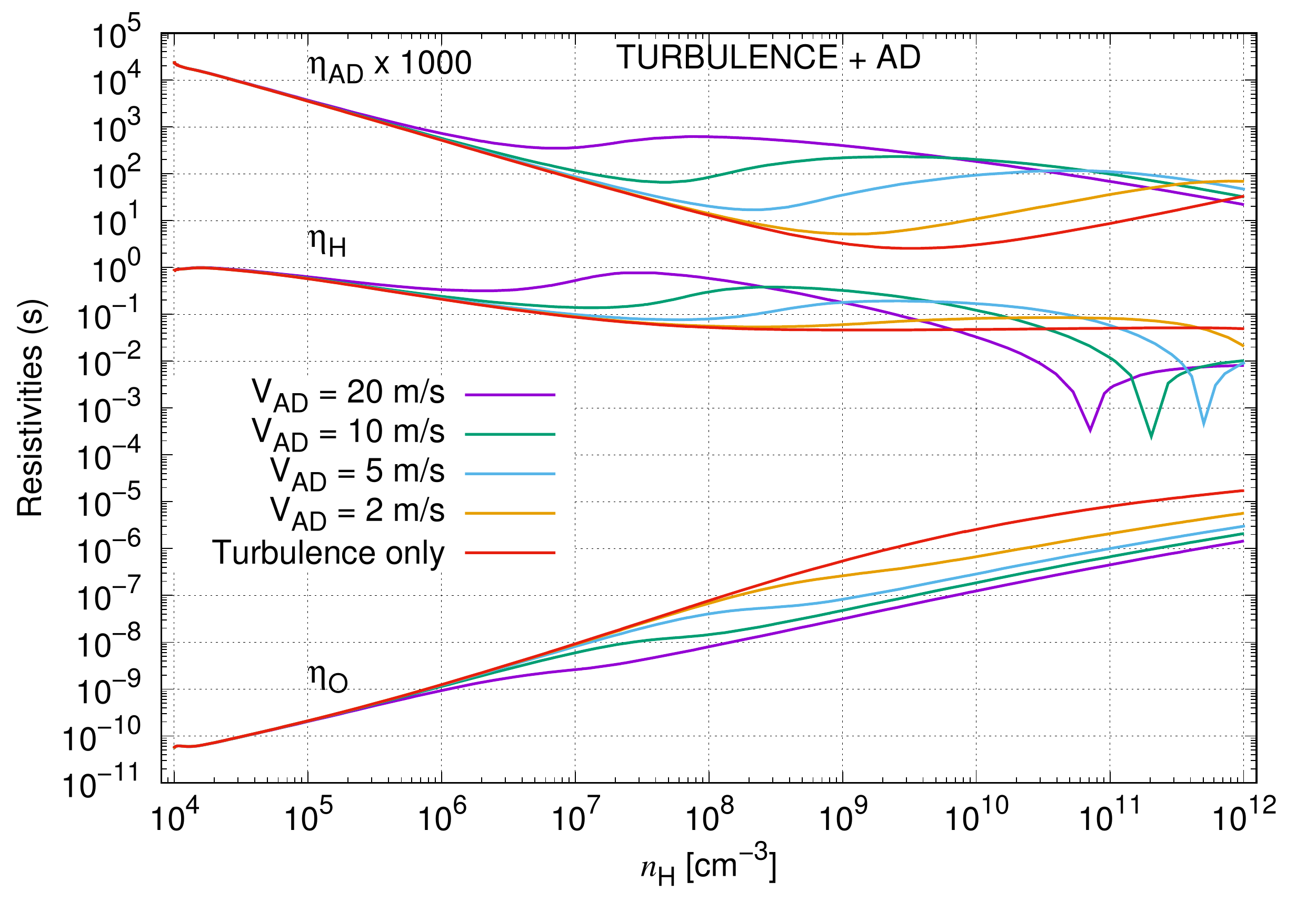}
\caption{Impact of the initial value of the ambipolar diffusion velocity on the evolution of the MHD resistivities with the density. \revise{Both turbulent and drift velocities are included.}
}
\label{fig:cmp_AD}
\end{figure}

Figure~\ref{fig:cmp_AD} summarizes our results concerning the influence of the efficiency in the removal of small grains on the profiles of the plasma resistivities. We present our results for 5 models starting with the same MRN distribution covered by ices, differing only by the initial (\ie\ in the parent cloud) value $V_{\rm AD}^{\rm ref}$ of the ambipolar diffusion velocity, from 0 to 20 m/s. The value $V_{\rm AD}^{\rm ref}$ of the ambipolar diffusion velocity is used here as a way to control the efficiency in the removal of small grains. It can also be considered as a way to quantify the balance in the competition between coagulation and fragmentation of grains in higher velocity impacts, \revise{that we remind is not included in this work.}
From Fig.~\ref{fig:cmp_AD}, we see that 
\revise{the more efficient the removal of small grains, the higher the Hall and ambipolar diffusion resistivities at intermediate densities ($10^6 < \nH < 10^9\,\cmc$), here again confirming the results obtained by \cite{zhao2016}. This trend, opposite to what was observed in the parent cloud (see Sect.~\ref{sec:noevol}), is observed whenever the plasma perpendicular conductivity is dominated by dust grains, and not ions.}
Our conclusions regarding the \revise{evolution of the} magnitude of the MHD resistivities during the collapse with dust coagulation therefore drastically depend on the detailled dynamics of small grains, and on the competition between coagulation and fragmentation (see Sect. \ref{sec:frag} for a discussion of the modeling of grain-grain fragmentation in core collapse).

\subsection{Influence of the Cosmic-Ray ionization rate}

\begin{figure}
\includegraphics[angle=-90,width=\hhsize]{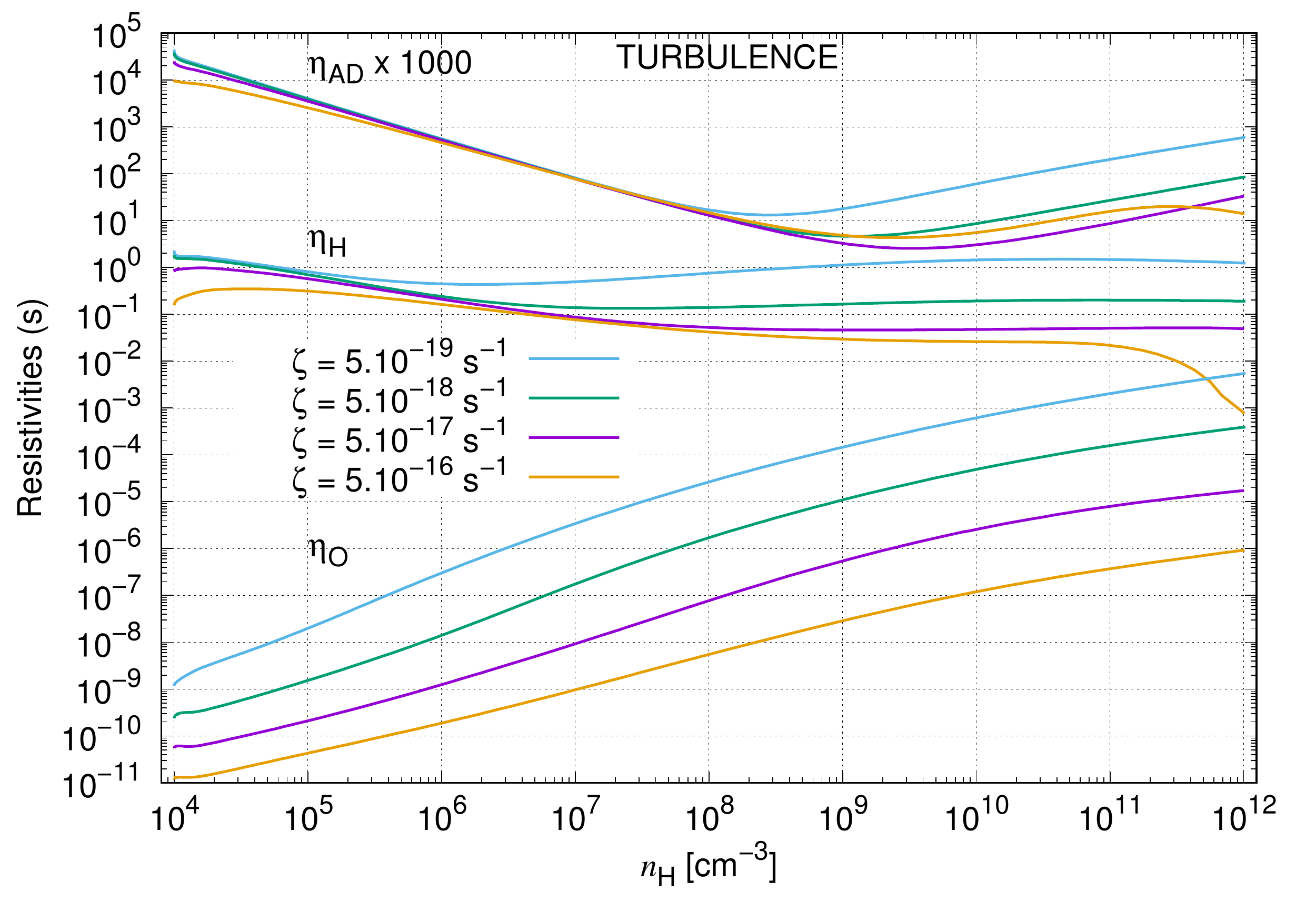}
\includegraphics[angle=-90,width=\hhsize]{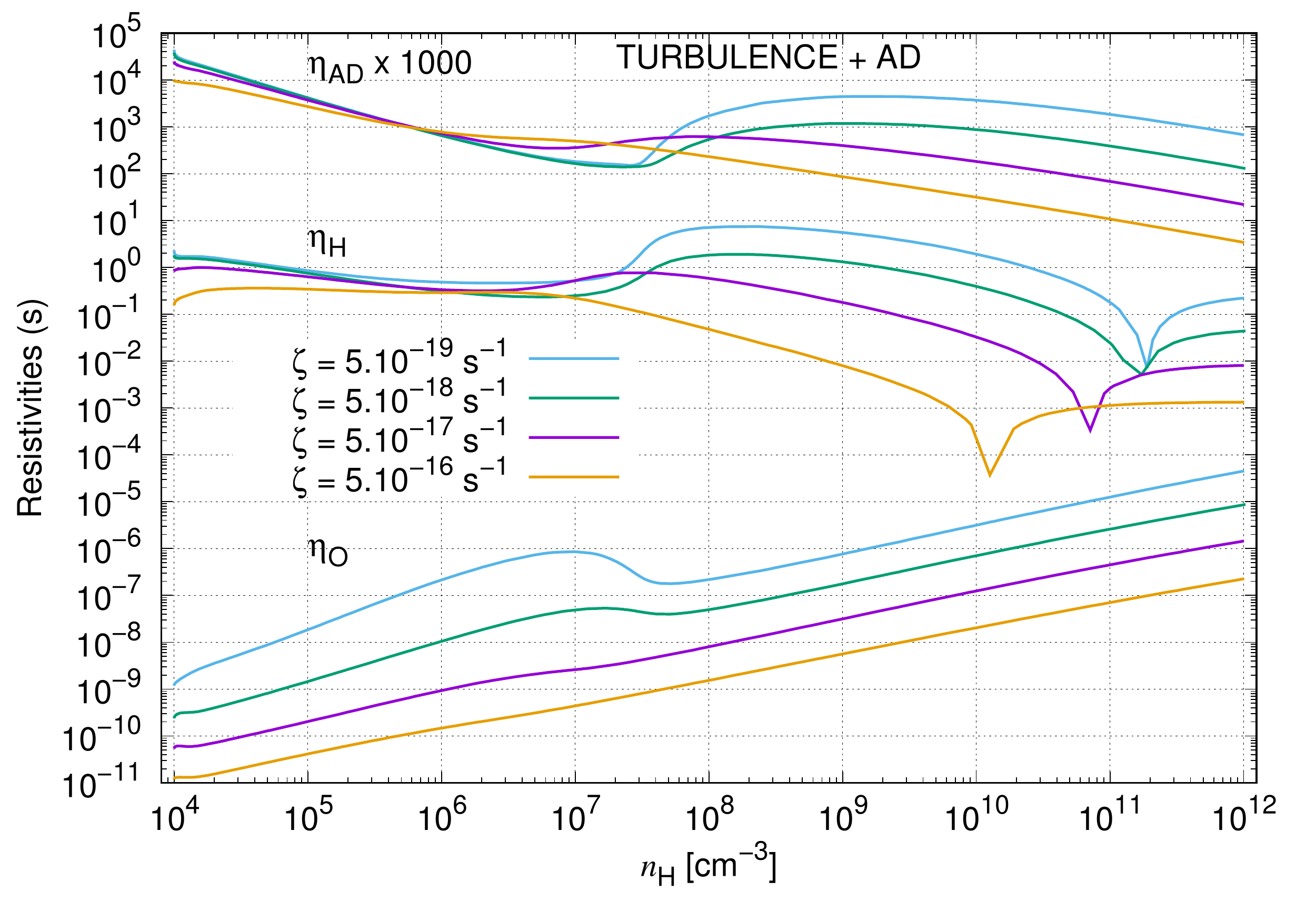}
\caption{Impact of the CR ionization rate $\zeta$ on the evolution of the MHD resistivities with the density for a model with relative velocities generated by turbulence \T, and turbulence and ambipolar diffusion with $V_{\rm AD}^{\rm ref}$ = 20\,m.s$^{-1}$ \B.
}
\label{fig:cmp_Zeta}
\end{figure}

The ionisation level, and therefore the MHD conductivities and resistivities, are not only controlled by the grain size distribution, but also by the cosmic-ray ionisation rate $\zeta$. 

In Fig.~\ref{fig:cmp_Zeta}, we present the evolution of the ionization level and MHD resistivities for four values\footnote{This interval of values ignores the possible contribution of local sources of cosmic-rays like supernova remnants \citep{Vaupre14} or young stars \citep{Ceccarelli14}. In these environments $\zeta$ can increase up to $\sim 10^{-14/-15}~\rm{s^{-1}}$.} of the cosmic-ray ionization rate ranging from $5.10^{-19}$ to $5.10^{-16}$\,\s, and two coagulation scenarios (turbulent velocities \textit{versus} turbulent and ambipolar diffusion velocities) corresponding to an inefficient \textit{versus} efficient removal of small grains, respectively. We recall that, for simplicity, the parameter $\zeta$ is kept constant through the collapse, a limit in our model that we discuss in Sect.~\ref{sec:CRnH}. Fig.~\ref{fig:cmp_Zeta} shows that, as expected, low values of the cosmic-ray ionization rate tend to produce high Hall and ambipolar diffusion MHD resistivities \revise{when the dust perpendicular conductivity is negligible compared to that of ions, but have almost no impact when the dust perpendicular conductivity dominates over that of ions. }

\begin{figure}
\includegraphics[angle=-90,width=\hhsize]{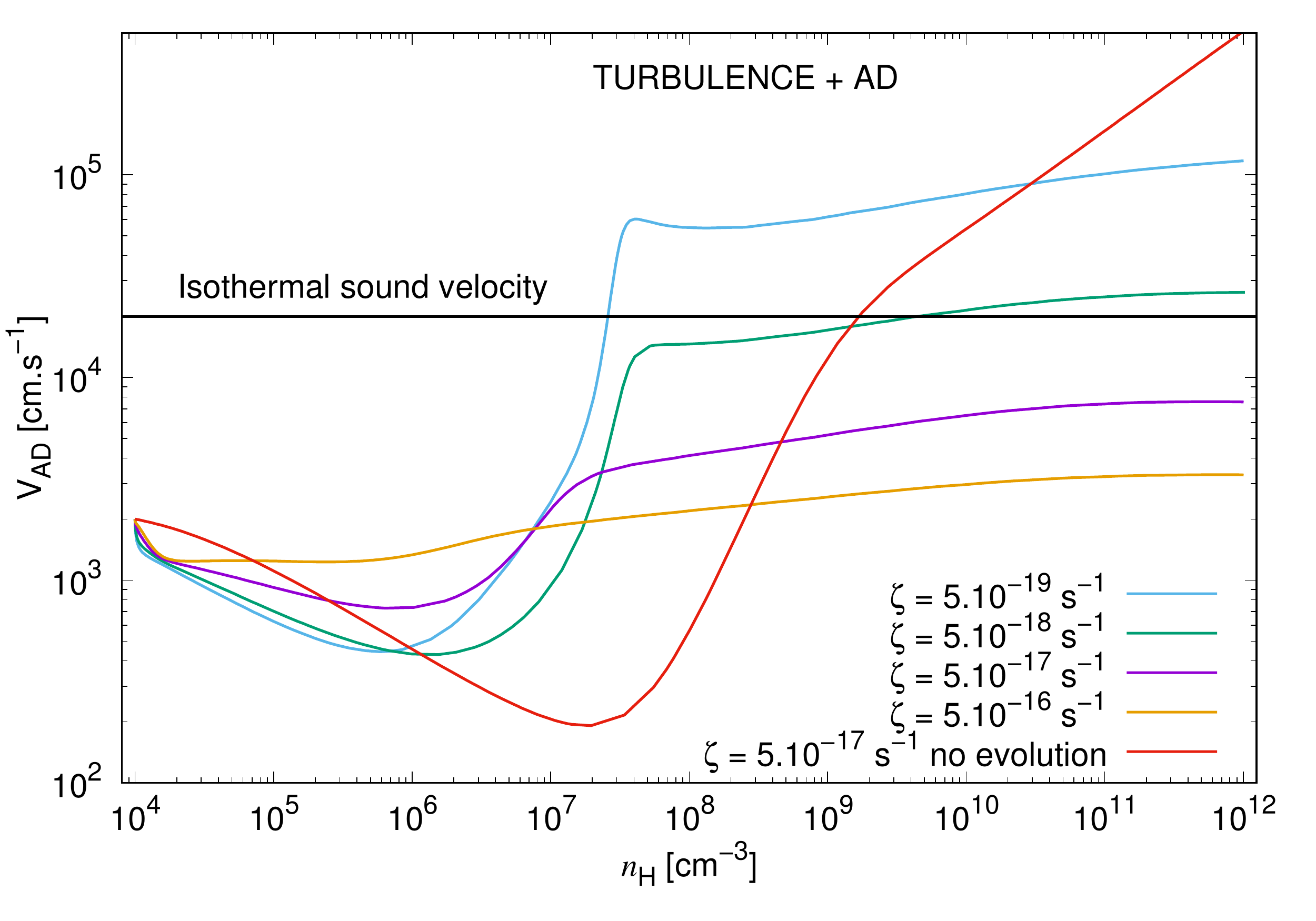}
\caption{Ambipolar diffusion velocity $\vAD$ as a function of the density, for different values of the CR ionization rate $\zeta$ and, for clarity, the same initial value in the parent cloud $\vADref= 20$\,m.s$^{-1}$. \revise{Calculations were done for our scenario with turbulent and ambipolar diffusion drift velocities. The scenario without dust evolution is added for reference.}}
\label{fig:VAD_Zeta}
\end{figure}

By modifying the ionization level, the variation of $\zeta$ in turn affects the coupling of the magnetic field with the gas, and therefore the ambipolar diffusion velocity. Figure~\ref{fig:VAD_Zeta} presents the evolution of $\vAD$ with the density for different values of the cosmic-ray ionization rate and, for clarity, for the same \revise{ initial value $V_{\rm AD}^{\rm ref}=20\,$m.s$^{-1}$ of the ambipolar diffusion velocity} in the parent cloud, meaning that we are interested by the relative, not absolute, evolution of $\vAD$ for different values of $\zeta$. \revise{We now build on our analysis of the dependence of $\vAD$ presented in Sect.~\ref{sec:Vdrift}.} For $\zeta=5\times10^{-16}$\,s$^{-1}$, the ionisation fraction is so high that the dust term at the denominator of Eq.~\eqref{eq:VAD} is negligible compared to the ion term and $\vAD$ is therefore almost constant. For lower values of $\zeta$, \revise{the dust term tends to dominate over the ion term, leading to} $\vAD \propto 1/\sqrt{\nH}$ at low density. Once the total dust cross-sections has significantly dropped through grain-grain coagulation and dust grains have started to decouple from the magnetic field lines ($\Gamma_k \simeq 1$, Eq.~\eqref{eq:VAD}), \revise{$\vAD$ \revise{starts to} increase}. At high density, the dust term has become negligible and $\vAD \propto 1/\sqrt{\zeta}$. \revise{The scenario without dust evolution follows the same trends at low density. At high density, the ion density remains constant and grains become neutral on average (Fig.~\ref{fig:stdnoevol}), leading to the asymptotic behaviour $\vAD \propto \sqrt{\nH}$.}

\section{Discussion}\label{sec:Discussion}

In this section, we discuss some limitations in our modeling that may affect our conclusions, and some perspectives of our work.

\subsection{Impact of fragmentation in grain-grain collisions}\label{sec:frag}

For reason of simplicity, our modeling of the evolution of the dust size distribution ignores the possibility that the colliding grains would fragment each other, partially or totally, in the collision. This is a strong hypothesis that is probably not justified. While the fragmentation of solid particles necessitates relative velocities of the order of a few km.s$^{-1}$ \citep{Tielens1994} which most probably do not exist in dense clouds \citep{OC07,YLD04}, the fragmentation of porous aggregates can happen at much lower velocities (a few m.\s) which are relevant for study of grain growth in dense clouds \citep{WR93,Ormel2009}. Note that, unlike for solids, the outcome of the collision of two aggregates \revise{is not a function of} the relative velocity but of the energy involved in the collision \revise{and on the sizes of the aggregates} \citep{Dominik97}.

If the fragmentation process is efficient under the conditions that govern the dynamics of dust grains in collapsing clouds, it will strongly enhance the abundance of the smaller grains of the size distribution, and therefore drastically affect the ionization balance and the MHD resistivities of the plasma (see Sect.~\ref{sec:Results}). \revise{The effect of fragmentation of small agregates made of very small monomers is expected to be so important for the resistivities of the collapsing cloud that it deserves a study in itself. Asymptotically, we expect the competition of fragmentation and coagulation to lead to evolutionary paths somewhere in between the "no evolution" and coagulation scenarios presented in this article.}

There is currently no model available describing the outcome of a collision between two aggregates composed of small (between 5 and 250 nm) monomers, in particular for the proportion of the grain mass that would be converted into very small fragments. The \DUSTDAP\ code can handle the fragmentation of grains, but the physics encoded is \revise{not relevant to the fragmentation of aggregates} but to the fragmentation of solids \revise{\citep[threshold velocity for fragmentation of the order of a km.s$^{-1}$,][]{Tielens1994}, as it must happen in interstellar shocks \citep{Guillet2009,Guillet2011}. This possibility is not used in this article}. A first approach would be to keep this framework but use a low value for the threshold velocity for fragmentation \revise{(of the order of a few m.s$^{-1}$)}, or use a threshold in the impact energy as recommended by \cite{Dominik97}. 



\subsection{Decrease of the cosmic-ray ionization rate $\zeta$ with the density}\label{sec:CRnH}

For reason of simplicity again, we kept the cosmic-ray ionization rate, $\zeta$, constant throughout the collapse, though it is known from observations that is tends to decrease with increasing column density \citep{Padovani2013}. The overall effect of an hypothetic decreasing value of $\zeta$ with the local density is to decrease the ionisation fraction and  to increase the MHD resistivities (Fig.~\ref{fig:cmp_Zeta}), though this is only an approximation as $\zeta$ depends on the total column density encountered by cosmic-rays during their propagation, and not on the local density. The increase of the resistivities that we infer from the follow-up of dust coagulation in cloud collapse should therefore be reinforced by a decrease of $\zeta$ during the collapse. Note however that the \DUSTDAP\ code computes all variables out-of-equilibrium, making it difficult to estimate from Fig.~\ref{fig:cmp_Zeta} the out-of-equilibrium response of the plasma to a drop of $\zeta$ over a timescale much smaller than the free-fall timescale. This aspect will be investigated in a future work.



\subsection{Grain dynamics}

The impact of grain-grain coagulation on the dust size distribution is very much dependent on the model assumed for the dynamics of dust grains. We have shown for example that the grain velocities induced by hydrodynamical turbulence and those generated by ambipolar diffusion result in very different size distribution, grain growth rate, and feedback on the plasma properties (Fig.~\ref{fig:models}). Using numerical simulations \cite{Pan2015} have reevaluated the models of grain turbulent dynamics for protostellar clouds and protoplanetary disks elaborated since the pioneering work of \cite{Volk80}. They demonstrate that these models generally overpredict the rms velocity of dust grains by a factor of 2. They also show that such dynamics  ignores the enhancement of the coagulation rate induced by increase of the dust-to-gas ratio \citep[an effect caused by the clustering of dust grains, confirmed by other numerical studies, see][]{Hopkins2016,Lee2017,Lebreuilly2019}. Luckily, these two sources of errors cancel out for certain grain sizes, revealing a overall good agreement between the \cite{Pan2015} simulation and the \cite{OC07} model.

Another source of uncertainty in the grain dynamics is the nature of the turbulence assumed for the model (or for the simulation): hydrodynamical (HD) or magnetohydrodynamical (MHD). The model of grain dynamics used here  \citep{OC07} is based on HD turbulence. \cite{YLD04} developped a model of grain dynamics under MHD turbulence. In this model, grains are still accelerated by gas vortices, but these vortices are not isotropic anymore because of the presence of the magnetic field, affecting the dependence of the rms grain velocity with the grain radius and gas density. Magnetic acceleration processes such as gyroresonance \citep[a coupling between magnetic waves and grain gyration,][]{YLD04} also bring their contribution to the total rms velocity of dust grains. The main difference between these MHD \citep{YLD04} and HD \citep{OC07} models of grain dynamics is the magnitude of the rms velocity, not so much its dependence on the grain radius and gas density. Indeed, the injection velocity of the MHD turbulence is assumed to be the Alfven velocity in \cite{YLD04}, which is almost one order of magnitude higher than the sound velocity used in the case of HD turbulence (see Sect.~\ref{sec:turbHD}). This factor difference has a straightfoward, proportional, effect on the coagulation rate. The physics developped in \cite{YLD04} is mainly focused on the warm and diffuse phase of the ISM, with an extension to the grain dynamics in dense clouds at the density of $\nH=10^4\,\cmc$. This density is our initial density before the collapse proceeds. For this reason, and also to facilitate comparisons of our results with other works of dust coagulation in cores, we chose to study the impact of the dynamics produced HD turbulence, not MHD turbulence. Still, our introduction of ambipolar diffusion is already a first step toward the entire inclusion of MHD aspects of grain dynamics, which we leave for a future work.



 


\subsection{Grain growth: comparison with other works}\label{sec:graingrowth}

In this section, we compare our results regarding the grain growth rate with other studies of dust coagulation in static and collapsing cores. 
Figure~\ref{fig:models} has shown that hydrodynamical turbulence can, according to our calculations, make grain grow up to $\sim 1-10\,\mu$m in a free-fall time. 
\cite{Flower2005} obtained similar results using another modified version of the Paris-Durham shock code and an approach based on a equivalent, single size, grain that grows in time. Other authors concentrating on grain growth in grain-grain processes toward the building-up of very large (mm) grains, found higher growth rates in their calculations \citep[\eg][]{WR93,Ossenkopf93,Ormel2009}. 
Coming after numerous detailed studies on the subject, we did not attempt in this paper to propose a realistic study of grain growth in cores. Our intent was to study the coupling between dust evolution and the ionization and resistivities of the plasma, concentrating on the evolution of the smallest grains ($a \ll 0.1\,\mu$m) of the size distribution that a study of grain growth can reasonably ignore. Still, we can make the following comments and comparison with these works. \cite{Ossenkopf93} included in its modeling the follow-up of the grain porosity, also adding the effect of gravitation on the grain dynamics and coulombian effects on the grain-grain collision rate, but worked at constant density, and ignored fragmentation. \cite{WR93} considered fractal grains, included other dynamical processes such as the dust settling along the disk and a modeling of collisional fragmentation of aggregates, and considered both static and collapsing cloud, but like \cite{Ormel2009} ignored the evolution of grains smaller than $0.1\,\mu$m. \cite{Ormel2009}, with a more physical description of the growth of aggregates through fragmentation, coagulation and compaction, started with a single size distribution of bare or ice-coated monomers of $0.1\,\mu$m. At constant density, they obtain little grain growth on a free-fall timescale. Over a longer timescale, higher growth rates are obtained as soon as grain enter into a regime of self-enhanced coagulation generated by the increasing grain porosity. Overall, these studies, which include a detailed description of the physics of evolution of dust agregates, generally find higher growth rates than in our study, as expected \citep{Ormel2009}. \revise{Whether this regime of high porosity exists or not in the presence of ambipolar diffusion with a size distribution of monomers like here is an open question. In any case, this higher growth rate, which is due to the effect of aggregate porosity that we ignored here, is most probably not as essential to the  calculation of the MHD resistivities of the plasma as the effect of fragmentation that we also ignored, because fragmentation more than grain growth controls the number density of the smaller grains} (see Sect.~\ref{sec:frag} for a discussion on the importance of dust fragmentation in the evaluation of the MHD resistivites of the collapsing cloud).




\subsection{Can the magnetic flux be regulated by the abundance of small grains ?}

The feedback of dust coagulation onto the evolution of the ambipolar diffusion velocity through the collapse (Eq.~\eqref{eq:VAD} and Fig.~\ref{fig:VAD_Zeta}) raises the possibility of the self-regulation of the magnetic flux in the parent cloud. The ambipolar diffusion velocity scales with the square of the magnetic field intensity. A higher value for the magnetic field intensity in the parent cloud therefore tends to increase the value of $\vAD$, which increases the coagulation rate between small grains (coupled to the magnetic field) and large grains (coupled to the gas). The removal of small grains in turn decreases the coupling of the magnetic field with the gas, leading to a loss of magnetic flux. The ambipolar diffusion velocity then decreases, which stops the removal of small grains. \revise{In particular, the process described here could limit the magnetic intensity inside dense cores}. 
This possibility, which is described here only qualitatively, could be investigated using MHD simulations \revise{by following the MHD dynamics and coagulation of a simplified grain size distribution composed of two fluids of grains (of 10 nm and 100 nm, for example) and building on the recipies for the grain charges and ionisation equilibrium in a dusty plasma from \cite{Ivlev16}.}

\section{Summary}\label{sec:summary}

In this article, extending the work by \cite{marchand2016}, we studied how the evolution of the dust size distribution may affect the MHD properties of a collapsing cloud. For this purpose, we have used the \DUSTDAP\ code, a numerical tool that was designed to follow the evolution of the dust size distribution and its feedback on the \revise{evolution of the ionisation, chemical content and} dynamics of interstellar shocks \citep{Guillet2011}. The code was amended to follow grain-grain coagulation in an element of volume undergoing isothermal compression in a free-fall collapse. 
For the simplicity of the analysis, we have for the moment ignored the possible production of small grains by fragmentation and craterization in grain-grain collisions \citep{Ormel2009}. The variation of the cosmic-ray ionization rate with the column density \citep{Padovani2013}, an important factor influencing the ionisation and resistivities of the plasma, was also ignored.

\revise{
We have considered two physical models for the grain size-dependent relative velocities responsible for grain-grain colllisions and coagulation in the collapsing cloud : stochastic velocities triggered by hydrodynamical turbulence \citep[following the recipe by][]{OC07}, and systematic velocities induced by ambipolar diffusion \citep[following our approach of grain dynamics in C-type shocks][]{Guillet2007}. In the former model small grains are coupled to the gas, while they are coupled to the magnetic field and therefore decoupled from the gas and from larger grains in the latter.

Here is what we find: in a first phase of the isothermal collapse ($\nH =  10^4-10^8\, \cmc$), grain-grain coagulation primarily removes small grains from the size distribution by sticking them onto larger grains. This depletion process, which happens faster and therefore sooner when ambipolar diffusion-induced velocities are self-consistently taken into account, does not significantly modifies the upper limit of the grain size distribution. In a second phase ($\nH =  10^8-10^{12}\, \cmc$), the mean grain size increases from $\sim 0.1\,\mu$m to $\sim10\,\mu$m owing to the large, turbulent-induced, relative velocities between grains of comparable sizes while ambipolar diffusion has no more impact on the grain dynamics.

We have further studied the impact of grain-grain coagulation onto the evolution of the ionisation, conductivities and resistivities of the plasma, and therefore on the coupling of the magnetic field with the collapsing gas.
The presence of small dust grains modifies the conductivities and resistivities of the plasma both directly and indirectly : directly through their own contribution to the conductivity, and indirectly through their control over the ionisation fraction and therefore over the conductivity of ions. If the total perpendicular conductivity is dominated by ions (as is typically the case in the parent cloud), removing small grains will decrease the Hall and ambipolar diffusion resistivities. If the total perpendicular conductivity is dominated by dust grains (as is the case during the collapse at densities lower than $\sim 10^7\,\cmc$ when starting with a MRN size distribution), removing small grains will on the contrary increase the MHD resistivities. 

Starting with an MRN size distribution in the parent cloud, the amplitude increase of the Hall and ambipolar diffusion resistivities induced by grain-grain coagulation depends on the model of dust dynamics. It is stronger when the drift velocities induced by ambipolar diffusion are taken into account than when only turbulent velocities are considered.
The ambipolar diffusion (resp. Hall) resistivity is comparable (resp. higher by one order of magnitude) in our model where dust grains coagulate during the cloud collapse than in a toy model without coagulation during the collapse where the removal of grains smaller than 0.1$\,\mu$m would have already happened in the parent cloud \citep[truncated-MRN size distribution,][]{zhao2016}. 
It is therefore likely that the disappearance of small grains due to the differential velocity induced by ambipolar diffusion will have a signficant impact on the formation of planet-forming disks. In particular it likely leads to the formation of somewhat larger disks, even when starting with a MRN size distribution in the parent cloud.  
Varying the uniform value assumed for the CR ionization rate $\zeta$ in the $5.10^{-19} - 5.10^{-16}\,$s$^{-1}$ range under the simplifying assumption that the ambipolar diffusion velocity is constant in the parent cloud, we find that the Ohm resisitivity (which is controled by free electrons) is, as expected, very sensitive to $\zeta$, but that the Hall and ambipolar diffusion resistivity are almost independent of it as long as the plasma total perpendicular  conductivity is dominated by dust grains.

Our study confirms that the MHD resistivities of the collapsing core are very sensitive to the abundance of small grains, and therefore to the mechanisms and grain dynamics that drive their production and removal. The possible fragmentation of aggregates in mutual collision being ignored in our analysis, the modeling of the competition between the coagulation and fragmentation of dust grains appears to be the next important step toward a consistent modeling of the feedback of dust evolution on the MHD dynamics of a collapsing core. 
}

\appendix

\section{Coagulation rate in Larson's collapse}\label{A-Larson}

\begin{figure}
\includegraphics[width=\hhsize]{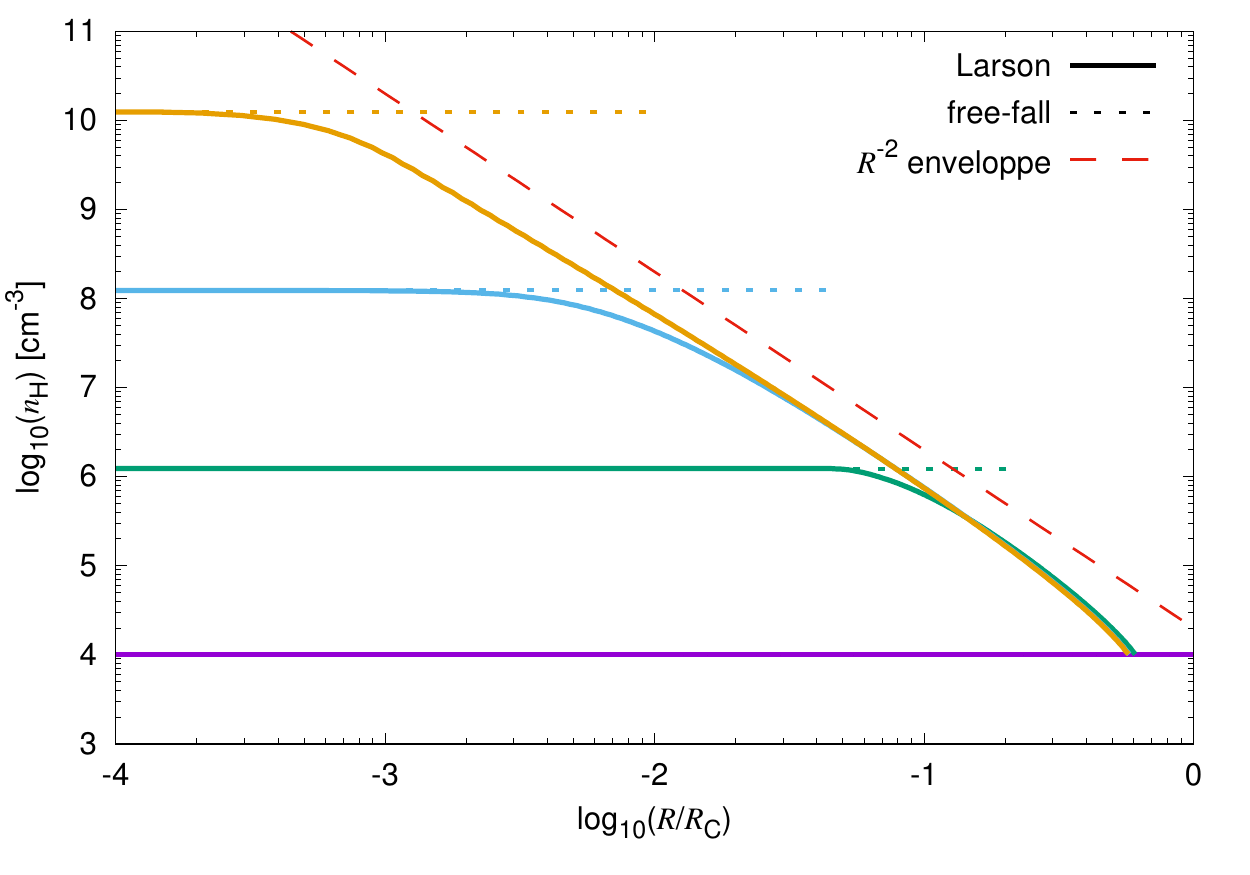}
\includegraphics[width=\hhsize]{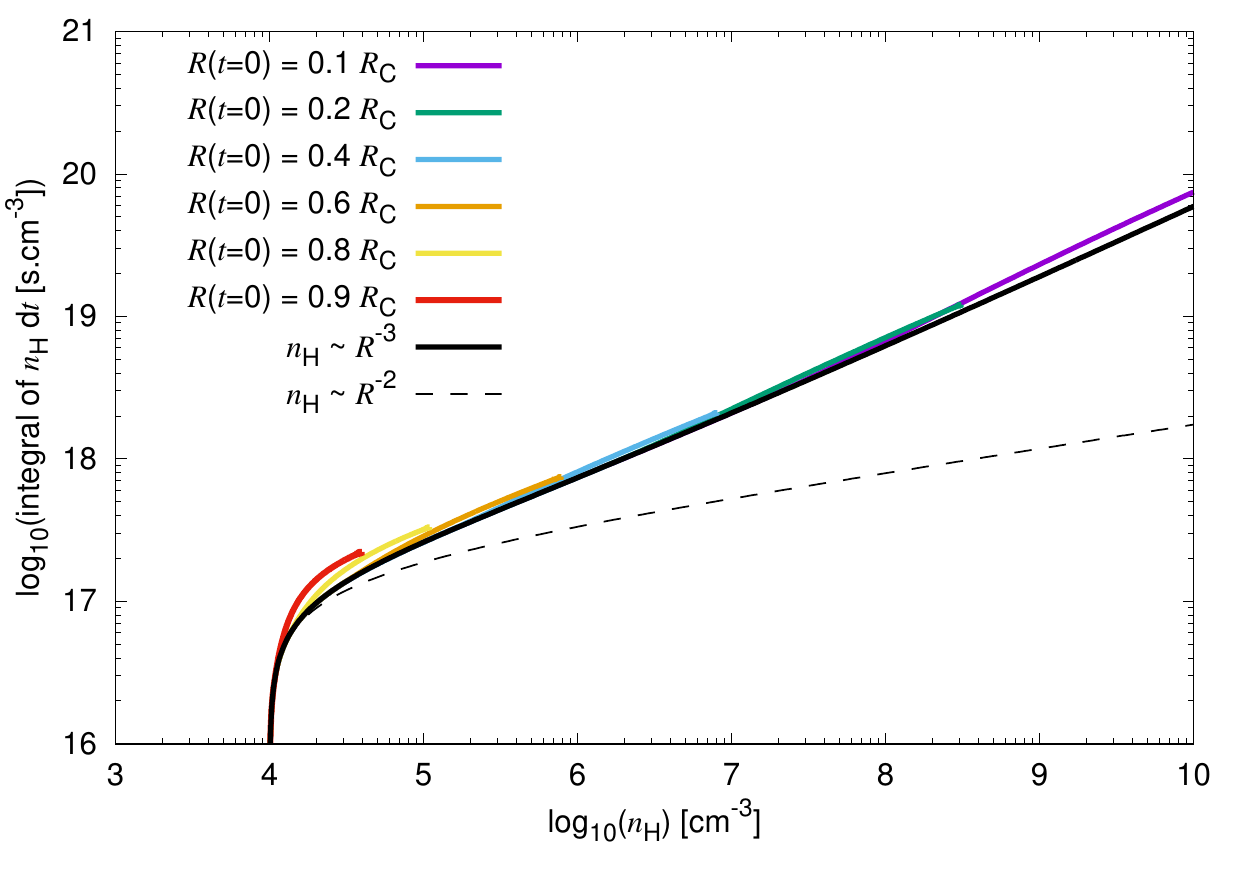}
\caption{
\revise{
\T\ Density profile of the collapsing cloud at four stages of its evolution for the Larson  \citep{Larson1969} scenario (thick lines), and for our simple collapse model of Sect.~\ref{sec:collapse}. 
\B\ $\int \nH\,\d t$ for the Larson collapse for gas cells distributed at different radius before the cloud collapse. Results for models of uniform compression assuming $\nH \propto R^{-3}$ (black thick line) and $\nH \propto R^{-2}$ (black thin line) are overplotted.}
}
\label{Larson}
\end{figure} 

\revise{
In this Section, we demonstrate that our model of a uniform compression of the cloud (Sect~\ref{sec:collapse}) is a good model to follow the dust size distribution induced by grain-grain coagulation through the cloud collapse \emph{up to a given density}, even if it is a bad model to describe the density structure of the cloud \emph{at a given time}.}


\revise{
To test our simple uniform compression model against a better representation of a cloud collapse, we compare the results we obtain with our model with those obtained for the isothermal collapse of a core sensitive to gravitational force and gas pressure \citep{Larson1969}.
We integrate Larson's equation in a Lagrangian approach. 
The cloud density profile is presented in the top panel of Fig.~\ref{Larson} at four epochs of the collapse. In the Larson scenario, we observe a uniform distribution of the gas density near the cloud center, surrounded by an envelop where the density scales as $R^{-2}$ that extends to a distance close to the initial radius $\RC$ of the cloud. On the contrary, in our simple free-fall model that ignores the effect of the pressure of the gas the density is uniform and the cloud dimension rapidly contracts.}

\revise{Our purpose in this article is to estimate the dust size distribution at any density during the collapse, not to describe the density distribution of the cloud. To check the reliability of our model for this purpose, we compare the total number of collisions experienced by a given grain particle in the Larson scenario with that obtained for our simple model.}

\revise{
The number of collisions experienced up to a time $\tau$ in the collapse by a given dust grain with other dust grains is primarily controled by the evolution of the local gas density, which increases by orders of magnitude during the collapse. The scaling of the collisional rate with the relative velocity between impinging particles is of secondary importance for our purpose because these velocities depend on the model used for the grain dynamics and only mildy evolves through the collapse ($\DV \propto \nH^{-1/4}$ in the model by \cite{Ormel2009}, see Fig.~\ref{Vdist-Turb}). To obtain a first order result, the total number of collisions experienced by a given particle during the collapse up to a time $\tau$ can be estimated through the proxy:
\begin{equation}
N(\tau) \equiv \int_0^\tau \nH(t) \,\d t\,.
\end{equation}
The bottom panel of Fig.~\ref{Larson} compares the evolution of $N(\tau)$ in the Larson scenario and for our model of a uniform collapse. For the Larson scenario, the evolution of $N(\tau)$ with the local density is presented for six initial positions of the grain at 0.9, 0.8, 0.6, 0.4, 0.2 and 0.1 times the initial radius $\RC$ of the cloud. We also present the evolution of $N(\tau)$ for our model of a uniform collapse, which by hypothesis does not depend on the initial position of the grain. 
The uniform compression scenario nicely reproduces the results obtained in the Larson scenario for all initial positions of the grain. This is clear for particles close to the center ($R/\RC < 0.6$), but also approximately for particles initially close to the boundary, within a factor 1.2. These results are surprising but easily explained. For the Larson scenario, a gas cell almost stops its compression once reached by the rarefaction wave (top panel of Fig.~\ref{Larson}): time goes on, but the gas density and the dust size distribution do not evolve much anymore. 
The dust size distribution in a Larson collapse is therefore close to the one obtained for a uniform collapse, as long as one compares the results at the same gas density, and not at the same time.}


\section{The coagulation algorithm}\label{A-smol}

The evolution of the grain size distribution by grain-grain coagulation is driven by the Smoluchowski equation \citep{Smol1916,Mizuno1988}:
\begin{eqnarray}
\frac{\d\rho(m,t)}{\d t} &  =  & - \int_{0}^{\infty}\,m\,K(m,m')\,n(m,t)\,n(m',t) \,\d m' \nonumber \\ & + & \frac{1}{2}\int_0^m m\,K(m'-m,m')\,n(m'-m,t)\,n(m',t) \,\d m' \nonumber \\
\end{eqnarray}
where $m$ is the grain mass, $n$ the density of grains, $\rho=m\times n$ the mass density of grains, and $K(m,m')$ the collision \emph{kernel} ($\cmc$ s$^{-1}$) between grains of mass $m$ and $m'$. 

\subsection{Modeling of the size distribution}

The grain size distribution is modeled by $\Nbins$ size bins, spread over the radius range $[\am:\ap]$, or equivalently over the mass range [$\mm:\mp]$, assuming a specific density $\mu$ common to all grain cores. The lower and upper grain mass in bin $k$, $\mkm$ and $\mkp$ respectively, follow a geometrical distribution in $k$ of common ratio $\eta$ \citep[see Appendix A from][]{Guillet2007}: 
\begin{eqnarray}
\eta & = & \exp^{-3\log{(\ap/\am)/\Nbins}} < 1 \\
\mkp & =  & \mp\,\eta^{k-1} \\
\mkm & = & \mp\,\eta^{k} 
\end{eqnarray}

Bin 1 contains the largest grains and bin $N$ contains the smallest grains, of the size distribution. We fix $\am = a_{N-} = 5$ nm, and $\ap = a_{1+} = 100\,\mu$m.
Each bin $k$ contains a mass density $\rhok$ of grain cores. 
At t = 0 bins where grain radius is larger than the MRN are empty. Silicate and carbon grains are mixed in a unique composite distribution of cores.

Following \cite{Mizuno1988}, the mass density of grains in bins $k-1$ and $k+1$ is used to derive the spectral index $\betak$ of the mass distribution within bin $k$:
\begin{equation}
\frac{\d n(m)}{\d m}\propto m^{\betak}\,,
\end{equation}
from which we can derive the average radius ($a_k$), cross-section ($\sigmak$) and mass ($m_k$) of grains in bin $k$. See Appendix A of \cite{Guillet2007} for more details on these calculations. The number of grains in bin $k$ is simply $\nk=\rhok/\mk$.

\subsection{Numerical implementation of the coagulation equation}

To discretize the coagulation equation, we use the following procedure. We will consider collisions between grains of bin $P$ (projectiles) and bin $T$ (targets), $P \ge T$ implying that projectiles can not be larger than targets. 
The coagulation rate between these two bins is 
\begin{equation}
\crate = \nT\,\nP\,K_{\rm P,T} = \nT\,\nP\,\pi \left(\aT+\aP\right)^2\Delta V_{\rm P,T}
\end{equation}
where $\Delta V_{\rm P,T}$ is the rms relative velocity between the projectile and target grains.

Grain-grain coagulation leads to a transfer of mass from the projectile and target bins to other bins at the rate $r$. Taking into account that each bin $k$ contains a size distribution of grains in the size range $[\mkm:\mkp]$, the smallest  coagulated grain, of mass $\mPm+\mTm$, must be transfered to the collector bin of index
\begin{equation}
\coll = 1+ \integer\left(\frac{\log{\left(\frac{\mPm+\mTm}{\mp}\right)}}{\log{\eta}}\right)\,,
\end{equation}
while the largest coagulated grains, of mass $\mPp+\mTp$, must be transfered to collector bins $C-1$ according to the same equation. There will be no transfer of mass from bins $P$ and $T$ to any other bins. 

The fraction $f$ of grain mass transfered to collector bin $C-1$, and therefore $1-f$ to collector $C$, is calculated as follows. We assume that all grains from the projectile bin are identifical to the \emph{average} grain of mass $\mP$. With respect to the size of the larger grains from the target bin, this is most of the time a reasonable assumption.
If $\mP+\mTm \ge \mCp$, all coagulated grains will be transfered to collector bin $C-1$ (and therefore none to collector bin $C$): $f=1$. If $\mP+ \mTp \le \mCp$, all coagulated grains will be transfered to collector bin $C$ (and none to collector bin $C-1$): $f=0$. In the general case, the fraction $f$ of coagulated grains transfered to collector bin $C-1$ is
\begin{eqnarray}
f & = & \frac
{\int_{\mPm}^{\mPp}\,\int_{\mCp-\mP}^{\mTp} (m+m')\, K(m,m')\, m{^\betaP}\,m'^{\betaT} \d m \,\d m'}
{\int_{\mPm}^{\mPp}\,\int_{\mTm}^{\mTp} (m+m')\, K(m,m')\, m{^\betaP}\,m'^{\betaT} \d m \,\d m'}\,.
\end{eqnarray}
There exists analytical expressions for $f$ whenever $K$ is a function of powers of $m$ and $m'$, which is usually the case.

The mass transfer rates from the projectile and target bins to the collector bins are therefore :
\begin{eqnarray}
\frac{\d\rhoP}{\d t}& = & - \mP \crate \\
\frac{\d\rhoT}{\d t}& = & - \mT \crate \\
\frac{\d\rhoC}{\d t}& = & \left(f-1\right) \left(\mP+\mT\right) \crate \\
\frac{\d\rhoP}{\d t}& = & f\,\left(\mP+\mT\right) \crate 
\end{eqnarray}
The total grain mass is conserved. This procedure is reproduced for any pair of projectile bin/target bin where $P \ge T$, \ie\ where the projectile is smaller than the target, or of the same size as.

\subsection{Tests of the coagulation algorithm}\label{TestSmolu}

The Smoluchowski equation possesses self-silimar solutions for the constant ($K(m,m') = 2)$, additive ($K(m,m')=m+m')$, and multiplicative ($K(m,m')=m \times m'$) kernels \citep[\eg][]{MenonPego}.

In log-log plots of the mass density, this self similarity appears clearly
\begin{eqnarray}
\frac{\d \log{\rho^C(m,t)}}{\d \log m} & = & \left(\frac{m}{t}\right)^2\,\exp^{-m/t} \\
\frac{\d \log{\rho^A(m,t)}}{\d \log m} & = & \sqrt{\frac{m\,\exp^{-2t}}{2\pi}}\,\exp^{-m\,\exp^{-2t}/2} 
\end{eqnarray}

At $t=0$, the self-similar solutions express
\begin{eqnarray}
\frac{\d \log^C{\rho(m,t=0)}}{\d m} & = & m\,\exp^{-m} \\
\frac{\d \log^A{\rho(m,t=0)}}{\d m} & = & \revise{\frac{\exp^{-m/2}}{\sqrt{{2\pi m}}}}
\end{eqnarray}
which can be integrated analytically to yield the mass density in each bin $k$:
\begin{equation}
\rhok^C  = \int_\mkm^\mkp m\,\exp^{-m} \,\d m = \left(1+\mkm\right)\exp^{-\mkm} - \left(1+\mkp\right)\exp^{-\mkp}  \\
\end{equation}
and
\begin{equation}
\rhok^A =  \int_\mkm^\mkp \sqrt{\frac{\exp^{-m/2}}{2\pi m}} \,\d m = \erf\left(\sqrt{\frac{\mkp}{2}}\right) - \erf\left(\sqrt{\frac{\mkm}{2}}\right)
\end{equation}
where $\erf$ is the error function.

\begin{figure*}
\includegraphics[width=\hhsize]{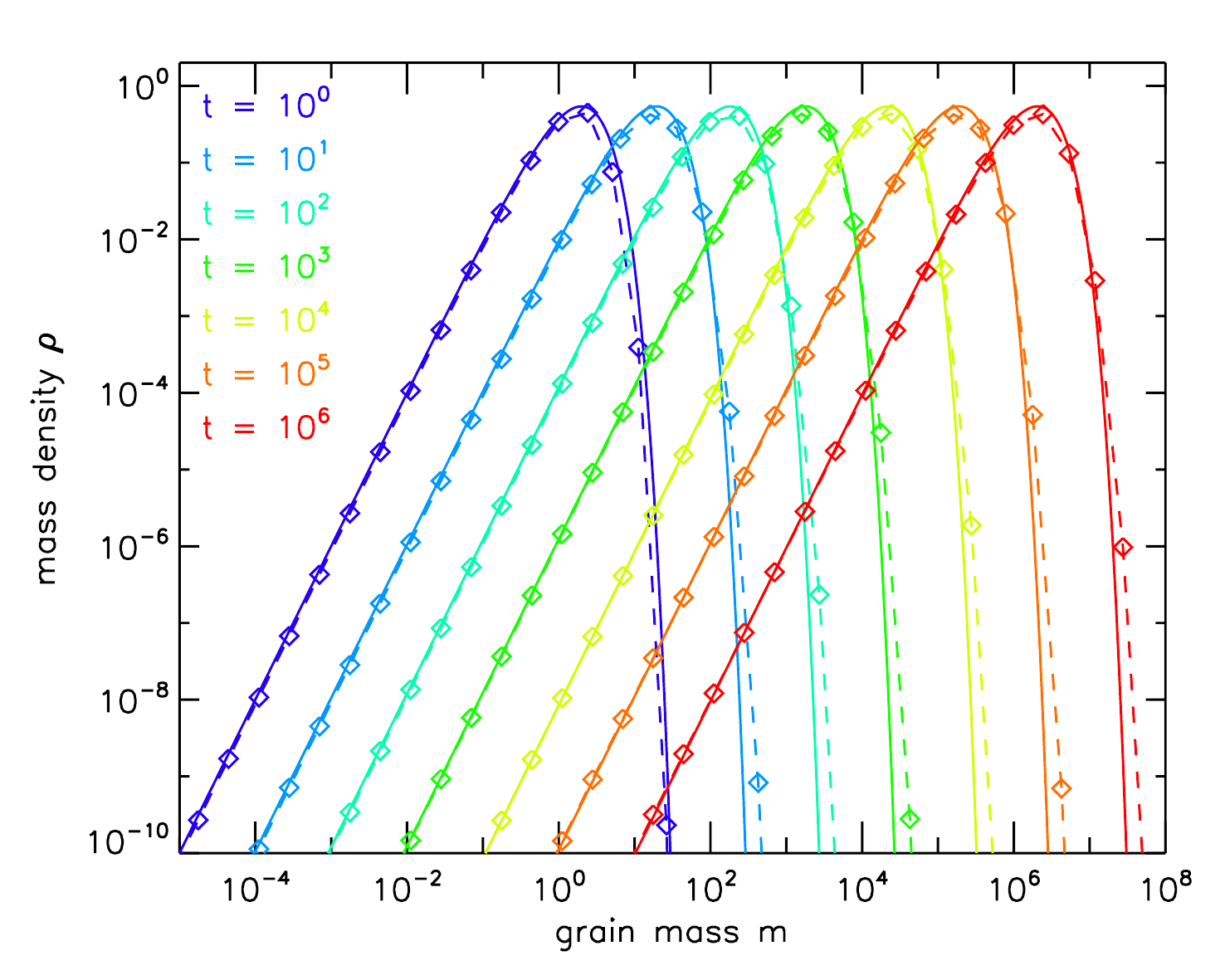}
\includegraphics[width=\hhsize]{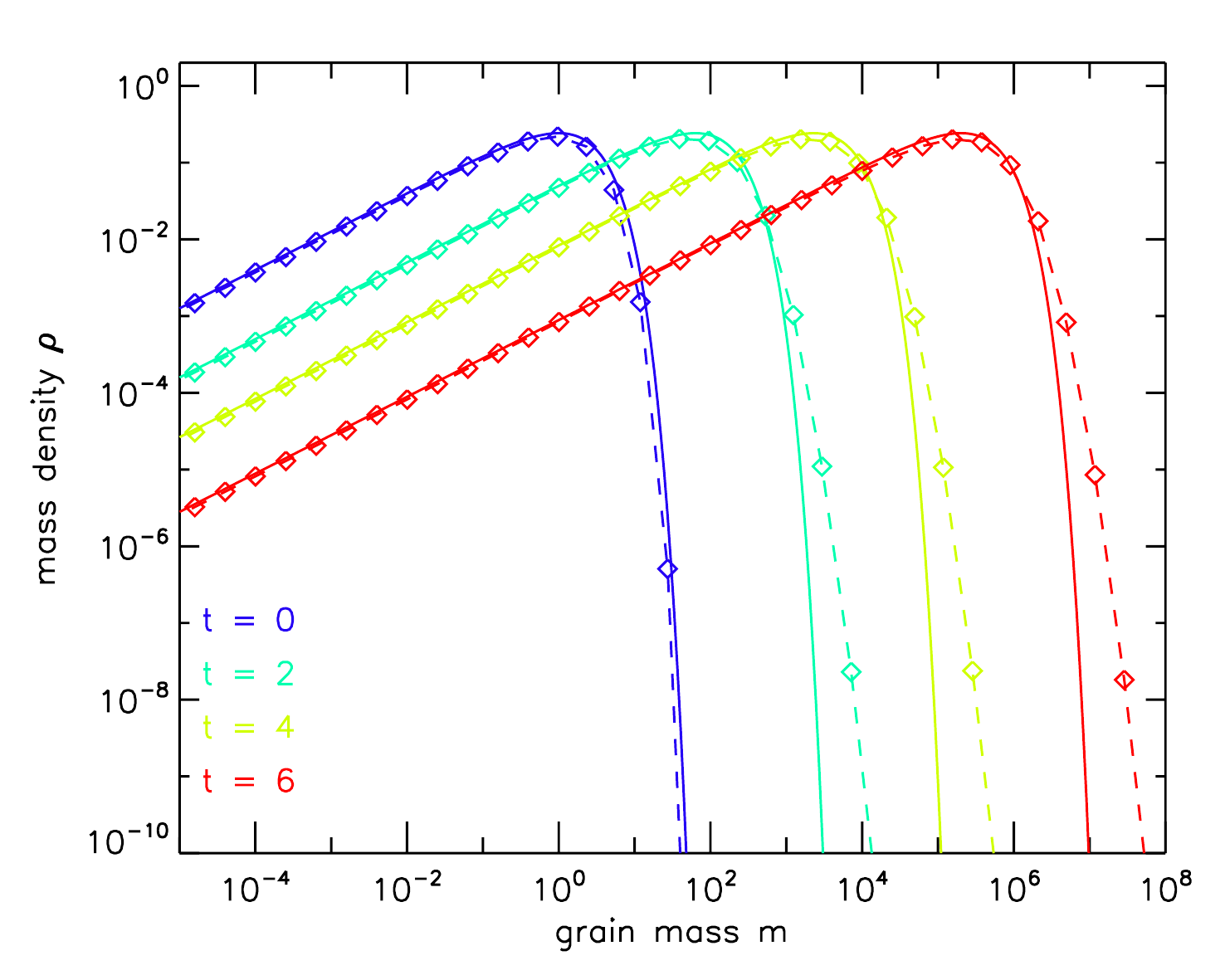}
\caption{Comparison between the analytical (solid) and numerical (dashed) solutions to the Smoluchowski equation, at various adimensioned time steps, in the case of a constant \L\ and additive \R\ kernels, with 35 size bins equally spaced in logarithmic scale over a range $[4\times10^{-6}:4\times10^8]$ in mass ($\eta = 0.4$). Diamonds indicate the mean mass and mass density in each bin. Grain growth is linear with time for the constant kernel, and exponential with time for the additive kernel.}
\label{smolCA}
\end{figure*} 

We test our coagulation algorithm by comparing our numerical results with analytical solutions for the constant and additive kernels (Fig.~\ref{smolCA}). We use 35 bins, equally spaced in logarithmic scale on the range of adimensioned masses $[4\times10^{-6}:4\times10^8]$ ($\eta = 0.4$)\footnote{With a smaller number of bins ($\eta > 0.4)$, the results from the code start to deviate from the analytical solution even at low radius.}. In both cases, our algorithm is able to reproduce analytical results over a $10^6$-fold increase of the average grain mass (a factor 100 in radius). The power-law of the smallest grains is well followed-up, which is particularly important in our study. Because of numerical errors, the exponential decay at large grain radius is overestimated in our simulation. This is however a small effect, which furthermore will have no impact on the ionisation and MHD properties of the gas which are dominated by the small grain population.

\subsection{Convergence tests}

\begin{figure}
\includegraphics[width=\hhsize]{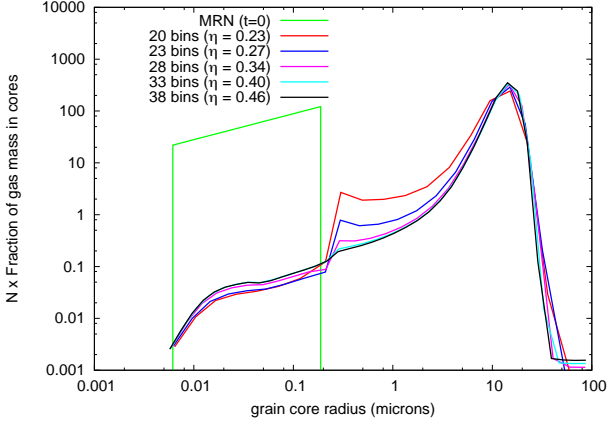}
\caption{Check for the convergence of our coagulation/accretion algorithm at $\nH=10^{12}\,\cmc$, varying the number of logarithmic bins in the 5 nm - 100 $\mu$m range.}
\label{convbins}
\end{figure} 

We test for the convergence of our coagulation algorithm using our standard model, but ignoring grain charge (which does not affect grain coagulation rates in our model). Convergence is obtained as soon as $\eta < 0.4$, as for the self-similar solutions to the Smoluchowski equation (see Sect. \ref{A-smol}). We will therefore use a total of 33 bins:13 bins for the MRN ($0.005\,\mu{\rm m}< a < 0.25\,\mu$m) and 20 empty bins for $0.25 < a < 100\,\mu$m. 

\section{Coagulation rates with a skewed maxwellian velocity distribution}\label{A-shifted}

Following the approach by \cite{Flower2005}, we consider the dynamics of particles having velocities $v_x$, $v_y$ and $v_z$ along $\xaxis$, $\yaxis$ and $\zaxis$ in the lab frame, respectively, each characterized by a normal distribution of zero expectation and same variance $\DV^2\revisem{/3}$. We observe the motion of those particles from a reference frame moving at a velocity $\mu > 0$ along the $\zaxis$ axis. 

Defining 
\revise{
\begin{equation}
\alpha \equiv \sqrt{\frac{3}{2}}\frac{1}{\DV}\,,
\end{equation}
}
the probability distribution functions along for $v_x$, $v_y$, and $v_z$ are
$f(v_x)= \frac{\alpha}{\sqrt{\pi}}\,\exp^{-\alpha^2 v_x^2}$
$f(v_y) = \frac{\alpha}{\sqrt{\pi}}\,\exp^{-\alpha^2 v_y^2}$
and $f(v_z) = \frac{\alpha}{\sqrt{\pi}}\,\exp^{-\alpha^2 \left(v_z-\mu\right)^2}$, respectively. 

The probability distribution of $\mathbf{v}$, the velocity vector measured in the frame of the observer, is $f(\mathbf{v})  =  f(v_x,v_y,v_z)$:
\begin{equation}
f(\mathbf{v}) 
 = \left(\frac{\alpha}{\sqrt{\pi}}\right)^3\,\exp^{-\alpha^2 \left(v^2+\mu^2\right)}\exp^{-2\alpha^2 v_z \,\mu}
\end{equation}

If $v$ is the velocity amplitude measured in the frame of the observer, and $(\theta,\phi)$ the spherical coordinates of the velocity vector, then by definition $v_z = v\cos{\theta}$.
To obtain the probability distribution of the velocity amplitude, $f(v)$, we integrate $f(\mathbf{v})$ over the spherical coordinates $(\theta,\phi)$:
\begin{eqnarray}
f(v) & = & \int_0^{\pi}\int_0^{2\pi} v^2 f\left(\mathbf{v}\right) \sin{\theta}\,\d\theta\,\d\phi \\
& =& 2\pi\left(\frac{\alpha}{\sqrt{\pi}}\right)^3\,\exp^{-\alpha^2 \left(v^2+\mu^2\right)} \int_0^{\pi} v^2\,\exp^{-2\alpha^2 v \mu\cos{\theta}}\,\sin{\theta}\,\d\theta \\
& =& 2\pi\left(\frac{\alpha}{\sqrt{\pi}}\right)^3\,\exp^{-\alpha^2 \left(v^2+\mu^2\right)}\left[\frac{v}{2\alpha^2\mu}\exp^{-2\alpha^2 v \mu\cos{\theta}}\right]_0^{\pi} \\
& =& \frac{1}{\sqrt{\pi}}\,\frac{\alpha}{\mu}\,v\,\exp^{-\alpha^2 \left(v^2+\mu^2\right)}\left(\exp^{2\alpha^2 v \mu}-\exp^{-2\alpha^2 v \mu}\right) \\
& =& \frac{1}{\sqrt{\pi}}\frac{\alpha}{\mu}\,v\,\left(\exp^{-\alpha^2 \left(v-\mu\right)^2}-\exp^{-\alpha^2 \left(v+\mu\right)^2}\right)
\end{eqnarray}

\revise{The mean collisional velocity among those collisions leading to grain-grain coagulation is}
\begin{equation}
\int_0^{\Vlim} v\,f(v)\,\d v = \frac{1}{\sqrt{\pi}}\frac{\alpha}{\mu}\int_0^{\Vlim} v^2\,\left(\exp^{-\alpha^2 \left(v-\mu\right)^2}-\exp^{-\alpha^2 \left(v+\mu\right)^2}\right) \\
\end{equation}
Let us first calculate:
\begin{eqnarray}
\int_0^{\Vlim} v^2\,\exp^{-\alpha^2(v+\mu)^2}\,\d v 
& = & 
\int_{\mu}^{\Vlim+\mu} \left(v^2+\mu^2-2v\mu\right)\exp^{-\alpha^2 v^2}\,\d v 
\end{eqnarray}

We have:
\begin{eqnarray}
\int \exp^{-\alpha^2 x^2}\,\d x &=& \frac{\sqrt{\pi}}{2}\,\frac{\erf{\left(\alpha x\right)}}{\alpha}\\
\int x\,\exp^{-\alpha^2 x^2}\,\d x &=& -\frac{1}{2\alpha^2} \exp^{-\alpha^2 x^2}\\
\int x^2\,\exp^{-\alpha^2 x^2}\,\d x &=& \frac{1}{2\alpha^2} \left(\frac{\sqrt{\pi}}{2}\,\frac{\erf{\left(\alpha x\right)}}{\alpha}-x\,\exp^{-\alpha^2 x^2}\right)
\end{eqnarray}

\begin{widetext}

Then
\begin{eqnarray}
\int_0^{\Vlim} v^2\,\exp^{-\alpha^2(v+\mu)^2}\,\d v 
& = & 
\,\frac{1}{2\alpha^2}
\left[
\left(2\mu-v\right)\exp^{-\alpha^2 v^2}
+\left(2\alpha^2\mu^2+1\right)\frac{\sqrt{\pi}}{2}\,\frac{\erf{\left(\alpha x\right)}}{\alpha} 
\right]_\mu^{\Vlim+\mu} 
\\ & = & 
\frac{1}{2\alpha^2}
\left(
\left(\mu-\Vlim\right)\exp^{-\alpha^2 (\Vlim+\mu)^2}
-\mu\exp^{-\alpha^2 \mu^2}
+\left(2\alpha^2\mu^2+1\right)\frac{\sqrt{\pi}}{2\alpha}
\left[
\erf{\left(\alpha\Vlim+\alpha\mu\right)} 
-\erf{\left(\alpha\mu\right)}
\right]
\right) 
\end{eqnarray}

Similarly,
\begin{eqnarray}
\int_0^{\Vlim}  v^2\,\exp^{-\alpha^2(v-\mu)^2}\,\d v & = & \frac{1}{2\alpha^2}
\left(
\left(-\mu-\Vlim\right)\exp^{-\alpha^2 (\Vlim-\mu)^2}
+\mu\exp^{-\alpha^2 \mu^2}
+\left(2\alpha^2\mu^2+1\right)\frac{\sqrt{\pi}}{2\alpha^2}
\left[
\erf{\left(\alpha\Vlim-\alpha\mu\right)} 
+\erf{\left(\alpha\mu\right)}
\right]
\right) 
\end{eqnarray}

As a consequence,
\begin{eqnarray}
&& \int_0^{\Vlim} v\,f(v)\,\d v  =   
\frac{1}{\sqrt{\pi}}\frac{\alpha}{\mu} \,\frac{1}{2\alpha^2} \times
\nonumber \\
&& 
\left[
-\left(\mu+\Vlim\right)\exp^{-\alpha^2 (\Vlim-\mu)^2}
-\left(\mu-\Vlim\right)\exp^{-\alpha^2 (\Vlim+\mu)^2}
+2\mu\exp^{-\alpha^2\mu^2}
+\left(2\alpha^2\mu^2+1\right)\frac{\sqrt{\pi}}{2\alpha}
\left(
\erf{\left(\alpha\Vlim-\alpha\mu\right)} 
-\erf{\left(\alpha\Vlim+\alpha\mu\right)} 
+2\erf{\left(\alpha\mu\right)}
\right) \right] 
\nonumber 
\\
&= & \frac{1}{\alpha\sqrt{\pi}} \left[
\exp^{-\alpha^2\mu^2}
-\frac{1}{2}
\left(
\exp^{-\alpha^2 (\Vlim-\mu)^2}\left[1+\frac{\Vlim}{\mu}\right]
+\exp^{-\alpha^2 (\Vlim+\mu)^2}\left[1-\frac{\Vlim}{\mu}\right]
\right)
\right)
+\left(\mu 
+\frac{1}{2\alpha^2\mu}\right)
\left(
\erf{\left(\alpha\mu\right)}
-\frac{\erf{\left(\alpha\Vlim+\alpha\mu\right)}-\erf{\left(\alpha\Vlim-\alpha\mu\right)}}{2}
\right]
\nonumber \\
&= & \frac{1}{\alpha} 
\left[
\frac{1}{\sqrt{\pi}}
\left(
\exp^{-\xi^2}
-\frac{1}{2}
\left(
\exp^{-\left(\chi-\xi\right)^2}
\left[1+\frac{\chi}{\xi}\right]
+\exp^{-\left(\chi+\xi\right)^2}
\left[1-\frac{\chi}{\xi}\right]
\right)
\right)
+\left(\xi+\frac{1}{2\xi}\right)\,h(\chi,\xi)
\right]
\end{eqnarray}

where $\xi=\alpha\mu$, $\chi = \alpha\Vlim$, and $h(\chi,\xi)= \erf{\left(\xi\right)} - \left(\erf{\left(\chi+\xi\right)} -\erf{\left(\chi-\xi\right)}\right)/2$.

\end{widetext}

Here are the two asymptotic cases:
\paragraph{No drift} 
When $\mu$ (and $\xi$) tend toward zero, $h(\chi,\xi)/\xi$ tends towards $\erf'(0) - \erf'(\chi)$ where $\erf'$ is the first derivative of $\erf$, \ie\ toward $\frac{2}{\sqrt{\pi}}\left(1-\exp^{-\chi^2}\right)$. We get:
\begin{eqnarray}
\int_0^{\Vlim} v\,f(v)\,\d v & = & 
\left[\frac{1}{\sqrt{\pi}}\left(1-\left(1+2\chi^2\right)\exp^{-\chi^2}\right)+\frac{1}{\sqrt{\pi}}\left(1-\exp^{-\chi^2}\right)\right] \nonumber \\
& = &  \frac{2}{\alpha\sqrt{\pi}} \left(1-\left(1+\chi^2\right)\exp^{-\chi^2}\right)\,,\nonumber \\
& =&   \sqrt{\frac{8}{3\pi}}\,\DV \left(1-\left(1+\chi^2\right)\,\exp^{-\chi^2}\right)\,, \label{eq:dVturb}
\end{eqnarray}
\revise{which translates into Eq.~(A.11) of \cite{Flower2005} once their convention is taken into account : $\delta v$ from their Eq.~(A.11) is in their convention the FWHM of the gaussian \emph{absolute} velocity components along $x$, $y$ and $z$ for each target and projectile, while $\DV/\sqrt{3}$ is under our convention the \emph{rms} of the gaussian for the \emph{relative} target-projectile velocity components. : $\Delta V = \Delta v\, / \sqrt{(4\ln{2})/3}$}.

\paragraph{Pure drift of velocity $\mu$ (no turbulence)} 
For the special case when $\DV = 0$ \revise{($\alpha, \xi, \chi \rightarrow +\infty$)}, we get:
\begin{equation}
\int_0^{\Vlim} v\,f(v)\,\d v   =   \frac{1}{\alpha} \xi\,h(\chi,\xi) =  \frac{1}{\alpha} \alpha\mu= \mu\,,
\end{equation}
as expected.

\begin{acknowledgements}

We thank the referee for her/his comments that helped improved the clarity of the manuscript. 
V.G thanks A. Ivlev and P. Lesaffre for stimulating discussions. This work was supported by the Programme National PCMI of CNRS/INSU.

\end{acknowledgements}

\bibliographystyle{aa}
\bibliography{biblio}

\end{document}